%% file: main.tex
\documentclass[lettersize,journal,hidelinks]{IEEEtran} %
\IEEEoverridecommandlockouts
\pdfoutput=1
\usepackage[utf8]{inputenc} 
\usepackage[T1]{fontenc}
\usepackage{url}
\usepackage{ifthen}
\usepackage{cite}
\usepackage[cmex10]{amsmath} %
\usepackage[dvipsnames]{xcolor}
\usepackage{amssymb}
\usepackage{multirow}
\usepackage{siunitx}
\usepackage[nolist]{acronym}
\usepackage{tikz}

\usetikzlibrary{spy,decorations.pathreplacing,calligraphy,calc,arrows, arrows.meta, dsp,matrix,patterns}
\usepackage{pgfplots}
\usepgfplotslibrary{groupplots,fillbetween}
\usepackage[font=footnotesize]{caption}
\usepackage{subcaption}
\usepackage{bm}
\usepackage{etoolbox} %
\tikzset{>=latex}
\usepackage[ruled]{algorithm2e}
\input{macros.tex}

\input{acronyms.tex}
\input{corporateColours.tex}

\input{plotcyclelists.tex}
\usepackage[dvipsnames]{xcolor}
\definecolor{mrgreen}{RGB}{101, 201, 0}

\newcommand\neu[1]{{\textcolor{black}{{#1}}}}

\newcommand\neur[1]{{\textcolor{black}{{#1}}}}

\def\msize{2.0pt}
\def\lw{0.8pt}

\begin{document}

\title{Joint Detection and Decoding:\\ A Graph Neural Network Approach}

\author{Jannis Clausius \IEEEmembership{Graduate Student Member, IEEE}, Marvin Rübenacke \IEEEmembership{Graduate Student Member, IEEE}, Daniel Tandler \IEEEmembership{Graduate Student Member, IEEE}, Stephan ten Brink \IEEEmembership{Fellow, IEEE}   
\thanks{
J. Clausius, M. Rübenacke, D. Tandler and S. ten Brink are with the Institute of Telecommunications, University of Stuttgart, 70569 Stuttgart, Germany (e-mail: clausius@inue.uni-stuttgart.de; ruebenacke@inue.uni-stuttgart.de;
tandler@inue.uni-stuttgart.de;
tenbrink@inue.uni-stuttgart.de)
                    Parts of this paper have been presented at 2024 IEEE Int. Symp. Inf. Theory (ISIT) \cite{jddisit}.\\
                  This work is supported by the German Federal Ministry of Education and Research (BMBF) within the project Open6GHub (grant no. 16KISK019).}}

\makeatletter
\patchcmd{\@maketitle}  %
{\addvspace{0.5\baselineskip}\egroup}
{\addvspace{-1.5\baselineskip}\egroup}
{}
{}
\makeatother

\maketitle
\begin{abstract}
Narrowing the performance gap between optimal and feasible detection in \ac{ISI} channels, this paper proposes to use \acp{GNN} for detection that can also be used to perform \ac{JDD}.
For detection, the \ac{GNN} is build upon the factor graph representations of the channel, while for \ac{JDD}, the factor graph is expanded by the Tanner graph of the \ac{PCM} of the channel code, sharing the \acp{VN}.
A particularly advantageous property of the \ac{GNN} is a) the robustness against cycles in the factor graphs which is the main problem for \ac{SPA}-based detection, and b) the robustness against \ac{CSI} uncertainty at the receiver.
Consequently, a fully deep learning-based receiver enables joint optimization instead of individual optimization of the components, so-called end-to-end learning.
Furthermore, we propose a parallel flooding schedule that also reduces the latency, which turns out to improve the error correcting performance. 
The proposed approach is analyzed and compared to state-of-the-art baselines for different modulations and codes in terms of error correcting capability and latency.
The gain compared to \ac{SPA}-based detection might be explained with improved messages between nodes and adaptive damping of messages.
For a higher order modulation in a high-rate \ac{TDD} scenario the \ac{GNN} shows a, at first glance, surprisingly high gain of 6.25\,dB compared to the best, feasible non-neural baseline.
\end{abstract}

\begin{IEEEkeywords}
Graph neural networks, detection, turbo detection, equalization, deep learning.
\end{IEEEkeywords}

\acresetall
\section{Introduction}

\IEEEPARstart{I}{nter-symbol} interference (ISI) can appear in many scenarios, e.g., band-limited channels after filtering at the receiver \cite{proakis2001digital} or single carrier transmission systems used in low power communication \cite{You2020}. 
In recent years, \ac{NN}-based algorithms entered the landscape of detection for \ac{ISI} channels besides the maturity of \ac{APP}-based and \ac{LMMSE}-based algorithms.
\ac{APP}-based approaches usually yield good performance, but suffer from high complexity.
In contrast, \ac{LMMSE}-based algorithms usually are of low complexity, but struggle with achieving the optimal \ac{MAP} performance.
The neural contender may offer a solution in between with good performance, yet feasible complexity.
While this class of detection algorithms have been introduced already 35 years ago \cite{gibson89mlpequalizers}, the increase in computing power has renewed interest in it. 
Model-based \acp{NN} appear to be particularly attractive, as they exploit domain knowledge  \cite{schlezinger2023model} and thereby promise lower complexity, better generalization, and require smaller training data sets than their purely data-driven counterparts.

In \cite{schlezinger2023model}, they are categorized into \emph{model-aided \acp{NN}} and \emph{\ac{NN}-aided inference}.
Model-aided \acp{NN} consist of handcrafted \acp{NN} for a given problem. They may share similarities with classical algorithms, e.g., \ac{CNN}-based detection \cite{huang21extrinsic,xu18nnjed} is similar to the \ac{LMMSE} algorithm and \ac{RNN}-based detection \cite{kechriotis94rnn,farsad2018neural,plabst24sicrnn} resembles the \ac{BCJR} \cite{bahl1974optimal} algorithm. 
In contrast, in \ac{NN}-aided inference, a classical algorithm is augmented by learned components, e.g., using \acp{MLP} as \acp{FN} in the \ac{BCJR} algorithm \cite{shlezinger2020bcjrnet}.
A particularly well suited basis for \ac{NN}-aided inference are factor graphs based on the \ac{FFG} \cite{forney72observationmodel} or the \ac{UFG} \cite{ungerboeck74observationmodel}.
Here, the \ac{SPA} yields a sub-optimal performance \cite{kschischang01factorgraph,colavolpe05bpdetection,colavolpe11ungerboeckdetection}, opening up the the possibility of neural augmentation.
For instance, in \cite{liu21nnfactornode}, a neural node was added to the factor graph, and in \cite{schmid22neuralBP, nachmani2016learning}, trainable weights were added. 
\neu{However, these neural augmentations might not suffice in cases of strong oscillation and amplification caused by severe \ac{ISI}.}

By proposing to replace nodes and edges by \acp{MLP}, we continue this path resulting in a \ac{GNN} \cite{gnn} right on the border between model-based \acp{NN} and \ac{NN}-aided inference.
\Acp{GNN} were successfully applied in channel decoding \cite{zhang2023factor, satorras2021neural, cammerer2022gnn} and \ac{MIMO} detection \cite{Scotti2020GraphNN,Cammerer2023ANR,Kosasih22GnnMIMO}.
We believe \acp{GNN} are a great fit for detection in \ac{ISI} channels.
Firstly, they promise robust computations to mitigate the negative effect of the vast amount of short cycles in the factor graphs for \ac{SPA}.
\neu{In particular, increasing robustness to amplification and oscillation. }
Secondly, in contrast to the \ac{MIMO} scenario, the sparse connectivity between nodes in the \ac{FFG} and \ac{UFG} reduces the complexity of the \ac{GNN} drastically, making it computationally attractive.
\neu{
For \ac{MIMO}, the \acp{GNN} use fully connected adjacency matrices \cite{Scotti2020GraphNN,Kosasih22GnnMIMO}.
This work aligns with \acp{GNN} based on factor graphs, as in  \cite{cammerer2022gnn,zhang2023factor}, but also considers the detection problem as part of the factor graph, enabling \ac{JDD} and various schedules for message passing.
Note that \acp{GNN} are not unique in leveraging the local structure of the \ac{ISI} problem for efficient detection. \Acp{RNN} and \acp{CNN} can do that as well \cite{huang21extrinsic,plabst24sicrnn}. }
\input{tables/table_1}

\neu{However, in contrast to \acp{RNN} or \acp{CNN}, \acp{GNN} can also leverage the sparse nature of an \ac{LDPC} code for decoding \cite{cammerer2022gnn} enabling \ac{JDD} and improving receiver performance over \ac{SDD}. }
\Ac{JDD} discriminates from \ac{SDD} by code/decoder aware detection or detection aware decoding.
One practical way to approach the asymptotic performance limit of \ac{JDD} is using \ac{SDD} components and allowing feedback from the decoder to the detector, resulting in \ac{TDD} \cite{Douillard95turbo}.
However, in the non-asymptotic regime (i.e., for finite block lengths), \ac{JDD} can be superior to \ac{TDD} by leveraging joint optimization instead of separate optimization of the components.
\neu{The concepts are visualized in Fig.~\ref{fig:systemmodel}.}
\ac{JDD} based on a model-free approach was investigated in \cite{ye17NNjed}, however, limited to ultra-short block lengths ($N=16$).
In contrast, the model-based approach in \cite{tsai20JEDbcjrNet} scales to longer block lengths, but is limited by the exponential nature of the \ac{BCJR} algorithm.

For short block lengths ($N=128$), we propose a true \ac{JDD} scheme that combines two \acp{GNN}; one for detection and one for decoding. 
Both \acp{GNN} are connected by their \acp{VN} to form a combined factor graph that allows information exchange between detection and decoding in both directions at the same time\neu{, called the flooding schedule}.
The joint \ac{GNN} constitutes a fully deep learning-based receiver that can be jointly optimized in an end-to-end fashion resulting in a \ac{JDD} scheme \cite{xu18nnjed,ye17NNjed}.
\neu{Since the connectivity in the graphs for detection and decoding are sparse, respectively, the joint graph is also sparsely connected. This enables the efficient use of \acp{GNN}.}
In \cite{henkel19jointequalLDPCdecoding}, such a combined graph was used with a decision feedback detector and an \ac{SPA} decoder.
For longer block lengths, we resort to a \ac{TDD} scheme where only detection is performed by a \ac{GNN} and decoding is performed by the conventional \ac{SPA}.
This means, the \ac{GNN} is trained to output \acp{LLR} and is neither aware of the code nor fine-tuned to the decoder.

Note that throughout the paper, we demonstrate \acp{NN} that are not optimized in terms of complexity.
The purpose is to show the performance limit that a neural architecture can achieve.
Thus, the \acp{NN} are over-parameterized enabling smooth training and reproducible results.
Our contributions are as follows:
\begin{itemize}
    \item We demonstrate that \acp{GNN} are capable of matching the \ac{MAP} detection performance for severe \ac{ISI} with \ac{BPSK} and outperforming state-of-the-art baselines with higher order modulation. \neu{Further, the robustness of the \ac{GNN} towards \ac{CSI} uncertainty at the receiver is demonstrated, i.e., noisy \ac{CIR}, and towards changing distributions of the \ac{CIR}. }
    \item We analyze the learned messages in the \ac{GNN} and give intuitive explanations for the performance gains.
    \item We propose \acp{GNN} for \ac{JDD} where the respective factor graphs are connected by their common \acp{VN} and allow joint processing.
    The proposed approach is compared to state-of-the-art baselines.
    \item Finally, we propose a flooding schedule for the \acp{GNN} on the joint factor graph instead of an iterative/sequential schedule, improving both performance and latency.
\end{itemize}

\begin{figure}[tbp]
  \centering
  \resizebox{.89\columnwidth}{!}{\input{./figures/systemmodel.tikz}}
  \caption{Block diagram of the system model with an \ac{ISI} channel (convolution with \acf{CIR} $\hv$ and addition of noise $\zv$) and \acfp{GNN} for \acf{SDD}, \acf{TDD} and \ac{JDD}.}
  \label{fig:systemmodel}
  \vspace{-0.6cm}
\end{figure}
\section{Preliminaries}

\input{tables/table_2}

\subsection{System Overview}

Fig.~\ref{fig:systemmodel} shows the block diagram of the considered system model. 
We assume 5G \ac{LDPC} channel encoding \neu{of the bits $\uv \in \{0,1\}^K$ with code dimension $K$ into codeword $\cv \in \{0,1\}^N$ } bits of length $N$ and $M$-\ac{QAM} \neu{into the transmission symbols $\xv \in \{0,1\}^{N_{\xv}}$ of length ${N_{\xv}}$}, where $M$ denotes the alphabet size.
Note that $M=2$ and $M=4$ results in \ac{BPSK} and \ac{QPSK}, respectively.
The \ac{ISI} channel is modeled as a \ac{TDL} with \ac{CIR} $\hv$  and \ac{AWGN}.
Hence, the discrete-time channel can be mathematically described by \cite{forney72observationmodel}
\begin{equation} \label{eq:channel_eq}
    y_i = \sum_{l = 0}^L h_l x_{i-l} + z_i
\end{equation}
with channel memory $L$, transmitted coded symbols $x_i$ and where $z_l \sim \mathcal{CN} (0,\sigma^2)$ denotes \ac{i.i.d.} Gaussian noise.
Furthermore, we assume that the symbols $x_i$ for $i \in [ -L+1,-1] \cup [N_\xv-1,N_\xv+L]$ are known at the receiver, i.e., fixed to zero, where $N_\xv$ denotes the transmission length.
Thus, we can rewrite equation (\ref{eq:channel_eq}) to
$ \yv = \mathbf{H} \widetilde{\xv} + \zv $
with the Toeplitz matrix $\mathbf{H} \in \CC ^{(N_\xv+L) \times (N_\xv+2L)}$ called the \emph{channel matrix}, equivalent transmit sequence $\widetilde{\xv} \in \CC^{N_\xv+2L}$ and $\zv  \in \CC ^{N_\xv+L}$. 
The goal of the receiver is to estimate the bit-wise \acp{APP} $P(u_i|\yv, \Hm)$ in the form of \acfp{LLR} $\ell_i = \log\frac{P(u_i=0|\yv, \Hm)}{P(u_i=1|\yv, \Hm)}$. 

\subsection{Sum-Product Algorithm}
Consider the factorization of a joint probability function $f(\mathcal{X})$ into a product of $J$ factors $g_j (\mathcal{X}_j)$, i.e., ${f(\mathcal{X}) = \prod_{1 < j \leq J } g_j (\mathcal{X}_j)}$,
where $\mathcal{X}$ corresponds to a set of random variables $\{x_1, \dots, x_N\}$ and $\mathcal{X}_j \subset \mathcal{X}$.
One can construct a corresponding factor graph consisting of \acp{VN} $\mathcal{V}$ and factor nodes $\mathcal{F}$  by assigning each variable $x_j$ a \ac{VN} $V_j \in \mathcal{V}$, each function $g_i$ a \ac{FN} $F_i \in \mathcal{F}$, and introducing an edge connecting between $V_j$ and $F_i$ if $x_j$ is an argument of $g_i$ \cite{kschischang01factorgraph}.
The marginals of $f$, i.e., $f(x_i)$ for each $x_i \in \mathcal{X}$, can be efficiently computed using the message passing algorithm \ac{SPA} on the factor graph.
The algorithm operates by sending messages from \acp{VN}/\acp{FN} to \acp{FN}/\acp{VN}, denoted as $m_{V_i \rightarrow F_j}$/$m_{F_j \rightarrow V_i} $.
In the log-domain, the messages are determined using 
\begin{align*}
   m_{V_i \rightarrow F_j} =& \sum_{F_k \in \mathcal{F}(V_i) \backslash \{F_j\}}  m_{F_k \rightarrow V_i} \\
   m_{F_j \rightarrow V_i}  =& \underset{\sim \{V_i\}}{\operatorname{\max} ^\star} \left( \ln{g_j(\mathcal{X}_j)} + \hspace{-2em} \sum_{ V_k \in \mathcal{V}(F_j) \backslash \{V_i \}} \hspace{-2em} m_{V_k \rightarrow F_j}  \right) 
\end{align*}
where $\mathcal{V}(F_j)$ and $\mathcal{F}(V_i)$ denote the \emph{neighborhood} of \acp{VN} of $F_j$ and of \acp{FN} of $V_i$, respectively and $\operatorname{\max}^\star_{\sim \{V_i\}}$ denotes the Jacobian logarithm \cite{Robertson95suboptimalMAP} applied over all messages except $m_{V_i \rightarrow F_j}$.
The Jacobian logarithm is defined as $\operatorname{\max}^\star(a_1,\dots,a_N) = \log(\operatorname{e}^{a_1}+\dots+\operatorname{e}^{a_N})$.

\subsection{Factor Graph-based Detection}
With an appropriate factorization of the joint symbol \ac{APP} $P(\xv| \mathbf{y,H})$, one can obtain an equivalent factor graph representation.
Here, the \acp{FN} correspond to $ \mathbf{H} \widetilde{\xv}$ and the \acp{VN} to the symbols in $\widetilde{\xv}$ (the first and last $L$ \acp{VN} are \emph{virtual} \acp{VN}).
\Ac{SPA} is used to calculate the symbol-wise \ac{APP} $P(x_i|\yv,\mathbf{H})$.
Using Bayes' theorem, we obtain
\begin{equation}\label{eq:prob}
     P(\xv | \yv) \propto P(\yv | \xv) P(\xv) \propto \exp \left(-\frac{\lVert \yv - \mathbf{H} \widetilde{\xv} \rVert^2 }{2\sigma^2} \right).
\end{equation}
Direct application of the chain rule yields
\begin{equation}
\label{eq:forney_factorization}
     P(\yv | \xv) \propto \prod_{i=1}^{N_\xv} \exp \left(-\frac{|| y_i - \sum_{l = 0}^L h_l \widetilde{x}_{i-l}||^2}{2\sigma^2} \right).
\end{equation}
The factor graph corresponding to the factorization of equation  (\ref{eq:forney_factorization}) was first proposed in \cite{forney72observationmodel} and is referred to as \ac{FFG} in the following.
\neu{An example for the \ac{FFG} is shown in Fig.~\ref{fig:ffg}. Intuitively, each \ac{FN} corresponds to a channel observation and is connected to the $L+1$ \acp{VN} that directly impact the channel observation. In other words, the rows of $\Hm$ correspond to \acp{VN}, the columns to \acp{FN} and are connected if the corresponding entry in $\Hm$ is non-zero.  }

By rewriting equation  (\ref{eq:prob}) and using the substitutions $\mathbf{G} := \mathbf{H}^\mathrm{H} \mathbf{H}$ and $\bm{\chi}:=\mathbf{H}^\mathrm{H} \yv$ we obtain an alternative factorization of the \ac{APP} \cite{colavolpe11ungerboeckdetection, ungerboeck74observationmodel}:
\begin{align}
     P(\yv | \xv) \propto& \prod_{i=1}^{N_\xv} \left[ F_i^{\mathrm{UFG}} (x_i) \prod_{\substack{{j=1} ,j \neq i}}^{N} \vspace{-2em}I_{i,j}(x_k,x_j)\right] \nonumber\\ 
     F_i^{\mathrm{UFG}} (x_i) :&= \exp \left( \frac{1}{2\sigma^2} (2\chi_i x_i - G_{i,i}|x_i|^2) \right) \label{eq:F_ufg} \\
    I_{i,j} (x_i,x_j) :&= \exp \left( - \frac{1}{2\sigma^2} G_{i,j}x_ix_j \right). \nonumber
\end{align}
The corresponding factor graph is referred to as \ac{UFG} in the following and is shown in Fig.~\ref{fig:ufg}.
\neu{In the \ac{UFG}, two \acp{VN} are connected via a \ac{FN}, if their corresponding symbols interfere with each other, i.e.  they are no more than $L+1$ positions apart.}
\neu{The distinctive feature compared to the \ac{FFG} is that a \ac{FN} only connects two \acp{VN} instead of $L+1$.}

\begin{figure}[tbp]
\centering
\input{./figures/ufg.tikz}
\caption{\neu{\Acf{UFG} for a channel with memory $L=2$. Input to the \acp{VN} are the estimates $\bm{\chi}:=\mathbf{H}^\mathrm{H} \yv$. The graph is sparse, since the \ac{FN} degree is $d_\mathrm{F}=2$ and the \ac{VN} degree is $d_\mathrm{V}=2L$.}
}
\label{fig:ufg}
\end{figure}

Applying \ac{SPA} to these factor graphs is not guaranteed to converge, since \ac{ISI} channels may contain short cycles in their graphs \cite{colavolpe05bpdetection}.
\neu{However, for \ac{ISI} channels with $L\ll N_\xv$ the graphs are sparse allowing for efficient graph-based processing. The number of edges divided by the maximal possible number of edges scales with $\frac{L}{N_\xv}$, disregarding constants.  }

\subsection{EXIT Charts}\label{sec:exit}
To visualize and analyze the behavior of iterative systems, i.e., \ac{TDD}, we use \Ac{EXIT} charts \cite{brink01exit}. In this case, we iterate between two components: a detector and a decoder. The information gain is quantified by the extrinsic mutual information $I_\mathrm{E}=I(\cv;\boldsymbol{\ell}_\mathrm{E})$ of the extrinsic \acp{LLR} $\boldsymbol{\ell}_\mathrm{E}$ and the transmitted codeword $\cv$ that is tracked over multiple iterations.
As long as for both components the information gain is larger than the \emph{a priori} information $I_\mathrm{A}$, the iterative process increases information. 
Assuming that we know the transfer characteristics $T(\cdot)$ of the detector $I_\mathrm{E} = T_\mathrm{Det}( I_\mathrm{A},\gamma)$, where $\gamma$ is the \ac{SNR}, and the decoder $I_\mathrm{E} = T_\mathrm{Dec}(I_\mathrm{A})$, the \ac{EXIT} chart plots $I_\mathrm{E}$ over $I_\mathrm{A}$ for the detector and $I_\mathrm{A}$ over $I_\mathrm{E}$ for the decoder, e.g., as in Fig.~\ref{fig:exit_chart}.
Decoding is successful if  $T_\mathrm{Dec}(I_\mathrm{A})\to 1$ before $T_\mathrm{Dec}$ and $T_\mathrm{Det}$ intersect.

In general, the transfer characteristic $T(I_\mathrm{A},\cdot)$ of a component can be derived analytically or empirically. 
For most detectors and decoders, an analytical description of $T(\cdot)$ is unknown, thus, we resort to an empirical one.
To empirically evaluate $I_\mathrm{E}=T(I_\mathrm{A},\cdot)$, we need to compute the extrinsic \acp{LLR} 
\begin{equation*}
    [ \boldsymbol{\ell}_{\mathrm{E}}]_i = [f(\boldsymbol{\ell}_{\mathrm{A},\sim i},\cdot)]_i
\end{equation*}
where $[\cdot]_i$ denotes the element with index $i$, $f(\cdot)$ is the processing of the component, e.g, a detector or a decoder, $\boldsymbol{\ell}_{\mathrm{A},\sim i}$ is the vector of \emph{a priori} \acp{LLR} with average mutual information $I_\mathrm{A}$ omitting index $i$.
In other words, the component $f(\cdot)$ needs to be evaluated once for each position $i$ with $\ell_i=0$ .
For true \ac{APP} components, the calculation of $\boldsymbol{\ell}_\mathrm{E}$ can be simplified to
\begin{equation}\label{eq:extrinsic_map}
    \boldsymbol{\ell}_\mathrm{E}=\boldsymbol{\ell}_\mathrm{T}-\boldsymbol{\ell}_\mathrm{A} = f(\boldsymbol{\ell}_\mathrm{A},\cdot)-\boldsymbol{\ell}_\mathrm{A},
\end{equation}
where $\boldsymbol{\ell}_\mathrm{T}$ refers to the total \acp{LLR}.

During inference of a considered system, $\boldsymbol{\ell}_\mathrm{A}$ are the extrinsic \acp{LLR} from the respective other component. 
However, we would like to characterize the component in question $T(I_\mathrm{A},\cdot)$ without any influence of the other component.
To circumvent this, the \emph{a priori} \acp{LLR} $\boldsymbol{\ell}_\mathrm{A}$ can be modeled as a \ac{BPSK} modulated side channel transmission, where the \ac{SNR} is chosen according to the desired \emph{a priori} information $I_\mathrm{A}$.
The resulting \acp{LLR} can be sampled from
\begin{align}
\label{eq:generate_prior}
\boldsymbol{\ell}_\mathrm{A} &\sim \mathcal{N}\left((1-2\uv)\cdot \mu(I_\mathrm{A}), 2\mu(I_\mathrm{A})\mathbf{I}\right)
\end{align}
where, $\uv\in\{0,1\}^K$ are the transmitted bits and  $\mu(I_\mathrm{A})\approx\frac{1}{2}\left(-\frac{1}{H_1}  \log_2 \left( 1-I_\mathrm{A}^{\frac{1}{H_3}}  \right)  \right)^\frac{1}{H_2}$ with constants $H_1=0.3073$, $H_2=0.8935$ and $H_3=1.1064$ \cite{bran05conv}.
Finally, the extrinsic \ac{BMI} can be estimated by \cite{hagenauer2002turbo}
\begin{equation*}
       I_\mathrm{E} =  1-\operatorname{E}_{\uv,\boldsymbol{\ell}_\mathrm{E}}\left\{\log_2(1+\operatorname{e}^{-(1-2\uv)\cdot \boldsymbol{\ell}_\mathrm{E}}) \right\}.
\end{equation*}

\subsection{Information Rates}

The achievable transmission rate of \ac{JDD} is given by the average mutual information \cite{shannon48theory}
\begin{equation*}
    R_\mathrm{JDD} = \frac{1}{K} I(\Um;\hat{\Um}) =  \frac{1}{K}  \left( H(\Um) - H(\Um|\hat{\Um}) \right),
\end{equation*}
where $H(\Um)$ is the joint entropy of the \neu{random variable $\Um$ associated with the uncoded bits} and $H(\Um|\hat{\Um})$ is the conditional entropy of $\Um$ given \neu{the estimate} $\hat{\Um}$.
Due to the complexity reduction of the divide and conquer approach, \ac{SDD} is commonly used in practice, which relates in the limit
\begin{align*}
    R^\star_\mathrm{JDD} &= \lim_{K\to\infty} \frac{1}{K}  \left( H(\Um) - H(\Um|\hat{\Um}) \right)\\
    &\geq   
    \lim_\mathrm{K\to\infty} \frac{1}{K} \sum_i \left( H(U_i) - H(U_i|\hat{\Um}) \right)\\
    &= R^\star_\mathrm{SDD}.
\end{align*}
If all $U_i$ and $U_j$ are conditionally independent knowing $\hat{\Um}$,  meaning $H(U_i|\hat{\Um},U_j)=H(U_i|\hat{\Um}) ~ \forall i\neq j$, then equality follows.

\ac{TDD} is a scheme to approach  $R^\star_\mathrm{JDD}$ with methods from \ac{SDD}. 
The area theorem \cite{ashikhmin04area} states that the area under the curve of the \ac{EXIT} characteristic of an optimal detector $T^\star_\mathrm{Det}(I_\mathrm{A})$ is equal to the \ac{JDD} rate
\begin{equation*}
    R^\star_\mathrm{JDD} \geq R^\star_\mathrm{TDD} = \int_0^1 T^\star_\mathrm{Det}(I_\mathrm{A})dI_\mathrm{A}.
\end{equation*}
While equality only holds for an optimal detector and if $I_\mathrm{A}$ comes from an binary erasure side channel, practice showed that \ac{AWGN} side channel information, which resembles soft feedback from a decoder, results in approximately the same characteristic.
Furthermore, to achieve $R^\star_\mathrm{JDD}$, an optimal channel code of an infinite length with
a matched \ac{EXIT} characteristic $T^\star_\mathrm{Dec}(I_\mathrm{A})$ is required.
Here, matched means that $ T^\star_\mathrm{Det}(I_\mathrm{A})- T^{-1}_\mathrm{Dec}(I_\mathrm{A}) \to 0^+$.
Since $T^\star_\mathrm{Det}(I_\mathrm{A})$ is weakly monotone increasing in $I_\mathrm{A}$
\begin{align*}
    R^\star_\mathrm{JDD} \geq R^\star_\mathrm{TDD} &= \int_0^1 T^\star_\mathrm{Det}(I_\mathrm{A})dI_\mathrm{A}\\
    &\geq \int_0^1 T^\star_\mathrm{Det}(0)dI_\mathrm{A} = T^\star_\mathrm{Det}(0) = R^\star_\mathrm{SDD}.
\end{align*}

Note that in the following, we take the point of view of a detector, meaning we assume \ac{i.i.d.} symbols $X_i$ resulting in $H(\mathbf{X})=\sum_i H(X_i)$.
Until here we assumed optimal components, e.g., \ac{MAP} detectors with output distribution $P_i=P_\mathrm{MAP}(X_i|\Ym, I_\mathrm{A})$. 
However, in this work we explore non-optimal detectors with output distribution $Q_i=P_\mathrm{Det}(X_i|\Ym,I_\mathrm{A})$.
The achievable rate  \cite{bocherer2017achievable,cammerer2019tcom} is then 
\begin{align*}
    R &= \frac{1}{N_\xv}\sum_i I(X_i;Q_i) = \frac{1}{N_\xv}\sum_i \left( H(X_i) - \mathcal{L}(X_i,Q_i) \right)\\
    &= \frac{1}{N_\xv}\sum_i \left( I(X_i;\Ym|I_\mathrm{A}) - D_\mathrm{KL}(P_i || Q_i) \right)
\end{align*}
where $\mathcal{L}(X_i,Q_i)$ is the cross-entropy function and $ D_\mathrm{KL}(\cdot || \cdot)$ is the Kullback-Leibler divergence.
For $I_\mathrm{A}=0$, it follows $R^\star_\mathrm{SDD}-R=\sum_i D_\mathrm{KL}(P_i || Q_i)$ and for turbo detection  $R^\star_\mathrm{TDD}-R= \sum_i \int_0^1 D_\mathrm{KL}(P_i || Q_i)dI_\mathrm{A}$, quantifying the rate loss in terms of the difference between the optimal output distribution $P_i$ and the output distribution of the detector $Q_i$.

Finally, as in practice bit-metric decoding is commonly used, e.g., in mobile communication standards, we are interested in the \ac{BMI} regarding the bits $C$.
This means the measure of interest for the bit estimates $Q^\mathrm{Bit}_i$ is 
\begin{align}
    R_\mathrm{SDD}^\mathrm{BMI} &= \frac{1}{N} \sum_i I(C_{i}|\Ym) \nonumber\\
    &=  1-\frac{1}{N} \sum_i \mathcal{L}_\mathrm{BCE}(C_i,Q^\mathrm{Bit}_i)\label{eq:loss_bmi}\\
    R_\mathrm{TDD}^\mathrm{BMI} &= \int_0^1 T(I_\mathrm{A})d I_\mathrm{A} \nonumber\\
    &= \int_0^1 1-\frac{1}{N} \sum_i \mathcal{L}_\mathrm{BCE}(C_i,Q^\mathrm{Bit}_i) d I_\mathrm{A} \label{eq:tdd_bmi},
\end{align}
where $\mathcal{L}_\mathrm{BCE}$ is the \ac{BCE}.

\vspace{2cm}
\section{\neu{Graph Neural Networks for Detection}}
\label{sec:GNN_eq}

\begin{figure}[tbp]
  \centering
  \input{./figures/ffg.tikz}
  \caption{\Acf{FFG} with \ac{GNN} elements are presented for an exemplary pair of \acf{VN}/\acf{FN} for a channel with memory $L=2$. Blue nodes represent \acfp{MLP}. 
  For \ac{LDPC} decoding the \acp{FN} are usually referred to as check nodes and the channel observations $\yv$ are input to the \acp{VN}.}
  \label{fig:ffg}
\end{figure}
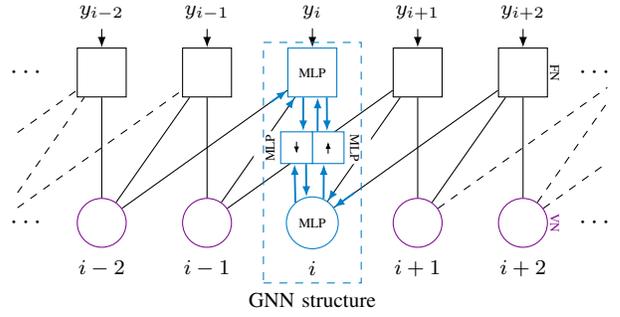

\neu{In this section, we describe the application of \acp{GNN} for detection of \ac{ISI} channels based on the \ac{FFG} and the \ac{UFG}. This resembles a \ac{SDD} scenario, where we consider the detection \acp{GNN} independently of the channel code assuming \ac{i.i.d.} inputs bits and no a priori information ($I_\mathrm{A}=0$) from the decoder. Thus, we evaluate the achievable rate, shown in equation (\ref{eq:loss_bmi}), as measure of  soft output quality of the detection, giving a good indicator on the performance, if one would use an outer channel code. }

\subsection{Graph Neural Network-based Detection}
The bipartite \ac{GNN} framework from \cite{cammerer2022gnn} can be directly applied to the \ac{FFG} and \ac{UFG} for detection. 
In each iteration of the \ac{GNN}, first the \acp{FN} are updated, then the edges towards \acp{VN}, followed by the \acp{VN} themselves, and finally the edges towards \acp{FN}.
As shown in Fig.~\ref{fig:ffg}, the \ac{GNN} assigns learnable parametric functions $f_{\boldsymbol{\theta}}(\cdot)$ to each \ac{VN}, each \ac{FN} and each directed edge.
Here, this parametric function is implemented by a \ac{MLP} with trainable weights $\boldsymbol{\theta}$.
The weights $\boldsymbol{\theta}$ are shared for each type of node or directed edge in the graph, i.e., there is a set $\boldsymbol{\theta}_V$ for \acp{VN}, $\boldsymbol{\theta}_F$ for \acp{FN}, $\boldsymbol{\theta}_{F\rightarrow V}$ for edges directed towards \acp{VN}, and $\boldsymbol{\theta}_{V\rightarrow F}$ for edges directed towards \acp{FN}.
Each \ac{MLP} represents a node state update function or an edge state update function.
The state of a node or edge is represented by a $d$-dimensional vector, where $d$ is the so called \emph{feature size}.
For simplicity, all nodes and edges share the same value of $d$.
\neu{The inference is summarized in Algorithm~\ref{alg:inference}.}
\begin{algorithm}[h]
\caption{GNN-Inference
}
	\SetAlgoLined
	\SetKwInOut{Input}{Input}
	\SetKwInOut{Output}{Output}
	\SetKwBlock{Repeat}{repeat}{}
	\SetKwFor{RepTimes}{For}{do}{end}
	\DontPrintSemicolon
	\Input{Channel observations $\yv$\\
	Trainable parameters $\boldsymbol{\theta}$:\\
	    - Input embedding $\Wm $\\
	    - Read-out $\vv $\\
	    - Variable to factor weights $\boldsymbol{\theta}_{V\rightarrow F}$\\
	    - Factor to variable weights $\boldsymbol{\theta}_{F\rightarrow V}$\\
	    - Variable attributes $\{\gv_{V_i}\}\forall V_i \in \mathcal{V}  $\\
	    - Factor attributes $\{\gv_{F_i}\}\forall F_i \in \mathcal{F}  $\\
	    - Message attributes $\{\gv_{V_i\rightarrow F_j},\gv_{F_j\rightarrow V_i}\}\forall (V_i,F_j) \in (\mathcal{V} ,\mathcal{F}(V_i))  $\\
	    }
	\Output{$\boldsymbol{\ell}$ \tcp{LLRs}} 
	\BlankLine
	\tcp{Embedding}
	\BlankLine
    $\sv_{V_i}^{(1)} \gets \mathbf{0}, \quad \forall V_i \in \mathcal{V}$\\
    $\sv_{F_i}^{(1)} \gets \Wm \cdot [\operatorname{real}( y_i),\operatorname{imag}( y_i)]^\mathrm{T}, \quad \forall F_i\in \mathcal{F}$\\
	\BlankLine
	\tcp{Message passing}
	\BlankLine
	\RepTimes{$t\in [1,...,N_\mathrm{It}]$}{
	\tcp{Factor to variable messages}
	\RepTimes{$(V_i,F_j) \in (\mathcal{V} ,\mathcal{F}(V_i))$}{
    $\mv_{F_j \rightarrow V_i} \gets f_{\boldsymbol{\theta}_{F\rightarrow V}}\left( \sv_{F_j}^{(t)}, \sv_{V_i}^{(t)} ,\gv_{F_j\rightarrow V_i} \right)$\\}
    \tcp{Variable nodes}
    \RepTimes{$V_i  \in \mathcal{V}$}{
    $\sv^{(t+1)}_{V_i} \gets  f_{\boldsymbol{\theta}_{V}} \left( \sv^{(t)}_{V_i}, \frac{1}{|\mathcal{F}(V_i)|}\sum_{F_j \in\mathcal{F}(V_i)} \mv_{F_j \rightarrow V_i} , \gv_{V_i}   \right)$\\}
    \tcp{Variable to factor messages}
    \RepTimes{$(V_i,F_j) \in (\mathcal{V} ,\mathcal{F}(V_i))$}{
$\mv_{V_i \rightarrow F_j} \gets f_{\boldsymbol{\theta}_{V\rightarrow F}}\left( \sv_{V_i}^{(t+1)} ,  \sv_{F_j}^{(t)}, \gv_{V_i\rightarrow F_j} \right)$\\}
\tcp{Factor nodes}
\RepTimes{$F_j  \in \mathcal{F}$}{
$\sv^{(t+1)}_{F_j} \gets  f_{\boldsymbol{\theta}_{F}} \left( \sv^{(t)}_{F_j} , \frac{1}{|\mathcal{V}(F_j)|}\sum_{V_i \in\mathcal{V}(F_j)} \mv_{V_i \rightarrow F_j} , \gv_{F_j}   \right)$\\}
    \BlankLine}
    \tcp{Read-out}
    \BlankLine
    \RepTimes{$V_i  \in \mathcal{V}$}{
	$\ell_i \gets \vv^\mathrm{T}\sv_{V_i}$}
	
	\label{alg:inference}
\end{algorithm}
The updated state $\sv^{(t+1)}_{V_i}$ in iteration $t+1$ of \ac{VN} $V_i$ is calculated by
\begin{equation}\label{eq:gnn_vn}
    \sv^{(t+1)}_{V_i} = f_{\boldsymbol{\theta}_{V}} \left( \sv^{(t)}_{V_i}, \frac{1}{|\mathcal{F}(V_i)|}\sum_{F_j \in\mathcal{F}(V_i)} \mv_{F_j \rightarrow V_i} , \gv_{V_i}   \right),
\end{equation}
 where $\mathcal{F}(V_i)$ is the set of connected \acp{FN}, $\mv_{F_j \rightarrow V_i}$ is the state of the edge from $F_j$ to $V_i$, and $\gv_{V_i}\in \RR^d$ is a trainable attribute of node $V_i$.
In a similar fashion, the update of \ac{FN} states $\sv^{(t+1)}_{F_j}$ is calculated by
\begin{equation}\label{eq:gnn_fn}
    \sv^{(t+1)}_{F_j} = f_{\boldsymbol{\theta}_{F}} \left( \sv^{(t)}_{F_j} , \frac{1}{|\mathcal{V}(F_j)|}\sum_{V_i \in\mathcal{V}(F_j)} \mv_{V_i \rightarrow F_j} , \gv_{F_j}   \right),
\end{equation}
where $\mathcal{V}(F_j)$ is the set of connected \acp{VN} to \ac{FN} $F_j$.
Between the \ac{VN} and \ac{FN} updates, the edges are updated according to
\begin{align*}
     \mv_{V_i \rightarrow F_j} &= f_{\boldsymbol{\theta}_{V\rightarrow F}}\left( \sv_{V_i} ,  \sv_{F_j}, \gv_{V_i\rightarrow F_j} \right),   \\ 
    \mv_{F_j \rightarrow V_i} &= f_{\boldsymbol{\theta}_{F\rightarrow V}}\left( \sv_{F_j}, \sv_{V_i} ,\gv_{F_j\rightarrow V_i} \right).
\end{align*}
While the node states $\sv$ and message states $\mv$ are updated during inference, the attributes $\gv$ are constant during inference, but are learned during training.
For initialization a normal distribution is chosen $\gv\sim\mathcal{N}(\mathbf{0},\mathbf{I})$.
In \cite{cammerer2022gnn}, the node and edge attributes were set to $\mathbf{0}$ as they did not improve the performance for decoding linear channel codes. 
However, in the case of detection, the edge attributes are crucial, since different edges fulfill different tasks, i.e., in classical \ac{SPA} detection the messages originating from a \ac{FN} are calculated with different functions.
Each edge function represents a different tap of the \ac{CIR}, thus, needs a marginalization over different variables.
The taps are the same for all \acp{FN}, meaning that not every edge needs an individual attribute vector. 
The periodicity of the edge type is $N_\mathrm{p} = \max( d_\mathrm{F}, d_\mathrm{V})$ where $d_\mathrm{F}$ and $d_\mathrm{V}$ are the factor node degree and variable node degree, respectively.
As a result, the number of individual feature vectors is $N_\mathrm{p}=L+1$ for the \ac{FFG} and $N_\mathrm{p}=2L$ for the \ac{UFG}.
This enables length flexibility in $N_\mathbf{x}$ of the \ac{GNN}, since nodes and edges can be added to the graph with the same weights as the existing ones.

Before and after inference on the graph, the channel outputs are projected to the feature space of the \ac{GNN} and the inference results are projected to \acp{LLR}, respectively.
Before inference, $\yv$ is linearly projected to the $d$ dimensional \ac{FN} state $\sv_{F_i} = \Wm \cdot [\operatorname{real}( y_i),\operatorname{imag}( y_i)]^\mathrm{T}$, where $\Wm \in \mathbb{R}^{d\times2}$ is a trainable projection matrix.
The \ac{VN} states are initialized with $\sv_{v_i} =\mathbf{0}$.
After inference, the \ac{VN} states are projected to \acp{LLR} by $\ell_i = \vv^\mathrm{T}\sv_{_i}$, where $\vv\in \mathbb{R}^{d}$ denotes a learnable projection vector.
All \acp{MLP} and trainable matrices are initialized by a normal Glorot distribution \cite{Glorot2010UnderstandingTD}.
Finally, a sigmoid function $\sigma_{\mathrm{sigmoid}}(\cdot)$ converts the \acp{LLR} to probabilities $q_i = \sigma_{\mathrm{sigmoid}}(\ell_i)$.
Note that the conversion from feature vectors to probabilities can be done after any number of iterations, allowing for adjustment to the desired latency, complexity, or performance.

\neur{Further, instead of using a normalized sum for the aggregation of messages in equation (\ref{eq:gnn_vn}) and (\ref{eq:gnn_fn}), one could use the self-attention mechanism resulting in a \ac{GAT} \cite{veličković2018graphattentionnetworks}.  }

\subsection{Factor Graph Neural Network-based Detection}
 A less complex variant of the \ac{GNN} is the \ac{FGNN} proposed in \cite{zhang2023factor}.
 The key difference is that the nodes only calculate the aggregation function and omit the \ac{MLP} processing shown in equation~(\ref{eq:gnn_vn}) and equation~(\ref{eq:gnn_fn}).
 While this reduces the expressive potential of the node functions, the \ac{FGNN} was shown to recover the max-log \ac{SPA} and a low-rank approximation of it.
 The other, smaller  difference is the consideration of the edge attributes. 
In a first step the node states are processed by a \ac{MLP} unaware of the edge attributes.
Secondly, the output is multiplied with a  feature extraction matrix specific to the edge, which is generated by a \ac{MLP}.
This results in message update functions
\begin{align*}
     \mv_{V_i \rightarrow F_j} &= f_{\boldsymbol{\theta}_{E}}\left(  \gv_{V_i\rightarrow F_j} \right) f_{\boldsymbol{\theta}_{V\rightarrow F}}\left( \sv_{V_i} ,  \sv_{F_j} \right),   \\ 
    \mv_{F_j \rightarrow V_i} &=f_{\boldsymbol{\theta}_{E}}\left(  \gv_{V_i\rightarrow F_j} \right) f_{\boldsymbol{\theta}_{F\rightarrow V}}\left( \sv_{F_j}, \sv_{V_i} \right),
\end{align*}
where $f_{\boldsymbol{\theta}_{E}}$ is a \ac{MLP} with weights $\boldsymbol{\theta}_{E}$ generating the feature extraction matrix. 
In \cite{zhang2023factor}, it is mentioned that an \ac{MLP} can be added to the nodes to increase expressiveness. 
Here, we do not do that, since the difference to the \ac{GNN} would be practically non existent.

\subsection{Channel Agnostic Embedding}
Previously, we used the linear embedding function 
$\sv_{F_i} = \Wm \cdot [\operatorname{real}( y_i),\operatorname{imag}( y_i)]_\mathrm{T}$
from the channel outputs $\yv$.
However, the learned embedding matrix $\Wm$ only implicitly depends on the \ac{CIR} $\hv$.
If $\hv$ is constant during training, $\Wm$ can overfit to the \ac{CIR}, but fails to work for varying \ac{CIR}. 
The goal is to have an embedding function that is aware of $\hv$ and processes the observations $\yv$ accordingly.
Here, we propose three solutions to the problem. 
\paragraph{Embedding of Log-Likelihoods}
The first solution is feeding channel agnostic log-likelihoods to the \ac{GNN}.
Simplifying equation (\ref{eq:forney_factorization}),
the $M^{L+1}$ log-likelihoods per observation $y_i$ for the \ac{FFG} can be calculated by 
\begin{equation*}
    \log\left(\mathrm{P}(y_i|\neur{\Tilde{x}_i,...,\Tilde{x}_{i-L}},\hv)
    \right) 
    =- \frac{|| y_i - \sum_{l = 0}^L h_l \Tilde{x}_{i-l}||^2}{2\sigma^2}.
\end{equation*}
Since the log-likelihoods are a sufficient statistic and contain implicitly the complete knowledge of the \ac{CIR}, using them is optimal but infeasible for large $M$ or $L$.
\neu{For the \ac{UFG}, initializing the \acp{VN} of the \ac{GNN} with equation (\ref{eq:F_ufg}) scales linear in $M$. }

\paragraph{Neural Channel State-aware Embedding}
The second solution is replacing the embedding matrix $\Wm$ by a small \ac{MLP} where the \ac{CIR} $\hv$ is an input
\begin{equation*}
    \sv_{F_i} = f_{\boldsymbol{\theta}_E} ( \operatorname{real}( y_i),\operatorname{imag}( y_i), \operatorname{real}(\hv), \operatorname{imag}(\hv)).
\end{equation*}
The idea is that $f_{\boldsymbol{\theta}_E}$ can approximate the log-likelihood $\log\left(\mathrm{P}(\neur{y_i|\Tilde{x}_i,...,\Tilde{x}_{i-L}},\hv)\right)$
with low complexity by a $d$-dimensional embedding vector. 
Especially for a high \ac{SNR},  $\log\left(\mathrm{P}(\neur{y_i|\Tilde{x}_i,...,\Tilde{x}_{i-L}},\hv)\right)$ is sparse in terms of entries greater than $0$.
Intuitively, either the observation is close to a symbol combination and is highly likely, or the likelihood vanishes. 
Since most symbol combinations lay directly upon the observation, the result is a sparse likelihood.
This facilitates the intuition that a low-dimensional approximation can be learned efficiently.
\neu{Note, for the \ac{UFG} $y_i$ can be replaced by $\chi_i$.}

\paragraph{Constant Channel Transformation}
The third idea is to transform the varying channel matrix $\Hm$ to a constant channel $\Tilde{\Hm}$ via the \ac{CCT} receive filter  \cite{falconer_adaptive_1973}
\begin{equation}\label{eq:cs_input}
    \Rm = \Tilde{\Hm} \left(\Hm^\mathrm{H}\Hm+\frac{1}{\sigma^2} \mathbf{I} \right)^{-1} \Hm^\mathrm{H}.
\end{equation}
If $\Tilde{\Hm}$ is constructed as in \cite{rusek12optimalshortening} and the memory of $\Tilde{\Hm}$ is smaller than the memory of $\Hm$, then this procedure is called \textit{channel shortening}. 
Here, we do not desire a smaller memory, but a constant $\Tilde{\Hm}$ which, subsequently, can be detected by the \ac{GNN}.
As all operations in equation (\ref{eq:cs_input}) are differentiable, $\Tilde{\Hm}$ can be directly learned during training. 
$\Tilde{\Hm}$ is initialized as a zero matrix with a band diagonal of ones with size $L+1$.
Finally, the new input to the embedding layer is $\Tilde{\yv} = \Rm\yv$.
This method scales well, as the complexity is independent on the memory $L$ or the alphabet size $M$, but is sub-optimal. 
The transformation is only optimal for Gaussian inputs and an optimized choice of $\Tilde{\Hm}$ based on $\Hm$. Here, neither is the case.

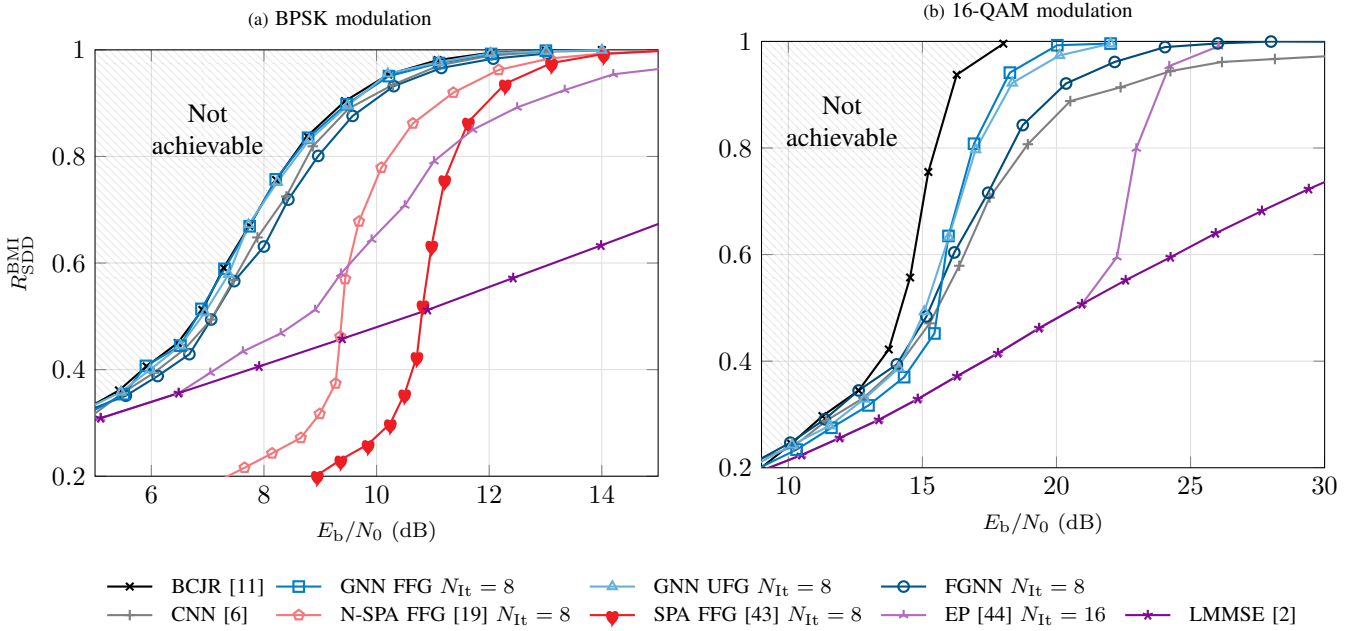
\begin{figure*}[t]
\begin{subfigure}[]{0.33\textwidth}
	\centering
	\caption{\footnotesize \ac{BPSK} modulation}
	\input{./tikz/bmi_uncoded_bpsk.tikz}
	\label{fig:bmi_uncoded_bpsk}
\end{subfigure}
\begin{subfigure}[]{0.33\textwidth}
	\centering
	\caption{\footnotesize $16$-QAM modulation}
	\input{./tikz/bmi_uncoded_qam16.tikz}
	\label{fig:bmi_uncoded_qam16}
\end{subfigure}
\begin{subfigure}[]{0.33\textwidth}
	\centering
	\caption{\footnotesize \neu{$64$-QAM modulation}}
	\input{./qam64/qam_64.tex}
	\label{fig:bmi_uncoded_qam64}
\end{subfigure}
	\caption{\footnotesize Achievable rate of the \ac{GNN} variants and various baselines over the Proakis-C channel.  Note that, due to complexity the 16\neu{/64}-QAM \ac{BCJR} curve uses an equivalent 4\neu{/8}-PAM. The \ac{GNN} performs in the gap between \ac{APP}-based and \ac{LMMSE}-based detection outperforming the other \ac{NN}-based approaches. \neu{For $64$-QAM most \ac{NN}-based approaches did not converge during training.}}
\end{figure*}

\vspace{2cm}
\subsection{Training}
\label{subsec:training_uncoded}
For training, the \ac{BCE} loss $\mathcal{L}_\mathrm{BCE}$ is employed 
\begin{equation*}
        \mathcal{L}_\mathrm{BCE}= \frac{1}{N}\sum_{i=1}^{N} [c_i\log_2q_i^{}
        +(1-c_i)\log_2(1-q_i^{})].
\end{equation*}
where $c_i$ is the transmitted bit and $q_i$ its estimate. 
Minimizing $\mathcal{L}_\mathrm{BCE}$ is equivalent to maximizing the \ac{BMI} $R_\mathrm{SDD}^{\mathrm{BMI}}$ (see equation (\ref{eq:loss_bmi})).
In \cite{lian19bploss} it was shown that in practice, minimizing $\mathcal{L}_\mathrm{BCE}$ minimizes the \ac{BER}.
Furthermore, the \ac{BCE} loss is used in a multi-loss fashion, meaning it is averaged over all $N_\mathrm{It}$ \ac{GNN} iterations $t\in [1,N_\mathrm{It}]$ as
\begin{equation}\label{eq:multi_loss}
        \mathcal{L}_\mathrm{Multi}= \frac{1}{N_\mathrm{It}} \sum_{t=1}^{N_\mathrm{It}}  \mathcal{L}^{(t)} _\mathrm{BCE}.
\end{equation}
This loss was first proposed in \cite{nachmani2016learning} and applied for \ac{GNN}-based decoding in \cite{cammerer2022gnn}.
In addition, we use the Adam \cite{kingma2014adam} optimizer.

\neu{The training data is artificially generated by the system shown in Fig.~\ref{fig:systemmodel}. The bits $\uv$ are uniformly randomly generated and transmitted over the simulated channel to serve as input data to the \ac{GNN}. Algorithm~\ref{alg:trainings_epoch} summarizes the training procedure.}
\begin{algorithm}[h]
\caption{Training epoch
}
	\SetAlgoLined
	\SetKwInOut{Input}{Input}
	\SetKwInOut{Output}{Output}
	\SetKwBlock{Repeat}{repeat}{}
	\SetKwFor{RepTimes}{For}{do}{end}
	\DontPrintSemicolon
	\Input{$\boldsymbol{\theta}$ \tcp{Trainable parameters}}
	\Output{$\boldsymbol{\theta}'$ \tcp{Updated train. parameters}}
	\BlankLine
	\tcp{Data generation}
	$\uv \gets \operatorname{random\_uniform\_integer}(\{0,1\}^K)$\\
	$ \yv \gets \operatorname{simulate\_transmission}(\uv)$ \\
	\tcp{GNN inference}
	$\hat{\uv} = \operatorname{GNN\_detector}(\yv; \boldsymbol{\theta})$ \\
	\tcp{Training}
	$\mathcal{L}_\mathrm{Multi} \gets \operatorname{Multi\_loss}(\uv, \hat{\uv} )$ \\
	$\boldsymbol{\theta}' \gets \operatorname{SGD}(\mathcal{L}_\mathrm{Multi},\boldsymbol{\theta})$
	\label{alg:trainings_epoch}
\end{algorithm}

\subsection{Results}

\input{tables/table_3}

\paragraph*{Setup}

For the evaluation of the proposed \ac{GNN} detector, a  transmission of $N_\xv=512$ \ac{BPSK}, $16$-\ac{QAM} and $64$-\ac{QAM} symbols over the Proakis-C channel ($\hv=[0.227, 0.460, 0.688, 0.460, 0.227]$) are simulated. The channel is chosen to represent a scenario with severe \ac{ISI}, where \ac{BPSK} signaling results in at least an estimated loss of \qty{2.01}{dB} and \qty{5.01}{dB} compared to an \ac{AWGN} channel, for \ac{JDD} and \ac{SDD}, respectively \cite{roumy99proakiscloss}. 
Note, the channel contains notches in the frequency representation posing a particular challenge for channel inverting detection, e.g., \ac{LMMSE} detection.
If not stated otherwise, we assume a fixed channel.
Therefore, the results demonstrate the algorithmic capability of \ac{GNN}-based detection rather than generalization, robustness and adaptability aspects.
The hyperparameters for the structure of the \ac{GNN} and the training are shown in the top part of Tab.~\ref{tab:table_example}. 
\neu{The parameters are based on \cite{cammerer2022gnn}. We noticed that increasing the size (layers and units) of the \acp{MLP} compared to \cite{cammerer2022gnn} enhances performance. Thus, we increased the parameters until we did not notice any more improvements in performance. For additional experiments regarding the training, we refer to the Appendix. }
\neur{As part of an ablation study, we investigated the use of \acp{GAT}, i.e., self-attention for message aggregation. However, this does not seem improve the performance. An example is shown in the appendix.}
For all figures showing achievable rates, each data points is optimized individually. 
That means retraining for \ac{NN}-based detectors and multiplying the output with a damping factor for classical detectors \cite{szczecinski2015bit}.

\paragraph*{Baselines}
The \acp{GNN} are compared to several baselines that can be divided into \ac{APP}-based, \ac{LMMSE}-based, and \ac{NN}-based detectors. The \ac{APP} baselines are the following:
\begin{itemize}
    \item The first \ac{APP}-based detector is the \ac{BCJR} algorithm \cite{bahl1974optimal,Forney1972MaximumlikelihoodSE}, implementing the optimal \ac{MAP} estimator.
    However, the problem is exponential complexity $O(M^L)$ in the number of bits per symbol $\log_2(M)$ and channel memory $L$.
        In the case of a $16$-QAM and the Proakis-C channel, the number of state transitions is $M^L\geq 10^6$.
    \item The second is the iterative \ac{SPA} on the \ac{FFG} and \ac{UFG}.
        On the Proakis-C channel, the \ac{SPA} on the \ac{UFG} does not converge.
        Furthermore, \ac{SPA} on the \ac{FFG} exhibits the same exponential complexity as \ac{BCJR} detection.
        While in general, more iterations result in better performance, the cycles in the factor graphs can result in an unstable behavior. Therefore, we apply a constant damping factor of $0.38$ to all edges.
        This allows performing more iterations ($N_\mathrm{It}=11$) before divergence and, thus, significantly improves the performance.
        Without the damping factor, \ac{SPA} diverges after $N_\mathrm{It}=5$ iterations.
        \item Better than guessing a constant damping factor is learning the damping weights per edge and per iteration, which is called \ac{N-SPA} \cite{schmid22neuralBP}.
        However, sharing the stability issues of \ac{SPA}, the training of the \ac{N-SPA} does not converge for $N_\mathrm{It}>7$.
\end{itemize}
Also, two \ac{LMMSE}-based baseline schemes are shown: 
\begin{itemize}
    \item The first is \ac{LMMSE} filtering \cite{proakis2001digital} with subsequent memoryless demapping.
    \item The second is called \ac{EP} \cite{santos2015block}, consisting of iterative message passing between a \ac{LMMSE} filter and a demapper for memoryless channels.
    \end{itemize}
\neur{Note that a combination of an optimal channel shortening filter and a \ac{BCJR} detector is discussed in \cite{rusek12optimalshortening}.
    We observed that the performance of this combination is usually between the \ac{LMMSE} and the \ac{EP} detector, thus, providing only little insight for our study.}
Finally, we compare to a \ac{NN}-based baseline:
\begin{itemize}
    \item The \ac{CNN} detector \cite{huang21extrinsic} can be interpreted as a trainable filter with subsequent \ac{APP} processing, e.g., an optimized channel shortening filter and an \ac{APP} equalizer with a small or no memory.
    The \ac{CNN} consists of multiple identical groups of layers, so-called \emph{blocks}.
It is compared with $B=3$ blocks for \ac{BPSK} and  $B=6$ blocks for \ac{QPSK}, $16$-\ac{QAM} \neu{and $64$-\ac{QAM}}.
Note that for \ac{QPSK} (with memory $L=6$) and $16$-QAM, we found that increasing the number of filters per layer by a factor of $8$ is necessary to ensure competitive performance.
\end{itemize}

\paragraph*{Description BPSK, $16$-QAM and $64$-QAM Rate}
Fig.~\ref{fig:bmi_uncoded_bpsk} shows the achievable \ac{SDD} rate of the \ac{GNN} detectors based on the \ac{FFG} and \ac{UFG}, the \ac{FGNN} based on the \ac{FFG}, and several baselines.
The other \ac{NN}-based detectors (\ac{FGNN} and \ac{CNN}) slightly deteriorate for medium and high rates from the \acp{GNN} by a gap of roughly \qty{0.5}{dB}.
The \ac{LMMSE}-based baselines (\ac{LMMSE} and \ac{EP}) perform close to the optimal \ac{MAP} performance of the \ac{BCJR} detector for low rates, however, display sub-optimal behavior for moderate and high rates. 

Fig.~\ref{fig:bmi_uncoded_qam16} shows the achievable \ac{SDD} rate using a $16$-\ac{QAM}.
Again, a \ac{BCJR} baseline displays the \ac{MAP} performance.
However, as the $16$-QAM is computationally infeasible, a $4$-PAM is used as surrogate which has identical performance for this special case where the \ac{CIR} is real-valued $\hv \in \RR$.
For the \acp{GNN} and other baselines, a $16$-\ac{QAM} is used. 
The neural approaches (\acp{GNN}, \ac{FGNN} and \ac{CNN}) perform better than the classical approaches.
For high rates the \ac{GNN} based on the \ac{FFG} performs best.
Finally, (neural) \ac{SPA} is not shown, due to infeasible complexity or non convergence for \ac{FFG} and \ac{UFG}, respectively.

\neu{
Fig.~\ref{fig:bmi_uncoded_qam64} displays the result for $64$-\ac{QAM}.
Here, the \acp{GNN} on the \ac{FFG} show slight performance gains compared to \ac{EP}, however, at a increased complexity. This indicates a limited scalability of the \acp{GNN} in terms of modulation order. The other \ac{NN}-based approaches did not converge during training. 
}

\begin{figure}[t]
	\centering
	\input{./tikz/latency_uncoded.tikz}
	\caption{\footnotesize \ac{BER} versus latency using \ac{BPSK} at \ac{SNR} $14\operatorname{dB}$ of different detectors over the Proakis-C channel. For iterative detectors, each marker corresponds to the performance after an iteration. }
	\label{fig:latency_uncoded}
\end{figure}
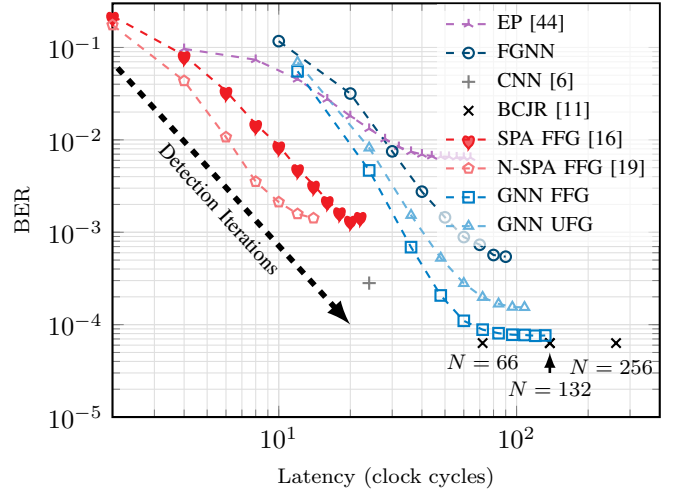

\paragraph*{Latency}
Fig.~\ref{fig:latency_uncoded} shows the \ac{BER} vs. latency in clock cycles.
We assume that every node or neural network layer is processed in one clock cycle.
This means $12$ cycles per iteration for the \acp{GNN}, $10$ for the \ac{FGNN}, $4$ for \ac{EP}, and $2$ for (neural) \ac{SPA}.
The latency of the \ac{BCJR} detector equates to $N_\xv+L+2$, and is, thus, dependent on the block length.
We assume a parallel computation of the forward and backward metric in $N_\xv+L$ cycles and $2$ additional cycles are needed to calculate the channel likelihoods and the bit estimates. 
Consequently, inherent parallelism of the \ac{GNN} shows larger latency gains over the \ac{BCJR} as the transmission length increases.

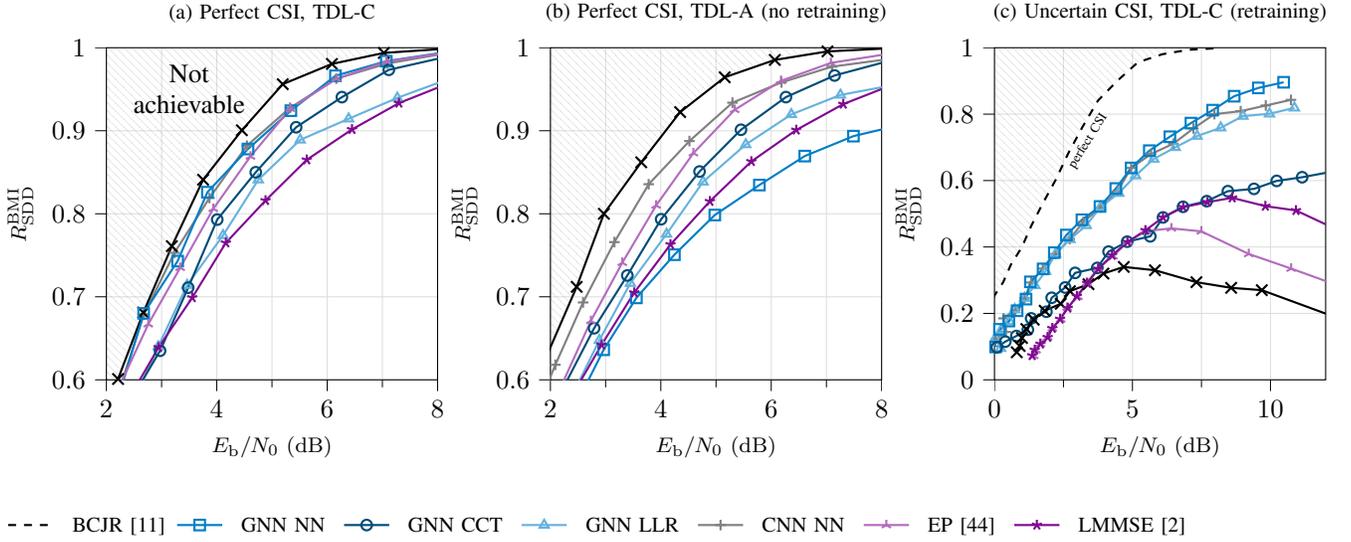
\begin{figure*}[t]
	\centering
	\input{./tikz/bmi_uncoded_qpsk_tdl_7.tikz}
	\caption{\footnotesize Achievable rate of the \ac{GNN} variants and various baselines using QPSK modulation trained over a 5G TDL-C channel ($7$ taps) with the channel embedding strategies: NN, CCT, and LLR. Note, $\hv$ is random for every transmission but follows a specified distribution. All detectors use $N_\mathrm{It}=8$ iterations. 
	Uncertain CSI means detection based on a noisy \ac{CIR} $\hv'=\hv+\nv'$ where $\nv'\sim \mathcal{N}(\mathbf{0},0.15\mathbf{I})$. The \ac{NN}-based detectors are robust against CSI uncertainty, but potentially degrade, if channel changes.} 
	\label{fig:bmi_uncoded_qpsk_tdl_7}
	\vspace{-1em}
\end{figure*}

\subsection{Analysis}

\neu{This section analyses how \acp{GNN} might improve over their basis, the \ac{SPA}. Note that this is not an analysis in the sense of explainable \acp{GNN} as in \cite{explainableGNN}. }
Fig.~\ref{fig:latency_uncoded} reveals two advantages of the \ac{GNN} compared to the classical \ac{SPA} or neural \ac{SPA} counterparts.
The first is the lower \ac{BER} after a single iteration, i.e., the first data point of the \ac{SPA} variants is higher than the first \ac{GNN} data point. 
The second is the larger decrease in \ac{BER} per iteration, as the slope of the \ac{GNN} is steeper. 
The reason  for the better initial \ac{BER} is the construction of sub-optimal messages  in the first iteration by the (neural) \ac{SPA} algorithm.
The result of \ac{VN} $V_j$ is the marginalization 
$\prod_{i =j-L}^j P(x_j|y_i)$. 
However, the optimal decision is 
\begin{align*}
    P(x_j|&y_j,\dots, y_{j-L}) = p(y_j,\dots,y_{j-L}|x_j)\frac{P(x_j)}{p(y_j,\dots,y_{j-L})} \\
    &\propto p(y_j| x_j) \cdot p(y_{j-1}|y_j, x_j) \cdot \dots  \\
    &  \qquad \cdot p(y_{j-L}| x_i,y_i,\dots,y_{i-L})\\
    & \propto  P(x_j|y_j)\cdot \frac{p(x_j,y_{j}|y_{i-1})}{p(x_j,y_{j})}\cdot \dots\\
    & \qquad \cdot \frac{p(x_j,y_{j-1},\dots,y_{j-L+1}|y_{i-L})}{p(x_j,y_{j-1},\dots,y_{j-L+1})} \\
    & =  P(x_j|y_j)\cdot P(x_j|y_{j-1})\cdot \underbrace{\frac{p(y_{j}|y_{j-1},x_j)}{p(x_j,y_{j})} }_{\text{omitted in SPA}}  \cdot ... \\
    &  \qquad \cdot P(x_{j}|y_{j-L}) \cdot 
    \underbrace{\frac{p(y_{j},\dots,y_{j-L+1}|y_{j-L},x_{j})}{p(x_j,y_{j-1},\dots,y_{j-L+1})}}_{\text{omitted in SPA}}.
\end{align*}  
We observe that the messages for an optimal decision contain information about the dependencies between the observations $y_{i},...,y_{i-L}$, which are omitted in the \ac{SPA}. 
Meaning the \ac{SPA} messages in the first iteration are only optimal if the observations are independent.
This is not the case for channels with memory resulting in an sub-optimal decision. 
As a consequence, the messages of the \ac{GNN}, which are abstract feature vectors, could capture the inter-dependencies of $y_{i},...,y_{i-L}$ leading to improved decisions early on.
Combining improved messages with adaptive damping may explain the larger decrease in \ac{BER} per iteration. 
In \cite{schmid2023local} it was demonstrated that adaptive damping of messages can be beneficial compared to fixed damping. 
Here, adaptive means that the damping factor is based on the message, which could be learned by the \acp{MLP} in the \ac{GNN}.
These intuitions and the line of reasoning is further supported by the following experiment.
We trained four variants of the \ac{GNN} on the \ac{FFG}: the full \ac{GNN} as described, the \ac{GNN} but with a summation as \ac{VN}, the \ac{GNN} but with a $\operatorname{\max} ^\star$ as \ac{FN}, and the \ac{GNN} but with equation (\ref{eq:extrinsic_map}) as edges.
Each replacement is the classical counterpart in the \ac{SPA}.
We observe that only the variant with the classical \ac{SPA} calculation on the edges deteriorates in performance.  
Meaning that the \ac{MLP} on the edge improves the performance complementary to the adaptive scaling.
Further,  the adaptive scaling, as in \cite{schmid2023local}, does no need to happen in the \ac{FN}, but can also be implemented in the \ac{VN} or on the edge.
However, a trainable \ac{FN} remains beneficial in terms of complexity on the \ac{FFG} as it may not require exponential complexity in $M$ or $L$.
Finally,  the \ac{VN} appears to not benefit from a \ac{MLP} in terms of performance nor complexity. 
Thus, a summation as the \ac{VN} update suffices.

\subsection{Robustness and Adaptivity}

One key aspect to analyze of a \ac{NN}-based detector is the robustness \cite{schlezinger20viterbinet}, i.e., the robustness to uncertainty in \ac{CSI} or a distribution change of the \ac{CIR}.
Fig.~\ref{fig:bmi_uncoded_qpsk_tdl_7} depicts results based on the proposed channel agnostic interfaces, which are based on \acp{LLR}, \ac{NN}, and a constant channel transformation.
The channel is a 5G \ac{TDL}-C model \cite{sionna} with \qty{20}{MHz} bandwidth, \qty{100}{ns} delay spread, and $7$ taps.
We evaluate three different scenarios:
\begin{enumerate}
    \item Perfect \ac{CSI}, where training and inference are performed on the same channel model.
    \item Perfect \ac{CSI}, where training is performed on \ac{TDL}-C and inference on \ac{TDL}-A. This scenario captures the change in distribution of the \ac{CIR}.
    Note that the baselines are not negatively influenced by the change as \ac{TDL}-A (with these parameters) allows for higher rates than \ac{TDL}-C.
    \item \ac{CSI} uncertainty: A noisy \ac{CIR} is passed to the detector.
    The noisy \ac{CIR} is constructed by $\hv'=\hv+\nv'$ where $\nv'\sim \mathcal{N}(\mathbf{0},0.15\mathbf{I})$.
    Note that the \acp{GNN} are retrained for this case.
\end{enumerate}
For the baselines we observe that the better the performance with genie \ac{CSI}, the worse it performs with noisy \ac{CSI}.
For the \acp{GNN} variants there seems not to be a clear trend
The \ac{NN}-based channel interface performs the best in the first scenario, but suffers the most in a change in distribution (second scenario) of the \ac{CIR}. 
However, it is the best model if retrained for the case of \ac{CSI} uncertainty.
This means that the \ac{NN}-based interface can adapt to scenario changes if retrained, but may be vulnerable if not. 
In general, the \ac{NN}-based detectors outperform all classical baselines in the case of \ac{CSI} uncertainty.
Similar to the \ac{LLR}-based variant, the \ac{CCT} is robust in terms of changes in \ac{CIR} distribution. 
However, the transformation bottelnecks the performance in case of \ac{CSI} uncertainty, as the \ac{GNN} cannot restore the lost information.
It is worth noting that the \ac{LLR}-based interface performs the worst in the first scenario.
This is probably due to the embedding, since the \ac{LLR} vector contains $16384$ values which are projected onto the feature space of the \ac{GNN} of size $d=16$.

\section{Graph Neural Networks for Joint Detection and Decoding}

\label{sec:jed}
In this section, we describe an \ac{NN} for \ac{JDD} and \ac{TDD} based on \acp{GNN}.
\neu{
For evaluating systems with feedback from a decoder, the appropriate performance metric is either the \ac{BER} or the \ac{BLER}.}
A fully deep learning-based receiver allows to jointly optimize the component networks rather than an individual optimization like in traditional receivers.
This is called end-to-end learning. 
We see the potential of this receiver mostly in the short block length regime, where sub-optimal \ac{SPA} decoding leaves room for improvement.
For longer block lengths, we propose to use the \acp{GNN} as the detection component in \ac{TDD}.
\neu{Note that in this case, the detector is trained independently of the channel code, but assumes a priori knowledge. As a result, the appropriate evaluation metric for detection is the  iterative achievable rate shown in equation (\ref{eq:tdd_bmi}).  }
Furthermore, using \acp{GNN} for the components prevents the \acp{NN} from the \emph{curse of dimensionality} (referring to exponential growth of points in high-dimensional objects) \cite{bellman61curseofdimensionality}, since the graph-based structures leverage the sparseness of the connections.
Moreover, the \acp{GNN} use high-dimensional processing the nodes and edges, leveraging the \emph{blessing of dimensionality} \cite{donoho00blessingofdimensionality}.

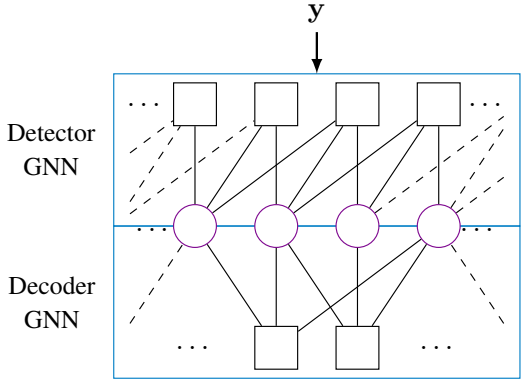
\begin{figure}[tbp]
  \centering
  \resizebox{.8\columnwidth}{!}{\input{figures/jed.tikz}}
  \caption{\ac{JDD} system with a decoder \ac{GNN} based on the Tanner graph of the channel code and an detector \ac{GNN} based on the \ac{FFG} of the channel. Note that the Tanner graph of the channel code code has been rearranged according to the interleaver to directly match the detector \acp{VN}.}
  \label{fig:combined_graph}
\end{figure}

The structure of the \ac{GNN} for \ac{JDD} is displayed in Fig.~\ref{fig:combined_graph}.
One set of \acp{FN} is based on the graph for detection as in Sec.~\ref{sec:GNN_eq} and one of \acp{FN} for decoding as in \cite{cammerer2022gnn}. 
Note that both graphs share the same \acp{VN}, building a connected graph. 
In the case of using the \ac{GNN} as a detector in \ac{TDD}, an embedding of the \emph{a priori} \acp{LLR} $\boldsymbol{\ell}_\mathrm{A}$ to the \acp{VN} is proposed.
Similar to the embedding of the \acp{FN} we propose a linear embedding
$\sv_{V_i} = \Wm_V \cdot \boldsymbol{\ell}_\mathrm{A}$.
Furthermore, interleaving and deinterleaving makes the \emph{a priori} information of the detector and decoder to appear more independent, thus, improving performance \cite{bicm98caire}.

\vspace{2cm}
\subsection{Schedule}
\begin{figure}[tbp]
  \centering
       \begin{subfigure}[b]{0.48\columnwidth}
       \centering
            \input{./figures/floodingschedule.tikz}
            \caption{ Flooding }
            \label{subfig:flooding}
        \end{subfigure}
        \hfill
        \begin{subfigure}[b]{0.48\columnwidth}
            \centering
            \input{./figures/sequentialschedule.tikz}
            \caption{ Sequential }
            \label{subfig:sequential}
        \end{subfigure}
  \caption{\footnotesize Schedules of the \ac{GNN}-based \ac{JDD}}
  \label{fig:schedule}
\end{figure}
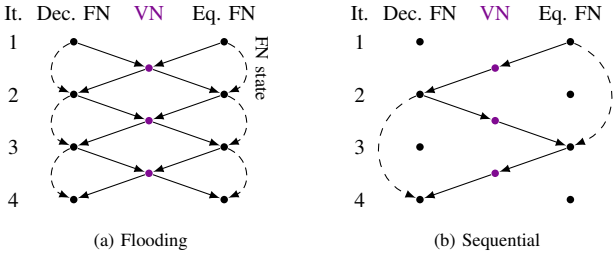

In contrast to the classical solution of using iterative \ac{BCJR} detection and \ac{SPA} decoding \cite{douillard95turboequalization}, the combined \ac{GNN} for joint detection and decoding enables an earlier use of detection with \emph{a priori} knowledge.
\ac{BCJR} runs are costly in complexity and latency, thus, a sufficient amount of iterations should be spend on \ac{SPA} decoding to generate a significant amount of \emph{a priori} information for the subsequent \ac{BCJR} run.
However, during this time, \ac{SPA} decoding does not benefit from new \emph{a priori} information from the detector. 
In case of the combined \ac{GNN}, the latency for detection and decoding are the same, thus, enabling schedules with earlier information exchange between the components.
This can result in improved error correcting performance at the same latency. 
Moreover, the combined \ac{GNN} can be updated in a flooding fashion. 
\neu{This means that all \acp{FN} (for detection and decoding) are updated in parallel and all \acp{VN} are updated in parallel. }
The difference to the iterative, sequential schedule is shown in Fig.~\ref{fig:schedule}.
The \ac{GNN} updates the \acp{FN} for detection and for decoding simultaneously and combines their beliefs after each iteration in their shared \acp{VN}.
\neu{Relating the description to Algorithm~\ref{alg:inference}, the only addition is that the set of \acp{FN} also contains the check nodes of the \ac{LDPC} code.}
Consequently, the same number of resources can be used as in the sequential, iterative case, but in half the time.
In every iteration, the component \acp{GNN} receive new \emph{a priori} information but also keep their respective \ac{FN} state.
In \cite{wiesmayr2022duidd}, it was found that keeping the state of the \acp{FN} is beneficial for iterative systems with a small number of inner iterations.

\subsection{Training}
The training is similar to the training process described in section \ref{subsec:training_uncoded} with some extensions.
For the \ac{JDD} \acp{GNN} based on the combined factor graph is trained at once. 
In the case of \ac{TDD}  with \ac{SPA} decoder, we employ \emph{training with Gaussian priors} \cite{pretrain20koike,clausius21serialAE,huang21extrinsic}. 
Note that the 5G \ac{LDPC} code contains punctured \acp{VN} that we include in the calculation of the loss.
The additional hyperparameters are shown below the horizontal line in Tab.~\ref{tab:table_example}.
The schedule of the combined \acp{GNN} is given in the form ($\#\mathrm{Outer~Iterations}$, [$\#\mathrm{Inner~Iterations}$]).
For deep \acp{GNN}, i.e., a large number of iterations, direct training led to sub-optimal performance. 
Therefore, we used a two stage training process in these cases.
First, we trained with the schedule given in Tab.~\ref{tab:table_example}, for which the \acp{GNN} trained reliable and consistent.
Second, the number of iterations is increased to the desired schedule and the model is finetuned. 
A flexible schedule is ensured by the weight-sharing over iterations in the \acp{GNN}.

For the training with Gaussian priors the \ac{GNN} for \ac{TDD} is trained independently of the decoder\neu{, as shown in Algorithm~\ref{alg:training_gaussian}}. 
The process is similar to evaluation of the \ac{EXIT} characteristic detailed in section \ref{sec:exit}.
Similarly, the priors are generated from equation (\ref{eq:generate_prior}) with a specified average mutual information $I_\mathrm{A}$.
The \acp{VN} of the \ac{GNN} are initialized with the \emph{a priori} \acp{LLR}.
In contrast to the \ac{EXIT} procedure, we are not interested in the extrinsic information $I_\mathrm{E}$, but the total information $I_\mathrm{T}$.
The loss in equation (\ref{eq:multi_loss}) is then calculated between the transmitted bits and the output of the \ac{GNN}
The extrinsic \acp{LLR} for inference can be calculated with equation (\ref{eq:extrinsic_map}).
An additional hyperparameter of the training with Gaussian priors is the distribution of $I_\mathrm{A}$. 
We choose $I_\mathrm{A}\sim \mathcal{U}(0,1)$, since the \ac{GNN} should generalize for all possible prior information. 
\begin{algorithm}[h]
\caption{Gaussian pre-training epoch for TDD
}
	\SetAlgoLined
	\SetKwInOut{Input}{Input}
	\SetKwInOut{Output}{Output}
	\SetKwBlock{Repeat}{repeat}{}
	\SetKwFor{RepTimes}{For}{do}{end}
	\DontPrintSemicolon
	\Input{$\boldsymbol{\theta}$ \tcp{Trainable parameters}}
	\Output{$\boldsymbol{\theta}'$ \tcp{Updated train. parameters}}
	\BlankLine
	\tcp{Data generation}
	$\uv \gets \operatorname{random\_uniform\_integer}(\{0,1\}^K)$\\
	$ \yv,\cv \gets \operatorname{simulate\_transmission}(\uv)$ \\
	\tcp{Gaussian prior generation}
	$I_\mathrm{A}\gets \operatorname{random\_uniform\_float}((0,1)) $\\
	$\boldsymbol{\ell}_\mathrm{A}\gets \operatorname{generate\_LLRs}(\cv,I_\mathrm{A})$\\
	\tcp{GNN inference}
	$\hat{\cv} = \operatorname{GNN\_detector}(\yv,\boldsymbol{\ell}_\mathrm{A}; \boldsymbol{\theta})$ \\
	\tcp{Training}
	$\mathcal{L}_\mathrm{BCE} \gets \operatorname{BCE\_loss}(\cv, \hat{\cv} )$ \\
	$\boldsymbol{\theta}' \gets \operatorname{SGD}(\mathcal{L}_\mathrm{Multi},\boldsymbol{\theta})$
	\label{alg:training_gaussian}
\end{algorithm}

\subsection{Results}
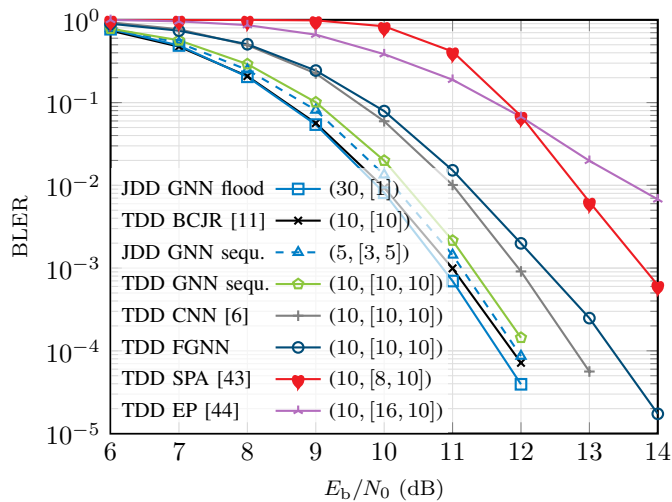
\begin{figure}[t]
	\centering
	\input{./tikz/ber_coded_bpsk.tikz}
  \vspace{-0.6cm}
	\caption{\footnotesize \ac{BER} of the bits $\uv$ using coded system ($N=132,~R_\mathrm{C}=0.5$, $5$G \ac{LDPC},  \ac{BPSK}) over \ac{SNR} of different detectors and decoders over the Proakis-C channel. The \ac{JDD}/\ac{TDD} schedules are given in ($\#\mathrm{Outer~Iterations}$, [$\#\mathrm{Inner~Iterations}$]). All \ac{TDD} systems use an \ac{SPA} decoder. The flodding \ac{GNN} outperforms the best \ac{TDD} baseline. %
}
	\label{fig:ber_coded}
	\vspace{-1em}
\end{figure}

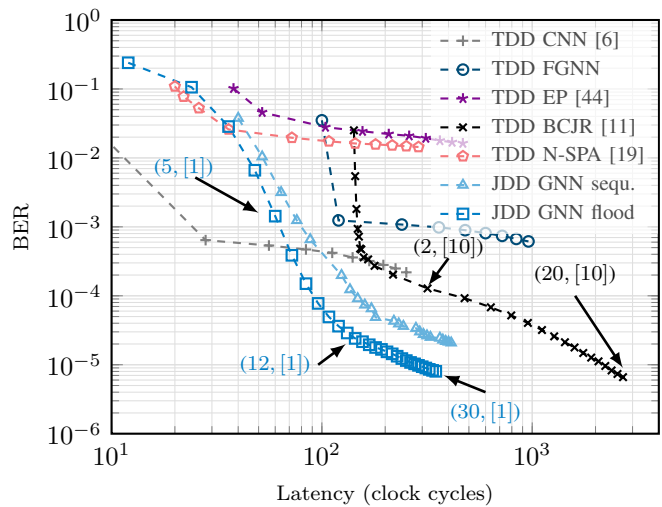
\begin{figure}[t]
	\centering
	\input{./tikz/latency_coded_bpsk.tikz}
	\caption{\footnotesize \ac{BER} of uncoded bits $\uv$ using coded system  ($N=132,~R_\mathrm{C}=0.5$, $5$G \ac{LDPC},  BPSK) versus latency at \ac{SNR} $12\operatorname{dB}$ of different detectors and decoders over the Proakis-C channel. The flooding \ac{GNN} approaches the \ac{TDD} \ac{BCJR} performance with $15\%$ of the latency. }
	\label{fig:latency}
	\vspace{-1em}
\end{figure}

\begin{figure*}[t]
\begin{subfigure}[]{0.5\textwidth}
	\centering
	\caption{\footnotesize BPSK modulation}
	\input{./tikz/bmi_iterativ_uncoded_bpsk.tikz}
	\label{fig:bmi_iterativ_uncoded_bpsk}
\end{subfigure}
\begin{subfigure}[]{0.5\textwidth}
	\centering
	\caption{\footnotesize $16$-QAM modulation}
	\input{./tikz/bmi_iterativ_uncoded_qam16.tikz}
	\label{fig:bmi_iterativ_uncoded_qam16}
\end{subfigure}
\caption{\footnotesize Achievable rate over the \ac{SNR} for \ac{TDD} and $16$-QAM modulation over the Proakis-C channel. Note, due to complexity the $16$-QAM \ac{BCJR} curve uses an equivalent $4$-PAM. The \ac{GNN} performs in the gap between \ac{APP}-based and \ac{LMMSE}-based detection outperforming the other \ac{NN}-based approaches. }
\end{figure*}
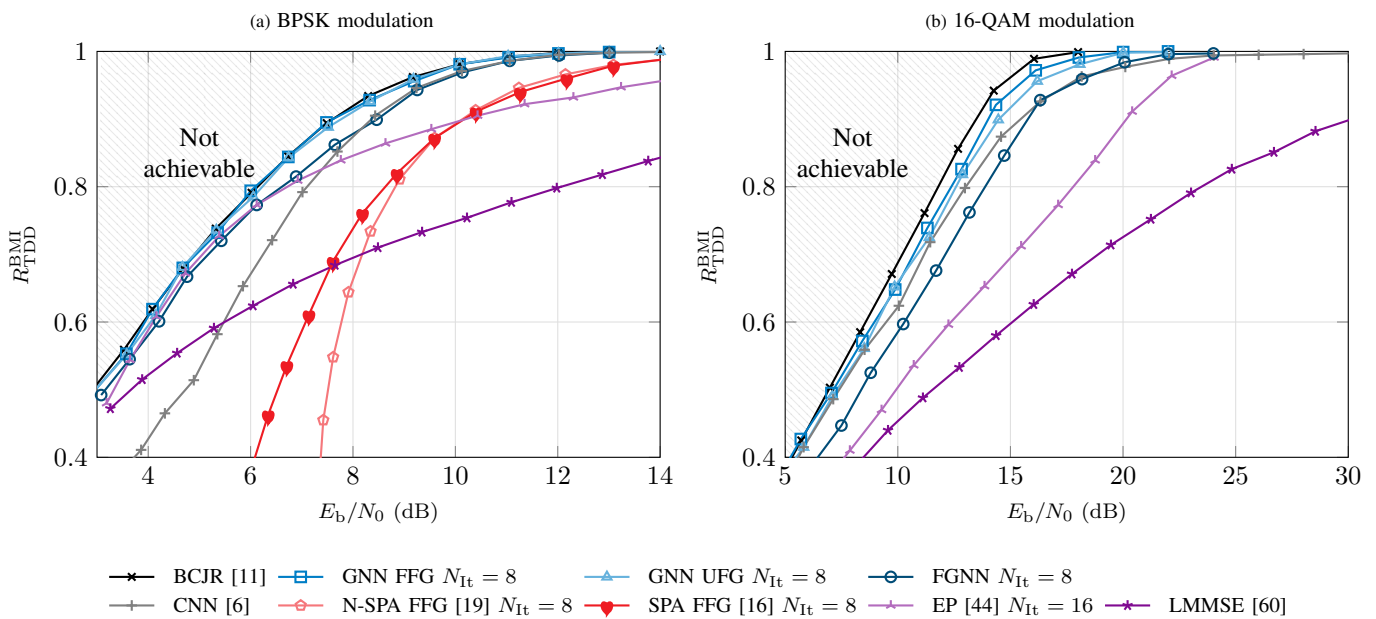

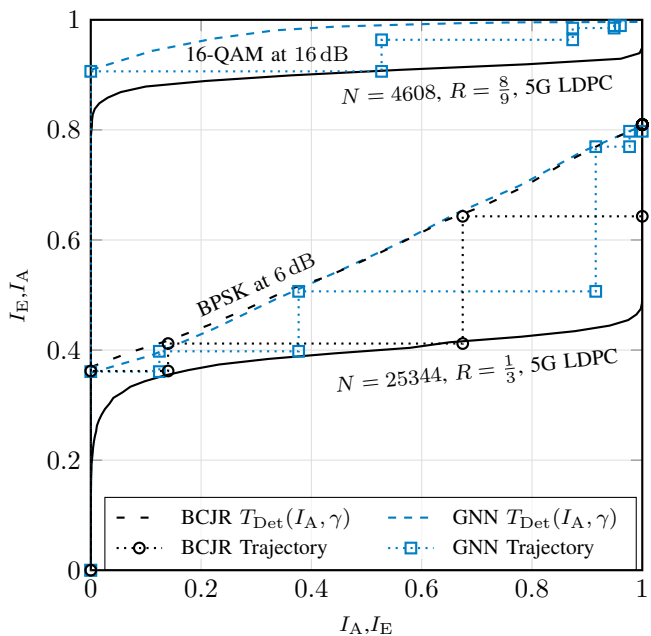
\begin{figure}[tbp]
  \centering
  \resizebox{1.0\columnwidth}{!}{\input{tikz/exit_chart.tikz}}
  \vspace{-1em}
  \caption{EXIT chart for a high rate scenario ($N=4608,~R_\mathrm{C}=\frac{8}{9}$, $5$G \ac{LDPC} with $N_\mathrm{It}=10$,  $16$-\ac{QAM}) and a low rate scenario ($N=25344,~R_\mathrm{C}=\frac{1}{3}$, $5$G \ac{LDPC} with $N_\mathrm{It}=10$,  BPSK) over the Proakis-C channel. The \ac{GNN} trajectory matches the prediction from the transfer characteristic $T_\mathrm{Det}(I_\mathrm{A},\gamma)$.}
  \label{fig:exit_chart}
  \vspace{-1em}
\end{figure}

\begin{figure}[]
	\centering
	\input{./tikz/ber_coded_16qam.tikz}
	\vspace{-1em}
	\caption{\footnotesize \ac{BER} of the bits $\uv$ using coded system ($N=4608,~R_\mathrm{C}=\frac{8}{9}$, $5$G \ac{LDPC},  $16$-\ac{QAM}) over \ac{SNR} of different detectors and decoders over the Proakis-C channel. The \ac{JDD}/\ac{TDD} schedules are given in ($\#\mathrm{Outer~Iterations}$, [$\#\mathrm{Inner~Iterations}$]). Note that, for complexity reasons, the \ac{BCJR} curve uses an equivalent $4$-PAM. The \ac{GNN} shows a gain of \qty{6}{dB} compared to EP, the best and feasible classical baseline.
 }
	\label{fig:ber_coded_16qam}
	\vspace{-1em}
\end{figure}
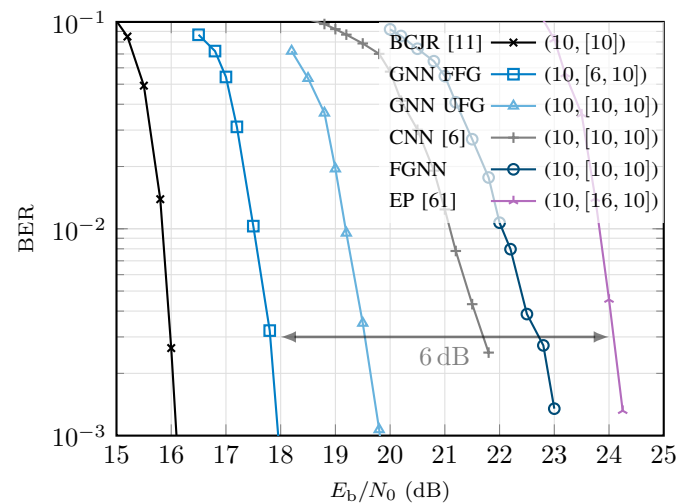

\paragraph*{JDD in the short block length regime}
For showcasing the potential of \acp{GNN} for \ac{JDD}, especially in low latency scenarios, we evaluate a transmission encoded by a 5G \ac{LDPC} code ($K=66, N=132, R_\mathrm{C}=0.5$) and \ac{BPSK} signaling over the Proakis-C channel.
In Fig.~\ref{fig:ber_coded}, the \ac{BER} of the bits $\uv$ is shown over the \ac{SNR}. 
The  schedules are given in the legend in the form ($\#\mathrm{Outer~Iterations}$, [$\#\mathrm{Inner~Iterations}$]).
The flooding schedule \ac{GNN} outperforms the \ac{TDD} \ac{BCJR}-\ac{SPA} baseline.
Unexpectedly, the flooding \ac{GNN} outperforms the sequential \ac{GNN} regardless of the latency constraints, suggesting that the flooding schedule may be advantageous during training. 
Finally, a \ac{GNN} trained with Gaussian priors in a \ac{TDD} framework is displayed. 
Regardless of the schedule, the \ac{JDD} \acp{GNN} outperform the \ac{TDD} \ac{GNN}.
Fig.~\ref{fig:latency} displays the \ac{BER} over the latency in clock cycles. 
Recall that  $12$, $2$, and $N+L+2$ clock cycles are needed for the flooding \ac{GNN}, \ac{SPA} and \ac{BCJR}, respectively.
In the low latency regime, the flooding \ac{GNN} outperforms the sequential \ac{GNN} and the \ac{BCJR}-SPA baseline, as it demonstrates a \ac{BER} three times lower than the sequential \ac{GNN} and ten times lower than the \ac{BCJR}-\ac{SPA} baseline.
The flooding \ac{GNN} is only outperformed by the \ac{BCJR}-SPA baseline after ca. seven times the latency of the \ac{GNN}.

\paragraph*{TDD in the long block length regime}
In the following we demonstrate the viability of \acp{GNN} for detection in \ac{TDD}.
Fig.~\ref{fig:bmi_iterativ_uncoded_bpsk} shows the achievable rate with \ac{TDD} (equation (\ref{eq:tdd_bmi})) compared to several baselines.
The result is similar to the $16$-QAM scenarion depicted in Fig.~\ref{fig:bmi_iterativ_uncoded_qam16}, specifically, that the \acp{GNN} perform close to the optimum. 
Especially in the high rate regime, \ac{NN}-based detection seems to be beneficial to the non-\ac{NN} baselines. 

\input{tables/table_4}

The validity of the rate is evaluated in the \ac{EXIT} chart depicted in Fig.~\ref{fig:exit_chart}.
The trajectory is transfer characteristic is shown for two cases.
The first case is a high rate case with a 5G \ac{LDPC} code ($N=4608$, $K=4032$, $\frac{E_\mathrm{b}}{N_0}=16\operatorname{dB}$) and $16$-QAM. 
The trajectory does not match the \ac{EXIT} characteristic  precisely.
A potential issue may be that the extrinsic \acp{LLR} from the decoder do not resemble Gaussian distributions, compared to the Gaussian priors during training.
The second case is a low rate 5G \ac{LDPC} code ($N=25344$, $K=8448$, $\frac{E_\mathrm{b}}{N_0}=6\operatorname{dB}$) and BPSK. 
In the second case, the \ac{BCJR} trajectory and transfer characteristic is shown for comparison.
Here, the \ac{EXIT} characteristic is matched more closely.
However, a small divergence occurs after three turbo iterations.
Note that the divergence is after decoding, meaning the \emph{a priori} information form the \ac{GNN} might be not extrinsic.
This problem may be attributed to the loops in the graph.
This aligns with the observation that equation (\ref{eq:extrinsic_map}) (which is used in \ac{TDD}) does not hold for \acp{GNN}, if  $R_\mathrm{SDD} \ll I_\mathrm{A}$.
In this case the \emph{a priori} \acp{LLR} dominate inference and the loops lead to too significant amplification.

Finally, we evaluate the \acp{GNN} in the \ac{TDD} framework in the high rate scenario ($N=4608$, $K=4032$) and $16$-QAM in Fig.~\ref{fig:ber_coded_16qam}.
Recall that the \ac{BCJR} has more than $10^6$ state transitions and is, thus, infeasible. 
Compared to the best, feasible non-\ac{NN} baseline, the \ac{GNN} based on the \ac{FFG} demonstrates a gain of more than \qty{6}{dB}.

\section{Discussion}

Before we discuss how \ac{NN}-based detection,  in particular \acp{GNN}, fit in to the landscape of detection, we detail the areas where classical detection leaves room for improvement. A summary is presented in Tab.~\ref{tab:my_label_new}.

Let us start with the ideal case of genie \ac{CSI}. 
While the \ac{APP}-based approaches (\ac{BCJR}, \ac{SPA} on \ac{FFG}) scale exponentially, the complexity for small alphabet size $M$ and memory $L$, i.e. $M^L \lessapprox 10^2$ is relatively small. 
Thus, only for $M^L \gtrapprox 10^2$ alternatives are required.
The complexity of \ac{LMMSE}-based detection is almost independent of $M$ and $L$, making them promising candidates.
The matrix inversion of complexity $\mathcal{O}(N^3)$ (or  $\mathcal{O}(N\log(N))$ for Toeplitz matrices \cite{martinsson_fast_2005}) can be avoided by a filter-type implementation. 
This can be a time-invariant filter for \ac{LMMSE} detection or a time varying filter for \ac{EP} \cite{santos_turbo_2018}.
However, \ac{LMMSE} detection performs highly sub-optimal for high rates.
Furthermore, \ac{EP} improves \ac{LMMSE} detection for high rates on the cost of iterative \ac{LMMSE} filtering with increased complexity of $\mathcal{O}(N^2)$ \cite{turboep17santos}. 
The first problem with \ac{LMMSE}-based approaches are notches in the frequency representation of the channel, as the \ac{LMMSE} tries inverting the channel.
The second problem is $p(\mathbf{y},\mathbf{x})$ being non-Gaussian.
The probability of the first issue decreases with increased channel memory $L$ (for stochastic \acp{CIR}), since the probability of destructive interference decreases.
The second issue can vanish for increasing $M$ assuming Gaussian-like constellation shaping.
To summarize, \ac{NN}-based detection may be beneficial if $M$ and $L$ are too large for \ac{APP}-based detection, but small enough so that $p(\mathbf{y},\mathbf{x})$ is not yet Gaussian, and the \ac{CIR} has notches in the frequency domain. 

Another area where \ac{NN}-based detection may shine, are imperfections. 
This could be a non-linear channel as in \cite{huang21extrinsic,plabst24sicrnn} or imperfect \ac{CSI} at the receiver as shown here or discussed in more detail in \cite{schlezinger20viterbinet,shlezinger2020bcjrnet}.
These imperfections are usually hard to consider in a classical detection algorithm, but can be easily learned by an \ac{NN}.

The last area worth mentioning is the possibility of a end-to-end optimized receiver. 
Instead of optimizing each component in the receiver it might be beneficial to optimize them jointly, which is enabled by the use of \acp{NN}.
This is demonstrated here for \ac{JDD}, and also considered in, e.g., \cite{doerner2022jointautoencoder, Cammerer2023ANR,han23SNNjed,wiesmayr2024design}.

The performance gains promised by \ac{NN}-based receivers, here \acp{GNN}, may come with increased complexity compared to \ac{LMMSE}-based approaches.
However, quantifying the complexity is not trivial. 
A \neu{meaningful} comparison \neu{can only be conducted} on an implementation level, due to the different type of functions and operations used.
For this, one would first need to optimize the \ac{NN} size (e.g., by pruning), quantize it, and choose a suitable implementation and hardware \cite{ney23cnn,freire23fpgacnn}.
Here, all the \acp{NN} (\ac{GNN}, \ac{CNN}, and \ac{FGNN}) are over parameterized and hence, showing the performance limit.
\neu{
Nonetheless, for an approximate complexity comparison between the \ac{NN}-based approaches and classical detectors, we assume that they are executed on a general hardware, e.g., GPU or TPU. 
Then, the metric of interest is the number of weights and the number of operations \cite{schmid22neuralBP}, as shown in Tab.~\ref{tab:numop}.
The operation count of the \ac{BCJR} and \ac{SPA}-\ac{UFG} is taken from \cite{schmid22neuralBP}. 
Similar to \ac{BCJR} detection, the operation count of \ac{SPA} on the \ac{FFG} amounts to $N_\mathrm{It}(M^{L+1}(L+2)+LM(2L+1))$, where $L$ is the memory and $M$ is the modulation order.
Next, the \ac{GNN} requires $N_\mathrm{It}2(2+L)((N_\mathrm{L}-2)N_\mathrm{U}(N_\mathrm{U}+2)+4N_\mathrm{U}d)$ operations, where $N_\mathrm{L}$ is the number of layers in each \ac{MLP}, $N_\mathrm{U}$ is the number of units in each \ac{MLP} and $d$ is the feature size.
Finally, we approximate counting the operations of \ac{EP} by counting the operations of the dominating \ac{LMMSE} step. Disregarding simplifications in the matrix inversion, the filter-type implementation from \cite{santos_turbo_2018} yields
\neur{$72L^3+60L^2+96L+10+N_\mathrm{It}(45L^3+66L^2+34L+6)$}
operations.
}

\input{tables/table_5}

\section{Conclusion}
The paper proposes and demonstrates the application of \acp{GNN} for detection and \ac{JDD}. 
In the case of detection, the \ac{GNN} based on the \ac{FFG} shows close-to-optimal performance.
Thus, \acp{GNN} may be regarded as an elegant way of bridging the performance gap between \ac{SPA} and \ac{BCJR} for detection.
Especially for high rates, the \ac{GNN} outperforms state-of-the-art baselines.
We also provided some intuition and reasoning that the gains might come from improved message passing between the nodes. 
In the case of \ac{JDD}, we demonstrated the end-to-end learning capabilities of \acp{GNN} for detection and decoding, leveraging sparse graphs and the blessing of dimensionality. 
With the proposed flooding schedule, the \ac{GNN} outperforms an iterative \ac{BCJR}-\ac{SPA} baseline with a substantially lower \ac{BER}.
Furthermore, we investigated the \ac{GNN} is a \ac{TDD} setup, where a gain of more than \qty{6}{dB} was demonstrated. 

\appendix \label{appendix}

\begin{figure}[h]
	\centering
	\input{./training/training.tex}
	\caption{\footnotesize \footnotesize \neu{Evaluation loss over epochs for \ac{JDD} training over the Proakis-C channel at \qty{10}{dB} with \ac{BPSK} modulation and a 5G-\ac{LDPC} code ($N=132,R_\mathrm{C}=0.5$). The \ac{GNN} is based on the \ac{FFG} and $N_\mathrm{It}=10$}}
	\label{fig:training}
\end{figure}
\neu{
In this section, we provide additional data regarding the training process. 
Fig.~\ref{fig:training} shows the smoothed evaluation loss over the training epochs for a \ac{JDD} scenario for several experiments.
The experiments vary the training hyperparameters and the structure hyperparameters.
Furthermore, several training runs with the parameters from Tab.~\ref{tab:table_example} are shown, indicating a stable and reproducible result.
The differences in the experiments are displayed in the legend by showing which parameter changes in contrast to Tab.~\ref{tab:table_example}.
Overall, we observe that alterations of the parameters lead to only small variations in the training process. A similar amount of variation is observed by replicating the training with the parameters from Tab.~\ref{tab:table_example}, but with different initiations.
Finally, we conclude that the training process is stable, reproducible and is only mildly influenced by the choice of hyperparameters.
}

\neur{To explore the performance limits of the \ac{GNN} approach, we performed additional experiments with \acp{GAT}. As shown in Fig.~\ref{fig:training}, the best training run indicates that more sophisticated architectures may not necessarily improve performance.}

\bibliographystyle{IEEEtran}
\bibliography{references}

\end{document}

%% file: macros.tex
\renewcommand{\vec}[1]{\mathbf{#1}}

\newcommand{\cv}{\vec{c}}

\newcommand{\gv}{\vec{g}}
\newcommand{\hv}{\vec{h}}

\newcommand{\mv}{\vec{m}}
\newcommand{\nv}{\vec{n}}

\newcommand{\sv}{\vec{s}}

\newcommand{\uv}{\vec{u}}
\newcommand{\vv}{\vec{v}}

\newcommand{\xv}{\vec{x}}
\newcommand{\yv}{\vec{y}}
\newcommand{\zv}{\vec{z}}

\newcommand{\Hm}{\vec{H}}

\newcommand{\Rm}{\vec{R}}

\newcommand{\Um}{\vec{U}}

\newcommand{\Wm}{\vec{W}}

\newcommand{\Ym}{\vec{Y}}

\newcommand{\CC}{\mathbb{C}}

\newcommand{\RR}{\mathbb{R}}

\newcommand{\LB}{\left(}
\newcommand{\RB}{\right)}

\renewcommand{\ln}[1]{\mathop{\mathrm{ln}}\LB #1\RB}

%% file: acronyms.tex
\begin{acronym}
 \acro{CSI}{channel state information}
 \acro{NBP}{neural belief propagation}
 \acro{N-SPA}{neural sum-product algorithm}
 \acro{FFG}{Forney factor graph}
 \acro{UFG}{Ungerboeck factor graph}
 \acro{PCM}{parity-check matrix}
 \acro{UE}{user equipment}
 \acro{UL}{uplink}
 \acro{BS}{basestation}
 \acro{TDD}{turbo detection and decoding}
 \acro{TDL}{tapped delay line}
 \acro{FDD}{frequency division duplex}
 \acro{ECC}{error-correcting code}
 \acro{MLD}{maximum likelihood decoding}
 \acro{HDD}{hard decision decoding}
 \acro{IF}{intermediate frequency}
 \acro{RF}{radio frequency}
 \acro{SDD}{soft decision decoding}
 \acro{NND}{neural network decoding}
 \acro{CNN}{convolutional neural network}
 \acro{ML}{maximum likelihood}
 \acro{GPU}{graphical processing unit}
 \acro{BP}{belief propagation}
 \acro{LTE}{Long Term Evolution}
 \acro{BER}{bit error rate}
 \acro{DER}{detection error rate}
 \acro{SNR}{signal-to-noise-ratio}
 \acro{ReLU}{rectified linear unit}
 \acro{BPSK}{binary phase shift keying}
 \acro{QPSK}{quadrature phase shift keying}
 \acro{AWGN}{additive white Gaussian noise}
 \acro{MSE}{mean squared error}
 \acro{LLR}{log-likelihood ratio}
 \acro{MAP}{maximum a posteriori}
 \acro{NVE}{normalized validation error}
 \acro{BCE}{binary cross-entropy}
 \acro{CE}{cross-entropy}
 \acro{BLER}{block error rate}
 \acro{SQR}{signal-to-quantisation-noise-ratio}
 \acro{MIMO}{multiple-input multiple-output}
 \acro{OFDM}{orthogonal frequency division multiplex}
 \acro{RF}{radio frequency}
 \acro{LOS}{line of sight}
 \acro{NLoS}{non-line of sight}
 \acro{NMSE}{normalized mean squared error}
 \acro{CFO}{carrier frequency offset}
 \acro{SFO}{sampling frequency offset}
 \acro{IPS}{indoor positioning system}
 \acro{TRIPS}{time-reversal IPS}
 \acro{RSSI}{received signal strength indicator}
 \acro{MIMO}{multiple-input multiple-output}
 \acro{ENoB}{effective number of bits}
 \acro{AGC}{automated gain control}
 \acro{ADC}{analog to digital converter}
 \acro{ADCs}{analog to digital converters}
 \acro{FB}{front bandpass}
 \acro{FPGA}{field programmable gate array}
 \acro{JSDM}{Joint Spatial Division and Multiplexing}
 \acro{NN}{neural network}
 \acro{IF}{intermediate frequency}
 \acro{LoS}{line-of-sight}
 \acro{NLoS}{non-line-of-sight}
 \acro{DSP}{digital signal processing}
 \acro{AFE}{analog front end}
 \acro{SQNR}{signal-to-quantisation-noise-ratio}
 \acro{SINR}{signal-to-interference-noise-ratio}
 \acro{ENoB}{effective number of bits}
 \acro{AGC}{automated gain control}
 \acro{PCB}{printed circuit board}
 \acro{EVM}{error vector mangnitude}
 \acro{CDF}{cumulative distribution function}
 \acro{MRC}{maximum ratio combining}
 \acro{MRP}{maximum ratio precoding}
 \acro{MRT}{maximum ratio transmission}
 \acro{DeepL}{deep-learning}
 \acro{DL}{deep learning}
 \acro{SISO}{single-input single-output}
 \acro{SGD}{stochastic gradient descent}
 \acro{CP}{cyclic prefix}
 \acro{MISO}{Multiple Input Single Output}
 \acro{LMMSE}{linear minimum mean square error}
 \acro{ZF}{zero forcing}
 \acro{USRP}{universal software radio peripheral}
 \acro{RNN}{recurrent neural network}
 \acro{GRU}{gated recurrent unit}
 \acro{LSTM}{long short-term memory}
 \acro{NTM}{neural turing machine}
 \acro{DNC}{differentiable neural computer}
 \acro{TCN}{temporal convolutional network}
 \acro{FCL}{fully connected layer}
 \acro{MLP}{multilayer perceptron}
 \acro{MANN}{memory augmented neural network}
 \acro{RNN}{recurrent neural network}
 \acro{DNN}{deep neural network}
 \acro{FIR}{finite impulse response}
 \acro{BPTT}{back-propagation through time}
 \acro{GAN}{generative adversarial network}
 \acro{ELU}{exponential linear unit}
 \acro{tanh}{hyperbolic tangent}
 \acro{BICM}{bit-interleaved coded modulation}
 \acro{OTA}{over-the-air}
 \acro{IM}{intensity modulation}
 \acro{DD}{direct detection}
 \acro{RL}{reinforcement learning}
 \acro{SDR}{software-defined radio}
 \acro{WGAN}{Wasserstein generative adversarial network}
 \acro{BMD}{bit-metric decoding}
 \acro{BMI}{bit-wise mutual information}
 \acro{LDPC}{low-density parity-check}
 \acro{IDD}{iterative demapping and decoding}
 \acro{IEDD}{iterative equalization, demapping and decoding}
 \acro{JDD}{joint detection and decoding}
 \acro{SDD}{separate detection and decoding}
 \acro{JSD}{Jensen-Shannon divergence}
  \acro{FLOP}{floating point operation}
 \acro{MMSE}{minimum mean square error}
 \acro{FFT}{fast Fourier transform}
 \acro{IFFT}{inverse fast Fourier transform}
 \acro{QAM}{quadrature amplitude modulation}
 \acro{EMD}{earth mover's distance}
 \acro{TDL}{tapped delay line}
 \acro{KL}{Kullback--Leibler}
 \acro{PRACH}{physical random access channel}
 \acro{URLLC}{ultra-reliable low-latency communication}
 \acro{ANOMA}{asynchronous non-orthogonal multiple access}
 \acro{FEC}{forward error correction}
 \acro{NOMA}{non-orthogonal medium access}
 \acro{MTC}{machine-type communications}
 \acro{mMTC}{massive machine-type communications}
 \acro{MCS}{modulation and coding scheme}
 \acro{PAPR}{peak-to-average power ratio}
 \acro{MAC}{medium access control}
 \acro{STO}{sampling time offset}
 \acro{STE}{straight-through estimator}
 \acro{PHY}{physical}
 \acro{CCE}{categorical cross-entropy}
 \acro{IoT}{Internet of Things}
 \acro{CCDF}{complementary cumulative distribution function}
 \acro{CRC}{cyclic redundancy check}
 \acro{ACLR}{adjacent channel leakage ratio}
\acro{MD}{missed detection}
 \acro{FA}{false alarm}
 \acro{FAR}{false alarm rate}
 \acro{BCJR}{Bahl--Cocke--Jelinek--Raviv}
 \acro{SCNC}{serially concatenated neural code}
 \acro{GNN}{graph neural network}
 \acro{VN}{variable node}
 \acro{CN}{check node}
 \acro{FN}{factor node}
 \acro{DUIDD}{deep-unfolded interleaved detection and decoding}
 \acro{ISI}{inter-symbol interference}
\acro{APP}{a posteriori probability}
\acro{EXIT}{extrinsic information transfer}
\acro{EP}{expectation propagation}
\acro{CS}{channel shortening}
\acro{FGNN}{factor graph neural network}
\acro{FCN}{fully connected network}
\acro{DCCB-SC}{deep
concatenated convolutional blocks with skip connections}
\acro{SPA}{sum-product algorithm}
\acro{CIR}{channel impulse response}
\acro{MAC}{multiply-accumulate}
\acro{CCT}{constant channel transformation}
\acro{GAT}{graph attention network}
\acro{i.i.d.}{independent and identically
distributed}
\end{acronym}

%% file: corporateColours.tex
\definecolor{mittelblau}{RGB}{0, 126, 198}
\definecolor{violettblau}{cmyk}{0.9, 0.6, 0, 0}
\definecolor{rot}{RGB}{238, 28 35}
\definecolor{apfelgruen}{RGB}{140, 198, 62}
\definecolor{gelb}{RGB}{1, 221, 0}
\definecolor{orange}{RGB}{244, 111, 33}
\definecolor{pink}{RGB}{237, 0, 140}
\definecolor{lila}{RGB}{128, 10, 145}
\definecolor{hellgrau}{RGB}{224, 224, 224}
\definecolor{mittelgrau}{RGB}{128, 128, 128}
\definecolor{dunkelgrau}{RGB}{80,80,80}
\definecolor{anthrazit}{RGB}{19, 31, 31}

%% file: plotcyclelists.tex
\pgfplotscreateplotcyclelist{corporate colours markers}{%
rot, every mark/.append style={fill=.!80!rot},mark=*\\%
mittelblau, every mark/.append style={fill=.!80!mittelblau},mark=square*\\%
apfelgruen, every mark/.append style={fill=.!80!apfelgruen},mark=triangle*\\%
orange, mark=star\\%
pink, every mark/.append style={fill=.!80!pink},mark=diamond*\\%
violettblau, every mark/.append style={fill=.!80!violettblau},mark=otimes*\\%
lila, mark=|\\%
gelb, every mark/.append style={fill=.!80!gelb},mark=pentagon*\\%
hellgrau, mark=text,text mark=p\\%
anthrazit, mark=text,text mark=a\\%
}

\pgfplotscreateplotcyclelist{corporate colours markers double}{%
rot, every mark/.append style={fill=.!80!rot},mark=*\\%
rot, dashed, every mark/.append style={fill=.!80!rot,solid},mark=*\\%
mittelblau, every mark/.append style={fill=.!80!mittelblau},mark=square*\\%
mittelblau, dashed, every mark/.append style={fill=.!80!mittelblau,solid},mark=square*\\%
apfelgruen, every mark/.append style={fill=.!80!apfelgruen},mark=triangle*\\%
apfelgruen, dashed, every mark/.append style={fill=.!80!apfelgruen,solid},mark=triangle*\\%
orange, mark=star\\%
orange, dashed, mark=star\\%
pink, every mark/.append style={fill=.!80!pink},mark=diamond*\\%
pink, dashed, every mark/.append style={fill=.!80!pink,solid},mark=diamond*\\%
violettblau, every mark/.append style={fill=.!80!violettblau},mark=otimes*\\%
violettblau, dashed, every mark/.append style={fill=.!80!violettblau,solid},mark=otimes*\\%
lila, mark=|\\%
lila, dashed, mark=|\\%
gelb, every mark/.append style={fill=.!80!gelb},mark=pentagon*\\%
gelb, dashed, every mark/.append style={fill=.!80!gelb,solid},mark=pentagon*\\%
}

\pgfplotsset{
  colormap/magma/.style={%
    /pgfplots/colormap={magma}{%
      rgb=(0.001462, 0.000466, 0.013866)
      rgb=(0.035520, 0.028397, 0.125209)
      rgb=(0.102815, 0.063010, 0.257854)
      rgb=(0.191460, 0.064818, 0.396152)
      rgb=(0.291366, 0.064553, 0.475462)
      rgb=(0.384299, 0.097855, 0.501002)
      rgb=(0.475780, 0.134577, 0.507921)
      rgb=(0.569172, 0.167454, 0.504105)
      rgb=(0.664915, 0.198075, 0.488836)
      rgb=(0.761077, 0.231214, 0.460162)
      rgb=(0.852126, 0.276106, 0.418573)
      rgb=(0.925937, 0.346844, 0.374959)
      rgb=(0.969680, 0.446936, 0.360311)
      rgb=(0.989363, 0.557873, 0.391671)
      rgb=(0.996580, 0.668256, 0.456192)
      rgb=(0.996727, 0.776795, 0.541039)
      rgb=(0.992440, 0.884330, 0.640099)
      rgb=(0.987053, 0.991438, 0.749504)
    },
  },
  colormap/inferno/.style={%
    /pgfplots/colormap={inferno}{%
      rgb=(0.001462, 0.000466, 0.013866)
      rgb=(0.037668, 0.025921, 0.132232)
      rgb=(0.116656, 0.047574, 0.272321)
      rgb=(0.217949, 0.036615, 0.383522)
      rgb=(0.316282, 0.053490, 0.425116)
      rgb=(0.410113, 0.087896, 0.433098)
      rgb=(0.503493, 0.121575, 0.423356)
      rgb=(0.596940, 0.154848, 0.398125)
      rgb=(0.688653, 0.192239, 0.357603)
      rgb=(0.775059, 0.239667, 0.303526)
      rgb=(0.851384, 0.302260, 0.239636)
      rgb=(0.912966, 0.381636, 0.169755)
      rgb=(0.956852, 0.475356, 0.094695)
      rgb=(0.981895, 0.579392, 0.026250)
      rgb=(0.987464, 0.690366, 0.079990)
      rgb=(0.973088, 0.805409, 0.216877)
      rgb=(0.947594, 0.917399, 0.410665)
      rgb=(0.988362, 0.998364, 0.644924)
    },
  },
  colormap/plasma/.style={%
    /pgfplots/colormap={plasma}{%
      rgb=(0.050383, 0.029803, 0.527975)
      rgb=(0.186213, 0.018803, 0.587228)
      rgb=(0.287076, 0.010855, 0.627295)
      rgb=(0.381047, 0.001814, 0.653068)
      rgb=(0.471457, 0.005678, 0.659897)
      rgb=(0.557243, 0.047331, 0.643443)
      rgb=(0.636008, 0.112092, 0.605205)
      rgb=(0.706178, 0.178437, 0.553657)
      rgb=(0.768090, 0.244817, 0.498465)
      rgb=(0.823132, 0.311261, 0.444806)
      rgb=(0.872303, 0.378774, 0.393355)
      rgb=(0.915471, 0.448807, 0.342890)
      rgb=(0.951344, 0.522850, 0.292275)
      rgb=(0.977856, 0.602051, 0.241387)
      rgb=(0.992541, 0.687030, 0.192170)
      rgb=(0.992505, 0.777967, 0.152855)
      rgb=(0.974443, 0.874622, 0.144061)
      rgb=(0.940015, 0.975158, 0.131326)
    },
  },
}

%% file: tables/table_1.tex
\begin{table}[tbp]
  \renewcommand{\arraystretch}{1.3}
  \caption{\neu{List of acronyms}}
  \label{tab:abb}
  \centering
  \setlength\tabcolsep{2pt}
  \begin{tabular}{l l|l l}
    BCE & binary cross-entropy &
    BMI & bit-wise mutual information\\
    CIR & channel impulse response&
    CCT & constant channel transformation\\
    CSI & channel state information&
    CNN & convolutional neural network\\
    FN & factor node&
    FFG & Forney factor graph \\
    FGNN & factor graph neural network&
    GNN & graph neural network\\
    ISI & inter-symbol interference&
    JDD & joint detection and decoding\\
    MLP & multilayer perceptron&
    NN & neural network\\
    RNN & recurrent neural network&
    SDD & separate detection and decoding\\
    SPA & sum-product algorithm&
    TDD & turbo detection and decoding \\
    UFG & Ungerboeck factor graph &
    VN & variable node \\
  \end{tabular}
\end{table}

%% file: figures/systemmodel.tikz
\begin{tikzpicture}[
    yscale=0.95,
    font=\footnotesize,
    block/.style={draw, rectangle, minimum size=1cm, align=center},
    gnn/.style={draw, rectangle, minimum size=1cm, align=center, draw=mittelblau,label=270:GNN},
	mapper/.pic = {
		\begin{axis}[xshift=-0.4cm, yshift=-0.4cm,
			width=0.8cm, height=0.8cm,
			scale only axis, hide axis, clip=false,
			xmin=-1.2, xmax=1.2, ymin=-1.2, ymax=1.2,
			domain=-1:1, samples=100, mark size=1pt]
			\addplot [color=black, mark=*] coordinates {(-0.707,0.707)};
			\addplot [color=black, mark=*] coordinates {(-0.707,-0.707)};
			\addplot [color=black, mark=*] coordinates {(0.707,0.707)};
			\addplot [color=black, mark=*] coordinates {(0.707,-0.707)};
			\addplot[thin] coordinates {(0,-1.2) (0,1.2)};
			\addplot[thin] coordinates {(-1.2,0) (1.2,0)};
		\end{axis}
	},
   tick/.pic = {
    		\draw[line width=0.5pt] (-0.4mm,-0.8mm) -- (0.4mm,0.8mm);
    	}
 ]
    \tikzstyle{arrow} = [thick,->]
    \tikzstyle{plus} =[circle,draw=black, fill=white, inner sep=1,minimum size=20pt, font=\normalsize];
\tikzset{
    between/.style args={#1 and #2}{
         at = ($(#1)!0.5!(#2)$)
    }
    }
    \coordinate (source) at (0.5,0.5);
    \node[block, draw=apfelgruen, label = 5G LDPC ](enc) at (2,0.5) {Enc.} ;
	\node[block, draw=apfelgruen, label= $M$-QAM](mapper)at (4,0.5) {Map.} ; 
    
    \node[block, minimum size=2cm, draw=rot, label={[,rotate=90, xshift=0.06cm]145:ISI channel}](chh) at (6.1,-0.95) {};
    \node[plus](chh) at (6,-0.5) {\large $*$};
    \node[plus](ch) at (6,-1.5) {$+$};
    \node[gnn ](eq) at (4,-2.5) {Det.};
    \node[gnn](dec) at (2,-2.5) {Dec.} ;
	\coordinate (sink) at (0.5,-2.5);

    \draw[arrow](source) -- node[above,yshift=2pt] {$\uv$} node[] (tickA) {} node[below] {\footnotesize $K$}(enc);
    
    \draw[arrow](enc) -- node[above,yshift=2pt] {$\cv$} node[] (tickB) {} node[below] {\footnotesize $N$}(mapper);
    
    \draw[arrow](mapper) -| node[above,yshift=2pt,xshift = -0.5cm] {$\xv$} node[,xshift = -0.5cm] (tickC) {} node[below, xshift = -0.5cm] {\footnotesize $N_\xv$}(chh);

     \draw[thick](chh) -- (ch);
     \draw[arrow](7,-0.5) --node[ above] {$\hv$} (chh);
     \draw[arrow](7,-1.5) --node[ above] {$\zv$} (ch);

    \draw[arrow]  (ch) |- node[above, xshift=-.5cm] {$\yv$} node[ xshift=-.5cm] (tickD) {} node[below, xshift=-.5cm] {\footnotesize $N+L$}(eq);
    
    \draw[arrow]  ([yshift=-0.2 cm]dec.east) -- node[below, align=center] {\footnotesize TDD} ([yshift=-0.2 cm]eq.west);
    
    \draw[arrow]  ([yshift=0.2 cm]eq.west) -- node[above, align=center] {\footnotesize SDD}([yshift=0.2 cm]dec.east);

    \draw[arrow]  (dec) -- node[above] {$\hat{\uv}$} node[] (tickE) {} node[below] {\footnotesize $K$} (sink);
    
    \pic at (tickA) {tick};
    \pic at (tickB) {tick};
    \pic at (tickC) {tick};
    \pic at (tickD) {tick};
    \pic at (tickE) {tick};
    
\draw[thick,black,decorate,decoration={brace,amplitude=8pt}] (1.5,-1.9)-- node[midway, below,yshift=20pt,]{Jointly optimized $\rightarrow$ JDD} (4.5,-1.9);
    
\end{tikzpicture}

%% file: tables/table_2.tex
\begin{table}[h]
  \caption{\neu{List of variables}}
  \label{tab:abb}
  \centering
  \setlength\tabcolsep{2pt}
  \begin{tabular}{l l | l l}
      $\mathcal{F}$ & set of \acp{FN} &
    $\cv$ & coded bits\\
    $\gv$ & attributes&
    $\hv$ & channel impulse response\\
    $\Hm$ & channel matrix&
    $I$ & mutual information\\
    $K$ & number of uncoded bits&
    $\mathcal{L}$ & loss function\\
    $L$ & channel memory&
    $\boldsymbol{\ell}$ & log-likelihood ratios \\
    $\mv$ & messages&
    $M$ & modulation alphabet size\\
    $N_\xv$ & number of symbols&
    $N$ & number of coded bits\\
    $N_\mathrm{It}$ & number of iterations&
    $R$ & achievable rate\\
    $R_\mathrm{C}$ & code rate &
    $\sv$ & states\\
    $\boldsymbol{\theta}$ & trainable variables&
    $\hat{\uv}$ & estimated uncoded bits\\
    $\uv$ & uncoded bits&
    $\mathcal{V}$ & set of \acp{VN}\\
    $\xv$ & symbols&
    $\yv$ & channel observation\\
  \end{tabular}
\end{table}

%% file: figures/ufg.tikz
\begin{tikzpicture}[
    xscale=1.4,
    squarenode/.style={draw, rectangle, minimum size=1.55em},
    nnnode/.style={draw, rectangle, minimum size=1.2em, inner sep=0pt, fill=white},
    circlenode/.style={draw, circle, minimum size=1.55em,lila},
    arrow/.style={->},
    nnarrow/.style={-{Latex[length=2pt 3 1]}, line width=0.7pt,  mittelblau},
    conn/.style={},
    connd/.style={dashed},
    ]

    \node (l2) at (2-0.5,0.8) {\footnotesize $\chi_{i-2}$};
    \node (l3) at (3-0.5,0.8) {\footnotesize $\chi_{i-1}$};
    \node (l4) at (4-0.5,0.8) {\footnotesize $\chi_{i}$};
    \node (l5) at (5-0.5,0.8) {\footnotesize $\chi_{i+1}$};
    \node (l6) at (6-0.5,0.8) {\footnotesize $\chi_{i+2}$};

    \node[nnnode, draw=white, rotate=-90] (text) at (6.30,-2) {\tiny FN};
    \node[nnnode, draw=white, text=lila, rotate=-90] (text) at (6.3,0.0) {\tiny VN};
       \foreach \x in {2,...,6} {
        \node [circlenode] (v\x) at ({\x-0.5},0) {};
        \draw [arrow] (l\x) -- (v\x);

        \node [squarenode] (cb\x) at (\x,-2) {};
        \node [squarenode] (cc\x) at (\x-0.5,-2) {};
           \draw [conn] (cb\x) -- (v\x);
             \draw [conn] (cc\x) -- (v\x);
     
    }
      \foreach \x in {2,...,6} {
      \pgfmathtruncatemacro{\nextx}{\x+1}
       \pgfmathtruncatemacro{\nextxx}{\x+2}
    \ifthenelse{\x = 2 \OR \x = 3 \OR \x = 4}{\draw [conn] (cb\nextx) -- (v\x);} {}
      \ifthenelse{\x = 2 \OR \x = 3 \OR \x = 4}{\draw [conn] (cc\nextxx.north) -- (v\x);} {}
    }
    
    \coordinate (v0) at (-0.5,0);
    \coordinate (v1) at (0.5,0);
    \coordinate (cb7) at (7,-2);
    \coordinate (cc7) at (6.5,-2);
    \coordinate (cb8) at (8,-2);
    \coordinate (cc8) at (7.5,-2);
    
    \draw[connd] ( $ (v1)!0.3!(cb2) $ ) -- (cb2);
    \draw[connd] ( $ (v1)!0.2!(cc3.north) $ ) -- (cc3.north);
    \draw[connd] ( $ (v0)!0.7!(cc2.north) $ ) -- (cc2.north);
          
    \draw[connd] (v5) -- ( $ (v5)!0.7!(cc7.north) $ );
  \draw[connd] (v6) -- ( $ (v6)!0.8!(cb7) $ );
   \draw[connd] (v6) -- ( $ (v6)!0.6!(cc8.north) $ );

    \draw [conn] (cb6) -- (v5);
    \node at (1,0) {$\cdots$};
    \node at (1,-2) {$\cdots$};
    \node at (6.7,0) {$\cdots$};
    \node at (6.7,-2) {$\cdots$};

    \node at (1.5,-2.6) {\footnotesize $i-2$};
    \node at (2.5,-2.6) {\footnotesize $i-1$};
    \node at (3.5,-2.6) {\footnotesize $i$};
    \node at (4.5,-2.6) {\footnotesize $i+1$};
    \node at (5.5,-2.6) {\footnotesize $i+2$};

\end{tikzpicture}

%% file: figures/ffg.tikz
\begin{tikzpicture}[
    xscale=1.4,
    squarenode/.style={draw, rectangle, minimum size=1.85em},
    nnnode/.style={draw, rectangle, minimum size=1.2em, inner sep=0pt, fill=white},
    circlenode/.style={draw, circle, minimum size=1.85em,lila},
    arrow/.style={->},
    nnarrow/.style={-{Latex[length=2pt 3 1]}, line width=0.7pt,  mittelblau},
    conn/.style={},
    connd/.style={dashed},
    ]

    \node (l2) at (2,0.8) {\footnotesize $y_{i-2}$};
    \node (l3) at (3,0.8) {\footnotesize $y_{i-1}$};
    \node (l4) at (4,0.8) {\footnotesize $y_{i}$};
    \node (l5) at (5,0.8) {\footnotesize $y_{i+1}$};
    \node (l6) at (6,0.8) {\footnotesize $y_{i+2}$};

    \node[nnnode, draw=white, rotate=-90] (text) at (6.30,0) {\tiny FN};
    \node[nnnode, draw=white, rotate=-90, text=lila] (text) at (6.30,-2) {\tiny VN};
    \foreach \x in {2,...,6} {
        
        \ifthenelse{\x = 4}{\node [squarenode, draw=mittelblau] (c\x) at (\x,0) {\tiny MLP};}{\node [squarenode] (c\x) at (\x,0) {};};
        
        \ifthenelse{\x = 4}{\node [circlenode, draw=mittelblau,text=black] (v\x) at (\x,-2) {\tiny MLP};   }{\node [circlenode] (v\x) at (\x,-2) {};   };
        \ifthenelse{\x = 4}{}{\draw [conn] (c\x) -- (v\x);};
        \draw [arrow] (l\x) -- (c\x);
    }

    \coordinate (v0) at (0,-2);
    \coordinate (v1) at (1,-2);
    \coordinate (c7) at (7,0);
    \coordinate (c8) at (8,0);

    \node at (1.3,0) {$\cdots$};
    \node at (1.3,-2) {$\cdots$};
    \node at (6.7,0) {$\cdots$};
    \node at (6.7,-2) {$\cdots$};

    \node at (2,-2.6) {\footnotesize $i-2$};
    \node at (3,-2.6) {\footnotesize $i-1$};
    \node at (4,-2.6) {\footnotesize $i$};
    \node at (5,-2.6) {\footnotesize $i+1$};
    \node at (6,-2.6) {\footnotesize $i+2$};

    \draw[connd] ( $ (v0)!0.6!(c2) $ ) -- (c2);
    \draw[connd] ( $ (v1)!0.2!(c2) $ ) -- (c2);
    \draw[connd] ( $ (v1)!0.1!(c3) $ ) -- (c3);

    \draw[conn] (v2) -- (c3);
    \draw[conn] (v2) -- (c4);
    \draw[conn] (v3) -- (c4);
    \draw[conn] (v3) -- (c5);
    \draw[conn] (v4) -- (c5);
    \draw[conn] (v4) -- (c6);
    \draw[conn] (v5) -- (c6);

    \draw[connd] (v5) -- ( $ (v5)!0.9!(c7) $ );
    \draw[connd] (v6) -- ( $ (v6)!0.8!(c7) $ );
    \draw[connd] (v6) -- ( $ (v6)!0.4!(c8) $ );

    \draw[nnarrow, mittelblau] ( $ (v2)!0.8!(c4) $ ) -- (c4);
    \draw[nnarrow, mittelblau] ( $ (v3)!0.8!(c4) $ ) -- (c4);

    \draw[nnarrow, mittelblau] ( $ (c6)!0.8!(v4) $ ) -- (v4);
    \draw[nnarrow, mittelblau] ( $ (c5)!0.8!(v4) $ ) -- (v4);

    \node[nnnode, draw=white, rotate=-90] (text) at (4.40,-1) {\tiny MLP};
    \node[nnnode, draw=mittelblau] (unn) at (4.15,-1) {\tikz[baseline=-0.5ex,                                 shorten <=2pt, shorten >=2pt] \draw[-{Latex[length=1.5pt 3 1]}, thin] (0,-.15) -- (0,.15);};

    \node[nnnode, draw=white, rotate=90] (text) at (3.60,-1) {\tiny MLP};
    \node[nnnode, draw=mittelblau] (dnn) at (3.85,-1) {\tikz[baseline=-0.5ex,                                 shorten <=2pt, shorten >=2pt] \draw[-{Latex[length=1.5pt 3 1]}, thin] (0,.15) -- (0,-.15);};

    \draw[nnarrow] (dnn.300) -- (v4.103);
    \draw[nnarrow, mittelblau] (v4.65) -- (unn.255);
    \draw[nnarrow, mittelblau] (v4.130) -- (dnn.263);

    \draw[nnarrow, mittelblau] (unn.123) -- (c4.282);
    \draw[nnarrow, mittelblau] (c4.300) -- (unn.94);
    \draw[nnarrow, mittelblau] (c4.250) -- (dnn.68);
    
    \node[rectangle,dashed, draw=mittelblau, minimum height=3.2cm, minimum width=1.3cm, label=below:\footnotesize GNN structure] (rect) at (4.0,-1.2) {};
\end{tikzpicture}

%% file: tikz/bmi_uncoded_bpsk.tikz
\begin{tikzpicture}[spy using outlines={rectangle, magnification=4, connect spies}]
		
		\pgfplotsset{compat=1.5}

		\begin{axis}[
			xmode=normal,
			ymode=normal,
			xlabel=\footnotesize $E_\mathrm{b}/N_0~(\mathrm{dB})$, %
			ylabel=\footnotesize $R_\mathrm{SDD}^\mathrm{BMI}$,
			xmin = 5,
			xmax=15,
			ymax=1,
			ymin=0.2,
			mark size=2.5pt,
			legend style={at={(axis cs:6,0.0001)},anchor=south west}, 
			grid=both,
			minor grid style={gray!25},
			major grid style={gray!25},
			width=\linewidth,
	        height=0.99\linewidth,
			cycle list name=corporate colours markers,
			legend cell align={left},
			line width=0.8pt, %
			]

            \addplot+ [black, mark options={solid},mark=x, line width=\lw, mark size= \msize, name path=map] 
			table[x expr=\thisrowno{0}-10*log10(1-(\thisrowno{3})) ,y expr=1-(\thisrowno{3}),col sep=comma]{./tikz/results/uncoded_rate_bpsk/uncoded_proakis_C_n132_QPSK_BCJR.txt};
			\label{plot:BCJR_bmi}
			
			\plot[name path=f15,thick,opacity=0,samples=100,domain=4:20.0] {5};
            \addplot[opacity=0.3,gray, pattern=north west lines] fill between[of=map and f15];
            \node[align=center] at (axis cs: 7,0.85) {\footnotesize Not\\\footnotesize achievable};

            \addplot+ [mittelblau!70!black, mark options={solid},mark=o, line width=\lw, mark size= \msize] 
			table[x expr=\thisrowno{0}-10*log10(1-\thisrowno{3}) ,y expr=1-\thisrowno{3},col sep=comma]{./tikz/results/uncoded_rate_bpsk/gnn_fgnn_iterativ_BPSK_proakisC_it_10_results.txt};
			\label{plot:fgnn_bmi}
            
            \addplot+ [mittelgrau, mark options={solid},mark=+, line width=\lw, mark size= \msize] 
			table[x expr=\thisrowno{0}-10*log10(1-\thisrowno{3}) ,y expr=1-\thisrowno{3},col sep=comma]{./tikz/results/uncoded_rate_bpsk/fancy_cnn_results.txt};
			\label{plot:fancy_cnn_bmi}

            \addplot+ [rot!60!white, mark options={solid},mark=pentagon, line width=\lw, mark size= \msize] 
			table[x expr=\thisrowno{0}-10*log10(1-(\thisrowno{3})) ,y expr=1-(\thisrowno{3}) ,col sep=comma]{./tikz/results/uncoded_rate_bpsk/uncoded_proakis_C_n132_QPSK_damp_038_NNBP_8.txt};
			\label{plot:nbp_ffg_8_bmi}

            \addplot+ [rot, mark options={solid},mark=heart, line width=\lw, mark size= \msize]
			table[x expr=\thisrowno{0}-10*log10(1-(\thisrowno{27})) ,y expr=1-(\thisrowno{27}),col sep=comma]{./tikz/results/uncoded_rate_bpsk/uncoded_proakis_C_n132_QPSK_damp_038_BP_10_supplement.txt};
			\label{plot:bp_ffg_8_bmi}

			\addplot+ [lila!60!white, mark options={solid},mark=Mercedes star, line width=\lw, mark size= \msize]
			table[x expr=\thisrowno{0}-10*log10(1-(\thisrowno{3})) ,y expr=1-(\thisrowno{3}),col sep=comma]{./tikz/results/uncoded_rate_bpsk/uncoded_proakis_C_n132_BPSK_EP_16_beta001_llrout_05_llrmax_4.txt};
			\label{plot:ep_24_bmi_uncoded}

			\addplot+ [lila, mark options={solid},mark=star, line width=\lw, mark size= \msize]
			table[x expr=\thisrowno{0}-10*log10(1-(\thisrowno{3})) ,y expr=1-(\thisrowno{3}),col sep=comma]{./tikz/results/uncoded_rate_bpsk/uncoded_proakis_C_n132_QPSK_LMMSE.txt};
			\label{plot:lmmse_bmi}

\addplot+ [mittelblau, mark options={fill=mittelblau, solid},mark=square, line width=\lw, mark size= \msize] 
			table[x expr=\thisrowno{0}-10*log10(1-(\thisrowno{3})) ,y expr=1-(\thisrowno{3}) ,col sep=comma]{./tikz/results/uncoded_bpsk/gnn_FFG_porakis_C_n_132_it_8_bmi_alvarado.txt};
			\label{plot:gnn_ffg_8_bmi}

            \addplot+ [mittelblau!60!white,mark options={solid},mark=triangle, line width=\lw, mark size= \msize] 
			table[x expr=\thisrowno{0}-10*log10(1-(\thisrowno{3})) ,y expr=1-(\thisrowno{3}) ,col sep=comma]{./tikz/results/uncoded_bpsk/gnn_ungerboek_proakis_C_n132_it_8_alvarado.txt};
			\label{plot:gnn_ung_8_bmi}

        \coordinate (legend) at (axis cs:5,-0.2);
        \coordinate (spypoint) at (axis cs:9.85,0.9);
        \coordinate (spyviewer) at (axis cs:5,0.8);
		\end{axis}
  \matrix [
draw=none,
fill=white,
fill opacity=1.0,
text opacity=1,
matrix of nodes,
align =left,
column sep = 0,
inner sep= 2,
anchor=south west,
font=\footnotesize,
column 1/.style={anchor=base west},
column 2/.style={anchor=base west},
column 3/.style={anchor=base west},
column 4/.style={anchor=base west},
column 5/.style={anchor=base west},
column 6/.style={anchor=base west},
column 7/.style={anchor=base west},
column 8/.style={anchor=base west},
column 9/.style={anchor=base west},
column 10/.style={anchor=base west},
column 11/.style={anchor=base west},
column 12/.style={anchor=base west},
mark options={solid},
] at (legend) {
	\ref{plot:BCJR_bmi} & BCJR \cite{bahl1974optimal}&  \ref{plot:gnn_ffg_8_bmi} & GNN FFG  $N_\mathrm{It}=8$ & \ref{plot:gnn_ung_8_bmi} & GNN UFG  $N_\mathrm{It}=8$ &\ref{plot:fgnn_bmi} & FGNN   $N_\mathrm{It}=8$ \\ \ref{plot:fancy_cnn_bmi} & CNN \cite{huang21extrinsic}&
	\ref{plot:nbp_ffg_8_bmi} & N-SPA FFG \cite{schmid22neuralBP}  $N_\mathrm{It}=8$ &
	\ref{plot:bp_ffg_8_bmi} & SPA FFG \cite{ffg05covalope} $N_\mathrm{It}=8$ &
	\ref{plot:ep_24_bmi_uncoded} & EP \cite{santos2015block}  $N_\mathrm{It}=16$ &
	\ref{plot:lmmse_bmi} & LMMSE \cite{proakis2001digital}\\
};

	\end{tikzpicture}

%% file: tikz/bmi_uncoded_qam16.tikz
\begin{tikzpicture}
		
		\pgfplotsset{compat=1.5}

		\begin{axis}[
			xmode=normal,
			ymode=normal,
			xlabel=\footnotesize $E_\mathrm{b}/N_0~(\mathrm{dB})$, %
			xmin = 9,
			xmax=30,
			ymax=1,
			ymin=0.2,
			mark size=2.5pt,
			legend style={at={(axis cs:6,0.0001)},anchor=south west}, 
			grid=both,
			minor grid style={gray!25},
			major grid style={gray!25},
			width=\linewidth,
	        height=0.99\linewidth,
			cycle list name=corporate colours markers,
			legend cell align={left},
			line width=0.8pt, %
			]

            \addplot+ [black, mark options={solid},mark=x, line width=\lw, mark size= \msize, name path = map] 
			table[x expr=\thisrowno{0}-10*log10(1-(\thisrowno{3})) ,y expr=1-(\thisrowno{3}),col sep=comma]{./tikz/results/uncoded_QAM16/uncoded_QAM16_proakis_C_BCJR.txt};
			\label{plot:BCJR_qam16_bmi}
			
			\plot[name path=f15,thick,opacity=0,samples=100,domain=4:20.0] {5};
            \addplot[opacity=0.3,gray, pattern=north west lines] fill between[of=map and f15];

			\addplot+ [lila!60!white, mark options={solid},mark=Mercedes star, line width=\lw, mark size= \msize]
			table[x expr=\thisrowno{0}-10*log10(1-(\thisrowno{3})) ,y expr=1-(\thisrowno{3}),col sep=comma]{./tikz/results/uncoded_QAM16/uncoded_n_500_proakis_C_n132_16QAM_EP_24_miv_1e08_maxv_100_llrmax_7_beta_00002_llrout_999.txt};
			\label{plot:ep_16_qam16_bmi}

			\addplot+ [lila, mark options={solid},mark=star, line width=\lw, mark size= \msize]
			table[x expr=\thisrowno{0}-10*log10(1-(\thisrowno{3})) ,y expr=1-(\thisrowno{3}),col sep=comma]{./tikz/results/uncoded_QAM16/uncoded_n_500_proakis_C_n132_16QAM_LMMSE.txt};
			\label{plot:lmmse_qam16_bmi}

            \addplot+ [mittelgrau, mark options={solid},mark=+, line width=\lw, mark size= \msize] 
			table[x expr=\thisrowno{0}-10*log10(1-(\thisrowno{3}))  ,y expr=1-(\thisrowno{3}),col sep=comma]{./tikz/results/uncoded_QAM16/cnn_results_per_snr.txt};
			\label{plot:cnn_qam16}
			
			\addplot+ [mittelblau!60!black, mark options={solid},mark=o, line width=\lw, mark size= \msize] 
			table[x expr=\thisrowno{0}-10*log10(1-(\thisrowno{3}))  ,y expr=1-(\thisrowno{3}),col sep=comma]{./tikz/results/uncoded_QAM16/use_5g_0_channel_proakis_c_n_128_Mb_4_const_qam_FGNN_light_Equalizer_shared_iter_readyresults_per_snr.txt};
			\label{plot:fgnn_qam16}

			\addplot+ [mittelblau, mark options={fill=mittelblau, solid},mark=square, line width=\lw, mark size= \msize] 
			table[x expr=\thisrowno{0}-10*log10(1-(\thisrowno{3})) ,y expr=1-(\thisrowno{3}) ,col sep=comma]{./tikz/results/uncoded_QAM16/gnn_share_edge_ITERATIVE_16QAM_proakisC_it_6_Units_96_F_48_snr_18_it_8_checkpoint_finetuneresults_per_snr.txt};
			\label{plot:gnn_ffg_8_bmi}

            \addplot+ [mittelblau!60!white,mark options={solid},mark=triangle, line width=\lw, mark size= \msize] 
			table[x expr=\thisrowno{0}-10*log10(1-(\thisrowno{3})) ,y expr=1-(\thisrowno{3}) ,col sep=comma]{./tikz/results/uncoded_QAM16/gnn_ungerboek_results_per_snr.txt};
			\label{plot:gnn_ung_qam16_bmi}

        \coordinate (legend) at (axis cs:15,-0.158);
        \coordinate (spypoint) at (axis cs:9.85,0.9);
        \coordinate (spyviewer) at (axis cs:5,0.8);
		\end{axis}
		\node[white,opacity=0.0] at (legend) {test};

	\end{tikzpicture}

%% file: qam64/qam_64.tex
\begin{tikzpicture}
		
		\pgfplotsset{compat=1.5}

		\begin{axis}[
			xmode=normal,
			ymode=normal,
			xlabel=\footnotesize $E_\mathrm{b}/N_0~(\mathrm{dB})$, %
			xmin = 15,
			xmax=30,
			ymax=1,
			ymin=0.2,
			mark size=2.5pt,
			legend style={at={(axis cs:6,0.0001)},anchor=south west}, 
			grid=both,
			minor grid style={gray!25},
			major grid style={gray!25},
			width=\linewidth,
	        height=0.99\linewidth,
			cycle list name=corporate colours markers,
			legend cell align={left},
			line width=0.8pt, %
			]

            \addplot+ [black, mark options={solid},mark=x, line width=\lw, mark size= \msize, name path = map] 
			table[x expr=\thisrowno{0}-10*log10(1-(\thisrowno{3})) ,y expr=1-(\thisrowno{3}),col sep=comma]{./qam64/use_5g_0_channel_proakis_c_n_66_Mb_3_const_pam_BCJR.txt};
			\label{plot:BCJR_qam16_bmi}
			
			\plot[name path=f15,thick,opacity=0,samples=100,domain=4:20.0] {5};
            \addplot[opacity=0.3,gray, pattern=north west lines] fill between[of=map and f15];
            \node[align=center] at (axis cs: 12,0.85) {Not\\achievable};

			\addplot+ [lila!60!white, mark options={solid},mark=Mercedes star, line width=\lw, mark size= \msize]
			table[x expr=\thisrowno{0}-10*log10(1-(\thisrowno{3})) ,y expr=1-(\thisrowno{3}),col sep=comma]{./qam64/uncoded_QAM64_proakis_C_EP7.txt};
			\label{plot:ep_16_qam16_bmi}

			\addplot+ [lila, mark options={solid},mark=star, line width=\lw, mark size= \msize]
			table[x expr=\thisrowno{0}-10*log10(1-(\thisrowno{3})) ,y expr=1-(\thisrowno{3}),col sep=comma]{./qam64/use_5g_0_channel_proakis_c_n_252_Mb_6_const_qam_LMMSE.txt};
			\label{plot:lmmse_qam16_bmi}

			\addplot+ [mittelblau, mark options={fill=mittelblau, solid},mark=square, line width=\lw, mark size= \msize] 
			table[x expr=\thisrowno{0}-10*log10(1-(\thisrowno{3})) ,y expr=1-(\thisrowno{3}) ,col sep=comma]{./qam64/results_per_snr.txt};
			\label{plot:gnn_ffg_8_bmi}

        \coordinate (legend) at (axis cs:15,-0.157);
        \coordinate (spypoint) at (axis cs:9.85,0.9);
        \coordinate (spyviewer) at (axis cs:5,0.8);
		\end{axis}
		\node[white, opacity=0.0] at (legend) {test};

	\end{tikzpicture}

%% file: tables/table_3.tex
\begin{table}[tbp]
  \renewcommand{\arraystretch}{1.3}
  \caption{Hyperparameters of the \acp{GNN} and training}
  \label{tab:table_example}
  \centering
  \setlength\tabcolsep{2pt}
  \begin{tabular}{c l c | l c}
  & NN Parameter & Value & Train. Parameter& Value \\
    \hline
    \hline
    \multirow{4}{*}{\rotatebox{90}{Detection}} & \# \ac{MLP} hidden layers & 2 & Learning rate & $10^{-4}$ \\
    &\# \ac{MLP} hidden units & 64 & Batch size & $256$ \\
    &activation & ReLU & Epochs Det. &  $5\cdot10^4$ \\
    & Feature size $d$ & 16 & SNR Det.&  \qty{10}{dB} to \qty{14}{dB} \\
    \cline{1-5}
    \multirow{2}{*}{\rotatebox{90}{JED}}  &Schedule Flooding & ($10$,\,$1$) & Epochs JDD& $1.6\cdot10^5$ \\
    &Schedule Sequential  & ($3$,\,[$3,5$]) & SNR JDD&  \qty{10}{dB} to \qty{13}{dB}\\
  \end{tabular}
  \vspace{-2em}
\end{table}

%% file: tikz/latency_uncoded.tikz
\begin{tikzpicture}
		
		\pgfplotsset{compat=1.5}

		\begin{axis}[
			xmode=log,
			ymode=log,
			xlabel=\footnotesize $\mathrm{ Latency~(clock~cycles)}$, %
			ylabel=\footnotesize $\mathrm{BER}$,
			xmin = 2,
			xmax=400,
			ymax=3*10^(-1),
			ymin=10^(-5),
			mark size=2.5pt,
			legend style={at={(axis cs:400,0.3)},anchor=north east,	draw=none,		fill=white,
            fill opacity=0.7,
            text opacity=1}, 
			grid=both,
			minor grid style={gray!25},
			major grid style={gray!25},
			width=\linewidth,
	        height=0.8\linewidth,
			cycle list name=corporate colours markers,
			legend cell align={left},
			line width=0.8pt, %
			]
			
			\addplot+ [lila!60!white, dashed,mark options={solid},mark=Mercedes star, line width=\lw, mark size= \msize] 
			table[x expr=\thisrowno{0}+0.00 ,y=ber,col sep=comma]{./tikz/results/latency_uncoded_bpsk/latency_uncoded_proakis_C_n132_BPSK_EP_16_beta001_llrout_05_llrmax_4_supplement.txt};
			\label{plot:ep_24_bmi}
            \addlegendentry{\footnotesize EP \cite{santos2015block} };
            
            \addplot+ [mittelblau!60!black, dashed,mark options={solid},mark=o, line width=\lw, mark size= \msize] 
			table[x expr=\thisrowno{0}+0.00 ,y=ber,col sep=comma]{./tikz/results/latency_uncoded_bpsk/latency_gnn_fgnn_iterativ_BPSK_proakisC_it_10.txt};
			\label{plot:ep_24_bmi}
            \addlegendentry{\footnotesize FGNN};

            \addplot[scatter,only marks,point meta=explicit symbolic,scatter/classes={
            c={mark=+,draw=mittelgrau}}]%
                table[x=x,y=y, meta=label] {
                x y label
                24 2.817e-04 c
                };
                \label{plot:latency_fancy_cnn}
                \addlegendentry{\footnotesize CNN \cite{huang21extrinsic}};

            \addplot[scatter,only marks,point meta=explicit symbolic,scatter/classes={
            c={mark=x,draw=black}}]%
                table[x=x,y=y, meta=label] {
                x y label
                72 6.317e-05 c
                138 6.317e-05 c
                138 6.317e-05 c
                262 6.317e-05 c
                };
                \label{plot:latency_bcjr}
            \addlegendentry{\footnotesize BCJR \cite{bahl1974optimal}};
			\node[] at (axis cs: 72, 0.00004) {\footnotesize $N=66$};
			\node[] at (axis cs: 138, 0.00002) {\footnotesize $N=132$};
            \draw[-latex, line width=1pt, color=black] (axis cs: 138, 3e-5) --node[below,yshift=0cm, xshift=.1cm, color=black]  {} (axis cs: 138, 5e-5);

   \node[] at (axis cs: 250, 0.000035) {\footnotesize $N=256$};

            \addplot+ [rot, dashed,mark options={solid},mark=heart, line width=\lw, mark size= \msize] 
			table[x expr=\thisrowno{0}+0.00 ,y=ber,col sep=comma]{./tikz/results/latency_uncoded_bpsk/latency_bp_proakis_c_it_5_damped038.txt};
			\label{plot:latency_bp}
            \addlegendentry{\footnotesize SPA \ac{FFG} \cite{colavolpe05bpdetection}};
            
            \addplot+ [rot!60!white, dashed,mark options={solid},mark=pentagon, line width=\lw, mark size= \msize] 
			table[x expr=\thisrowno{0}+0.00 ,y=ber,col sep=comma]{./tikz/results/latency_uncoded_bpsk/latency_neural_bp_proakis_c_it_7.txt};
			\label{plot:latency_nbp}
            \addlegendentry{\footnotesize N-SPA \ac{FFG} \cite{schmid22neuralBP}};

            \addplot+ [mittelblau, dashed,mark options={solid},mark=square, line width=\lw, mark size= \msize] 
			table[x expr=\thisrowno{0}+0.00 ,y=ber,col sep=comma]{./tikz/results/latency_uncoded_bpsk/latency_gnn_ffg_uncoded.txt};
			\label{plot:latency_gnn_ffg_uncoded}
            \addlegendentry{\footnotesize GNN \ac{FFG}};

            \addplot+ [mittelblau!60!white, dashed,mark options={solid},mark=triangle, line width=\lw, mark size= \msize] 
			table[x expr=\thisrowno{0}+0.00 ,y=ber,col sep=comma]{./tikz/results/latency_uncoded_bpsk/latency_gnn_ung_uncoded.txt};
			\label{plot:latency_gnn_ung_uncoded}
            \addlegendentry{\footnotesize GNN \ac{UFG}};

   \draw[-latex, dashed,line width=0.75mm, color=black] (axis cs: 2.1, 6e-2) --node[below,yshift=0cm, xshift=.0cm, color=black, sloped,  align=center]  { \footnotesize Detection \footnotesize Iterations} (axis cs: 20, 1e-4);
        
		\end{axis}

	\end{tikzpicture}

%% file: tikz/bmi_uncoded_qpsk_tdl_7.tikz
\begin{tikzpicture}[spy using outlines={rectangle, magnification=4, connect spies}]
		
		\pgfplotsset{compat=1.5}
		
\begin{groupplot}[
    group style={
        group size= 3 by 1, 
        horizontal sep=1.5cm,
    },
	width=0.33\linewidth,
	height=6cm,
	scale=1,
	minor x tick num=1,
	xmajorgrids,
	xminorgrids=true,
	ymajorgrids,
	yminorticks=true,
    grid=both,
    tick align=outside,
    tick pos=left,
    minor grid style={gray!25},
	major grid style={gray!25},		
    xlabel={\(\displaystyle E_\mathrm{b} \, / \, N_0\) in dB},
    xtick style={color=black},
    ymajorgrids,
    ytick style={color=black},
]
\nextgroupplot[
    title={\footnotesize (a) Perfect CSI, TDL-C},
			xmode=normal,
			ymode=normal,
			xlabel=\footnotesize $E_\mathrm{b}/N_0~(\mathrm{dB})$, %
			ylabel=\footnotesize $R_\mathrm{SDD}^\mathrm{BMI}$,
			xmin = 2,
			xmax=8,
			ymax=1,
			ymin=0.6,
			mark size=2.5pt,
			legend style={at={(axis cs:6,0.0001)},anchor=south west}, 
			grid=both,
			minor grid style={gray!25},
			major grid style={gray!25},
			cycle list name=corporate colours markers,
			legend cell align={left},
			line width=0.8pt, %
			]

            \addplot+ [mittelgrau, mark options={solid},mark=+, line width=\lw, mark size= \msize]
			table[x expr=\thisrowno{0}-10*log10(1-(\thisrowno{3})) ,y expr=1-(\thisrowno{3}),col sep=comma]{./tikz/results/uncoded_qpsk_tdl_7/fancy_cnn_results_per_snr_genie_h.txt};
			\label{plot:cnn_qpsk_bmi}

            \addplot+ [black, mark options={solid},mark=x, line width=\lw, mark size= 3pt, name path=map] 
			table[x expr=\thisrowno{0}-10*log10(1-(\thisrowno{3})) ,y expr=1-(\thisrowno{3}),col sep=comma]{./tikz/results/uncoded_qpsk_tdl_7/use_5g_0_channel_5g_tdl_b_7_n_256_Mb_2_const_qam_bcjr.txt};
			\label{plot:BCJR_tdl_bmi}
			
			\plot[name path=f15,thick,opacity=0,samples=100,domain=0:8.0] {5};
            \addplot[opacity=0.3,gray, pattern=north west lines] fill between[of=map and f15];
            \node[align=center] at (axis cs: 3.5,0.95) {Not\\achievable};

            \addplot+ [mittelblau!60!white, mark options={fill=mittelblau!60!white, solid},mark=triangle, line width=\lw, mark size= \msize] 
			table[x expr=\thisrowno{0}-10*log10(1-(\thisrowno{3})) ,y expr=1-(\thisrowno{3}) ,col sep=comma]{./tikz/results/uncoded_qpsk_tdl_7/use_5g_0_channel_5g_tdl_b_7_n_128_Mb_2_const_qam_GNN_Equalizer_LLR_inresults_per_snr_genie_h.txt};
			\label{plot:gnn_llr_genie_h}

            \addplot+ [mittelblau, mark options={fill=mittelblau, solid},mark=square, line width=\lw, mark size= \msize] 
			table[x expr=\thisrowno{0}-10*log10(1-(\thisrowno{3})) ,y expr=1-(\thisrowno{3}) ,col sep=comma]{./tikz/results/uncoded_qpsk_tdl_7/use_5g_0_channel_5g_tdl_b_7_n_128_Mb_2_const_qam_GNN_Equalizer_NN_in_iter_ready_results_per_snr_genie_h.txt};
			\label{plot:gnn_nn_genie_h}

            \addplot+ [mittelblau!60!black, mark options={fill=mittelblau!60!black, solid},mark=o, line width=\lw, mark size= \msize] 
			table[x expr=\thisrowno{0}-10*log10(1-(\thisrowno{3})) ,y expr=1-(\thisrowno{3}) ,col sep=comma]{./tikz/results/uncoded_qpsk_tdl_7/use_5g_0_channel_5g_tdl_b_7_n_128_Mb_2_const_qam_GNN_Equalizer_Shared_Edge_Attr_VNSUM_CS_iter_ready_results_per_snr_genie_h.txt};
			\label{plot:gnn_cs_genie_h}

			\addplot+ [lila!60!white, mark options={solid},mark=Mercedes star, line width=\lw, mark size= \msize]
			table[x expr=\thisrowno{0}-10*log10(1-(\thisrowno{3})) ,y expr=1-(\thisrowno{3}),col sep=comma]{./tikz/results/uncoded_qpsk_tdl_7/uncoded_5gtdl_7_qpsk_n396_ep_llr_7_it_8_beta_005_llr_099.txt};
			\label{plot:ep_qpsk_bmi_hgenie}

			\addplot+ [lila, mark options={solid},mark=star, line width=\lw, mark size= \msize]
			table[x expr=\thisrowno{0}-10*log10(1-(\thisrowno{3})) ,y expr=1-(\thisrowno{3}),col sep=comma]{./tikz/results/uncoded_qpsk_tdl_7/uncoded_5gtdl_7_LMMSE_qpsk_n264.txt};
			\label{plot:lmmse_qpsk_bmi}

        \coordinate (legend) at (axis cs:0,0.4);
        \coordinate (spypoint) at (axis cs:9.85,0.9);
        \coordinate (spyviewer) at (axis cs:5,0.8);

\nextgroupplot[
    title={\footnotesize(b) Perfect CSI, TDL-A (no retraining)},
			xmode=normal,
			ymode=normal,
			xlabel=\footnotesize $E_\mathrm{b}/N_0~(\mathrm{dB})$, %
			ylabel=\footnotesize $R_\mathrm{SDD}^\mathrm{BMI}$,
			xmin = 2,
			xmax=8,
			ymax=1,
			ymin=0.6,
			mark size=2.5pt,
			legend style={at={(axis cs:6,0.0001)},anchor=south west}, 
			grid=both,
			minor grid style={gray!25},
			major grid style={gray!25},
			cycle list name=corporate colours markers,
			legend cell align={left},
			line width=0.8pt, %
			]

            \addplot+ [mittelgrau, mark options={solid},mark=+, line width=\lw, mark size= \msize]
			table[x expr=\thisrowno{0}-10*log10(1-(\thisrowno{1})) ,y expr=1-(\thisrowno{1}),col sep=comma]{./tikz/results/uncoded_qpsk_tdl_7/cnn_results_proakis_A.txt};

            \addplot+ [black, mark options={solid},mark=x, line width=\lw, mark size= 3pt, name path=map] 
			table[x expr=\thisrowno{0}-10*log10(1-(\thisrowno{3})) ,y expr=1-(\thisrowno{3}),col sep=comma]{./tikz/results/uncoded_qpsk_tdl_7/use_5g_0_channel_5g_tdl_b_7_n_256_Mb_2_const_qam_bcjr_tdl_A.txt};
			\label{plot:BCJR_tdl_bmi}
			
			\plot[name path=f15,thick,opacity=0,samples=100,domain=0:8.0] {5};
            \addplot[opacity=0.3,gray, pattern=north west lines] fill between[of=map and f15];

			\addplot+ [mittelblau!60!white, mark options={fill=mittelblau!60!white, solid},mark=triangle, line width=\lw, mark size= \msize] 
			table[x expr=\thisrowno{0}-10*log10(1-(\thisrowno{1})) ,y expr=1-(\thisrowno{1}) ,col sep=comma]{./tikz/results/uncoded_qpsk_tdl_7/use_5g_0_channel_5g_tdl_b_7_n_128_Mb_2_const_qam_GNN_Equalizer_LLR_in_results_channel_A.txt};
			\label{plot:gnn_llr_genie_h_proakis_A}

			\addplot+ [mittelblau, mark options={fill=mittelblau, solid}, mark=square, line width=\lw, mark size= \msize] 
			table[x expr=\thisrowno{0}-10*log10(1-(\thisrowno{1})) ,y expr=1-(\thisrowno{1}) ,col sep=comma]{./tikz/results/uncoded_qpsk_tdl_7/results_channel_A.txt};
			\label{plot:gnn_nn_noisy_h}

			\addplot+ [mittelblau!60!black, mark options={fill=mittelblau!60!black, solid},mark=o, line width=\lw, mark size= \msize] 
			table[x expr=\thisrowno{0}-10*log10(1-(\thisrowno{1})) ,y expr=1-(\thisrowno{1}) ,col sep=comma]{./tikz/results/uncoded_qpsk_tdl_7/mmse_gnn_results_channel_A.txt};
			\label{plot:gnn_cs_tdl_A}

			\addplot+ [lila!60!white, mark options={solid},mark=Mercedes star, line width=\lw, mark size= \msize]
			table[x expr=\thisrowno{0}-10*log10(1-(\thisrowno{3})) ,y expr=1-(\thisrowno{3}),col sep=comma]{./tikz/results/uncoded_qpsk_tdl_7/use_5g_0_channel_5g_tdl_b_7_n_256_Mb_2_const_qam_EP8_tdl_A.txt};

			\addplot+ [lila, mark options={solid},mark=star, line width=\lw, mark size= \msize]
			table[x expr=\thisrowno{0}-10*log10(1-(\thisrowno{3})) ,y expr=1-(\thisrowno{3}),col sep=comma]{./tikz/results/uncoded_qpsk_tdl_7/use_5g_0_channel_5g_tdl_b_7_n_256_Mb_2_const_qam_LMMSE_tdl_A.txt};

\nextgroupplot[
    title={\footnotesize (c) Uncertain CSI, TDL-C (retraining)},
			xmode=normal,
			ymode=normal,
			xlabel=\footnotesize $E_\mathrm{b}/N_0~(\mathrm{dB})$, %
			ylabel=\footnotesize $R_\mathrm{SDD}^\mathrm{BMI}$,
			xmin = 0,
			xmax=12,
			ymax=1,
			ymin=0.0,
			mark size=2.5pt,
			legend style={at={(axis cs:6,0.0001)},anchor=south west}, 
			grid=both,
			minor grid style={gray!25},
			major grid style={gray!25},
			cycle list name=corporate colours markers,
			legend cell align={left},
			line width=0.8pt, %
			]

			\addplot+ [mittelgrau, mark options={solid},mark=+, line width=\lw, mark size= \msize]
			table[x expr=\thisrowno{0}-10*log10(1-(\thisrowno{3})) ,y expr=1-(\thisrowno{3}),col sep=comma]{./tikz/results/uncoded_qpsk_tdl_7/cnn_results_per_snr_noisy_h.txt};
			\label{plot:cnn_qpsk_bmi_dash}

            \addplot+ [black, dashed, mark options={solid},mark=none, line width=\lw, mark size= 3pt, name path=map] 
			table[x expr=\thisrowno{0}-10*log10(1-(\thisrowno{3})) ,y expr=1-(\thisrowno{3}),col sep=comma]{./tikz/results/uncoded_qpsk_tdl_7/use_5g_0_channel_5g_tdl_b_7_n_256_Mb_2_const_qam_bcjr.txt};
			\label{plot:BCJR_tdl_bmi}
			
			\plot[name path=f15,thick,opacity=0,samples=100,domain=0:8.0] {5};
            \addplot[opacity=0.3,gray, pattern=north west lines] fill between[of=map and f15];
            \node[rotate=61] at (axis cs: 2.7, 0.62) {\tiny BCJR with perfect CSI};

			\addplot+ [mittelblau!60!white, mark options={fill=blue, solid},mark=triangle, line width=\lw, mark size= \msize] 
			table[x expr=\thisrowno{0}-10*log10(1-(\thisrowno{3})) ,y expr=1-(\thisrowno{3}) ,col sep=comma]{./tikz/results/uncoded_qpsk_tdl_7/use_5g_0_channel_5g_tdl_b_7_n_128_Mb_2_const_qam_GNN_Equalizer_LLR_in_results_per_snr.txt};
			\label{plot:gnn_llr_noisy_h}

            \addplot+ [mittelblau, mark options={fill=mittelblau, solid},mark=square, line width=\lw, mark size= \msize] 
			table[x expr=\thisrowno{0}-10*log10(1-(\thisrowno{3})) ,y expr=1-(\thisrowno{3}) ,col sep=comma]{./tikz/results/uncoded_qpsk_tdl_7/use_5g_0_channel_5g_tdl_b_7_n_128_Mb_2_const_qam_GNN_Equalizer_NN_in_iter_ready_results_per_snr.txt};
			\label{plot:gnn_nn_noisy_h}

            \addplot+ [mittelblau!60!black, mark options={fill=mittelblau!60!black, solid},mark=o, line width=\lw, mark size= \msize] 
			table[x expr=\thisrowno{0}-10*log10(1-(\thisrowno{3})) ,y expr=1-(\thisrowno{3}) ,col sep=comma]{./tikz/results/uncoded_qpsk_tdl_7/use_5g_0_channel_5g_tdl_b_7_n_128_Mb_2_const_qam_GNN_Equalizer_Shared_Edge_Attr_VNSUM_CS_iter_ready_results_per_snr.txt};
			\label{plot:gnn_cs_noisy_h}
			
			\label{plot:gnn_cs_tdl_A}

			\addplot+ [black, mark options={solid},mark=x, line width=\lw, mark size= 3pt] 
			table[x expr=\thisrowno{0}-10*log10(1-(\thisrowno{3})) ,y expr=1-(\thisrowno{3}),col sep=comma]{./tikz/results/uncoded_qpsk_tdl_7/use_5g_0_channel_5g_tdl_b_7_n_256_Mb_2_const_qam_bcjr_noisy_h.txt};
			\label{plot:BCJR_tdl_bmi_noisy_h}

			\addplot+ [lila!60!white, mark options={solid},mark=Mercedes star, line width=\lw, mark size= \msize]
			table[x expr=\thisrowno{0}-10*log10(1-(\thisrowno{3})) ,y expr=1-(\thisrowno{3}),col sep=comma]{./tikz/results/uncoded_qpsk_tdl_7/use_5g_0_channel_5g_tdl_b_7_n_264_Mb_2_const_qam_EP_hnoise_015.txt};
			\label{plot:ep_qpsk_bmi}

			\addplot+ [lila, mark options={solid},mark=star, line width=\lw, mark size= \msize]
			table[x expr=\thisrowno{0}-10*log10(1-(\thisrowno{3})) ,y expr=1-(\thisrowno{3}),col sep=comma]{./tikz/results/uncoded_qpsk_tdl_7/use_5g_0_channel_5g_tdl_b_7_n_264_Mb_2_const_qam_LMMSE_hnoise_015.txt};
			\label{plot:lmmse_qpsk_hnoise_bmi}

\end{groupplot}

  \matrix [
draw=none,
fill=white,
fill opacity=0.7,
text opacity=1,
matrix of nodes,
align =left,
column sep = 0,
inner sep= 2,
anchor=south west,
font=\footnotesize,
column 1/.style={anchor=base west},
mark options={solid},
] at (legend) {
	\ref{plot:BCJR_tdl_bmi_noisy_h} & BCJR \cite{bahl1974optimal}& 
	\ref{plot:gnn_nn_genie_h} & GNN NN &  
	\ref{plot:gnn_cs_genie_h} & GNN CCT & 
	\ref{plot:gnn_llr_genie_h} & GNN LLR &
	\ref{plot:cnn_qpsk_bmi} & CNN NN &
	\ref{plot:ep_qpsk_bmi_hgenie} & EP \cite{santos2015block} & 
	\ref{plot:lmmse_qpsk_bmi} & LMMSE \cite{proakis2001digital}&  \\
};

	\end{tikzpicture}

%% file: figures/jed.tikz
\begin{tikzpicture}[
    yscale=.75,
    rotate=90,
    block/.style={draw, rectangle, minimum size=1cm, align=center},
    squarenode/.style={draw, rectangle, minimum size=1.5em},
    circlenode/.style={draw, circle, minimum size=1.5em, lila, fill=white},
    arrow/.style={->,thick},
    noarrow/.style={thick},
    conn/.style={},
    connd/.style={dashed},
    ]

\draw[mittelblau] (-.5,-2.5) rectangle (2,2.5);
\draw[mittelblau] (2,-2.5) rectangle (4.5,2.5);

    \node [squarenode] (dc1) at (0,.5) {};
    \node [squarenode] (dc2) at (0,-.5) {};
    \coordinate (dc3) at (0,-2.5);
    \coordinate (dc4) at (0,2.5);
    
    \node [circlenode] (v1) at (2,1.5) {};
    \node [circlenode] (v2) at (2,.5) {};
    \node [circlenode] (v3) at (2,-.5) {};
    \node [circlenode] (v4) at (2,-1.5) {};
    \draw[conn] (dc1) -- (v1);
	\draw[conn] (dc1) -- (v2);
	\draw[conn] (dc1) -- (v4);
	\draw[conn] (dc2) -- (v2);
    \draw[conn] (dc2) -- (v3);
	\draw[conn] (dc2) -- (v4);

 \draw[connd] ( $ (dc3)!0.2!(v4) $ ) -- (v4);
 \draw[connd] ( $ (dc4)!0.2!(v1) $ ) -- (v1);

    \coordinate (vm1) at (2,3.5);
    \coordinate (v0) at (2,2.5);

    \node at (4,2.1) {$\dots$};
    \node at (4,-2.1) {$\dots$};
    \node at (0,-1.5) {$\dots$};
    \node at (0,1.5) {$\dots$};
    \node at (1.95,2.0) {$\dots$};
    \node at (1.95,-2.0) {$\dots$};
    
    \node [squarenode] (ec1) at (4,1.5) {};
    \node [squarenode] (ec2) at (4,0.5) {};
    \node [squarenode] (ec3) at (4,-0.5) {};
    \node [squarenode] (ec4) at (4,-1.5) {};
    \coordinate (ec5) at (4,-2.5);
    \coordinate (ec6) at (4,-3.5);
    
    \draw[conn] (v1) -- (ec1);
	\draw[conn] (v1) -- (ec2);
	\draw[conn] (v1) -- (ec3);
	\draw[conn] (v2) -- (ec2);
    \draw[conn] (v2) -- (ec3);
    \draw[conn] (v2) -- (ec4);
    \draw[conn] (v3) -- (ec3);
    \draw[conn] (v3) -- (ec4);
    \draw[conn] (v4) -- (ec4);
    
    \draw[connd] ( $ (vm1)!0.6!(ec1) $ ) -- (ec1);
    \draw[connd] ( $ (v0)!0.2!(ec1) $ ) -- (ec1);
    \draw[connd] ( $ (v0)!0.1!(ec2) $ ) -- (ec2);

    \draw[connd] (v3) -- ( $ (v3)!0.9!(ec5) $ );
    \draw[connd] (v4) -- ( $ (v4)!0.8!(ec5) $ );
    \draw[connd] (v4) -- ( $ (v4)!0.4!(ec6) $ );

    \coordinate (input) at (4.5,0);
    \node (y) at (5.5,0) {$\yv$};

    \draw[arrow] (y) -- (input);

    \node[anchor=east,align=center] at (0.75,2.6) {\small Decoder\\\small GNN};
    \node[anchor=east,align=center] at (3.25,2.6) {\small Detector\\\small GNN};
	
\end{tikzpicture}

%% file: figures/floodingschedule.tikz
\begin{tikzpicture}[
    yscale=0.7,
    dot/.style={circle, draw=none,fill=black,inner sep=1pt},
    syncnode/.style={circle, draw=none,fill=lila,inner sep=1pt},
    arrow/.style={->},
    ]
    
    \node at (2,-.5) {\footnotesize Eq. FN\vphantom{q}};
    \node[lila] at (1,-.5) {\footnotesize VN\vphantom{q}};
    \node at (0,-.5) {\footnotesize Dec. FN\vphantom{q}};

    \node at (-.8,-.5) {\footnotesize It.\vphantom{q}};
    \foreach \x in {1,...,4} {
        \node at (-.8,-\x) {\footnotesize \x};
        \node [dot] (d\x) at (0,-\x) {};
        \node [dot] (e\x) at (2,-\x) {};
    }
    \foreach \x in {1,...,3} {
        \node [syncnode] (s\x) at (1,-\x-.5) {};
        \draw[arrow] (e\x) -- (s\x);
        \draw[arrow] (d\x) -- (s\x);
        \pgfmathtruncatemacro{\xp}{\x+1}
        \draw[arrow] (s\x) -- (e\xp);
        \draw[arrow] (s\x) -- (d\xp);

        \draw[arrow, densely dashed] (d\x) to[out=210,in=150]  (d\xp); 
        \draw[arrow, densely dashed] (e\x) to[out=330,in=30]  (e\xp);
    }
    \node[align=center, rotate=270] at (2.5,-1.5) {  \scriptsize FN state};
\end{tikzpicture}

%% file: figures/sequentialschedule.tikz
\begin{tikzpicture}[
    yscale=0.7,
    dot/.style={circle, draw=none,fill=black,inner sep=1pt},
    syncnode/.style={circle, draw=none,fill=lila,inner sep=1pt},
    arrow/.style={->},
    ]
    
    \node at (2,-.5) {\footnotesize Eq. FN\vphantom{q}};
    \node[lila] at (1,-.5) {\footnotesize VN\vphantom{q}};
    \node at (0,-.5) {\footnotesize Dec. FN\vphantom{q}};

    \node at (-.8,-.5) {\footnotesize It.\vphantom{q}};
    \foreach \x in {1,...,4} {
        \node at (-.8,-\x) {\footnotesize \x};
        \node [dot] (d\x) at (0,-\x) {};
        \node [dot] (e\x) at (2,-\x) {};
    }
    \foreach \x in {1,...,3} {
        \node [syncnode] (s\x) at (1,-\x-.5) {};
    }   
    \draw[arrow] (e1) -- (s1);
    \draw[arrow] (s1) -- (d2);
    \draw[arrow] (d2) -- (s2);
    \draw[arrow] (s2) -- (e3);
    \draw[arrow] (e3) -- (s3);
    \draw[arrow] (s3) -- (d4);    
    
    \draw[arrow, dashed] (d2) to[out=210,in=150] (d4); 
    \draw[arrow, dashed] (e1) to[out=330,in=30]  (e3);
    
\end{tikzpicture}

%% file: tikz/ber_coded_bpsk.tikz
\begin{tikzpicture}
		
        \pgfplotsset{compat=1.10}

		\begin{axis}[
			xmode=normal,
			ymode=log,
			xlabel=\footnotesize $E_\mathrm{b}/N_0~(\mathrm{dB})$, %
			ylabel=\footnotesize $\mathrm{BLER}$,
			xmin = 6,
			xmax=14,
			ymax=1*10^(-0),
			ymin=10^(-5),
			mark size=2.5pt,
			legend style={at={(axis cs:8,0.1)},anchor=south west}, 
			grid=both,
			minor grid style={gray!25},
			major grid style={gray!25},
			width=\linewidth,
	        height=0.8\linewidth,
			cycle list name=corporate colours markers,
			legend cell align={left},
			line width=0.8pt, %
			xtick distance=1.0
			]

            \addplot+ [black,mark options={solid},mark=x, line width=\lw, mark size= \msize] 
			table[x expr=\thisrowno{0}+0.00 ,y=bler,col sep=comma]{./tikz/results/coded_bpsk/coded_proakis_C_BCJR_5GLDPC_k66_n132_SPA_10_IDD_10.txt};
			\label{plot:coded_proakis_C_BCJR_5GLDPC_k66_n132_SPA_20_IDD_10_supplement_best}

            \addplot+ [mittelblau,mark options={fill=mittelblau, solid},mark=square, line width=\lw, mark size= \msize] 
			table[x expr=\thisrowno{0}+0.00 ,y=bler,col sep=comma]{./tikz/results/coded_bpsk/gnn_JED_Flood_5G_proakis_C_k66_n132.txt};
			\label{plot:gnn_flood_best}

            \addplot+ [mittelblau,dashed, mark options={ solid},mark=triangle, line width=\lw, mark size= \msize] 
			table[x expr=\thisrowno{0}+0.00 ,y=bler,col sep=comma]{./tikz/results/coded_bpsk/gnn_JED_Seq_5G_proakisC_k66_n132_sched_35_35_35_finetune_35_35_35_35_35.txt};
			\label{plot:gnn_seq_best}
            
            \addplot+ [apfelgruen, mark options={ solid},mark=pentagon, line width=\lw, mark size= \msize] 
			table[x expr=\thisrowno{0}+0.00 ,y=bler_10,col sep=comma]{./tikz/results/coded_bpsk/gnn_iterativ_BPSK_proakisC_it_10_finetune_snr7_9_idd_20_equ_10_spa_10_supplement.txt};
			\label{plot:gnn_seq_bp_best}

			\addplot+ [mittelgrau, mark options={ solid},mark=+, line width=\lw, mark size= \msize] 
			table[x expr=\thisrowno{0}+0.00 ,y=bler,col sep=comma]{./tikz/results/coded_bpsk/use_5g_0_channel_proakis_c_n_132_Mb_1_const_pam_CNN_iterativ.txt};
			\label{plot:fancy_cnn_bp_best}

            \addplot+ [mittelblau!60!black, mark options={ solid},mark=o, line width=\lw, mark size= \msize] 
			table[x expr=\thisrowno{0}+0.00 ,y=bler,col sep=comma]{./tikz/results/coded_bpsk/gnn_fgnn_iterativ_BPSK_proakisC_it_10_spa_10_idd_20_k66_n132.txt};
			\label{plot:fgnn_bp_best}

            \addplot+ [rot,mark options={ solid},mark=heart, line width=\lw, mark size= \msize] 
			table[x expr=\thisrowno{0}+0.00 ,y=bler,col sep=comma]{./tikz/results/coded_bpsk/coded_proakis_C_5GLDPC_k66_n132_BP_8_SPA_10_IDD_10.txt};
			\label{plot:ffg_best}
            
            \addplot+ [lila!60!white,mark options={ solid},mark=Mercedes star, line width=\lw, mark size= \msize] 
			table[x expr=\thisrowno{0}+0.00 ,y=bler,col sep=comma]{./tikz/results/coded_bpsk/coded_proakis_C_5GLDPC_k66_n132_EP_16_beta_005_llrmax_4_llrout_059_SPA_10_IDD_10.txt};
			\label{plot:ep_best}

        \coordinate (legend) at (axis description cs:0.0,0.0);

		\end{axis}

  \matrix [
draw=none,
fill=white,
fill opacity=0.7,
text opacity=1,
matrix of nodes,
align =left,
column sep = -5.,
inner sep= 2,
anchor=south west,
font=\footnotesize,
column 1/.style={anchor=base west},
column 2/.style={anchor=base west},
column 3/.style={anchor=base west},
mark options={solid},
] at (legend) {
	    JDD GNN flood & \ref{plot:gnn_flood_best} ($30,[1]$) \\
	TDD BCJR \cite{bahl1974optimal}& \ref{plot:coded_proakis_C_BCJR_5GLDPC_k66_n132_SPA_20_IDD_10_supplement_best} ($10,[10]$)\\
    JDD GNN sequ. & \ref{plot:gnn_seq_best} ($5,[3,5]$)\\
    TDD GNN sequ. & \ref{plot:gnn_seq_bp_best} ($10,[10,10]$) & \\
    TDD CNN \cite{huang21extrinsic}& \ref{plot:fancy_cnn_bp_best} ($10,[10,10]$) & \\
    TDD FGNN & \ref{plot:fgnn_bp_best} ($10,[10,10]$) & \\
    TDD SPA \cite{ffg05covalope}& \ref{plot:ffg_best} ($10,[8,10]$)  &  \\
    TDD EP \cite{santos2015block} & \ref{plot:ep_best} ($10,[16,10]$)  &  \\
};

	\end{tikzpicture}

%% file: tikz/latency_coded_bpsk.tikz
\begin{tikzpicture}
		
		\pgfplotsset{compat=1.5}

		\begin{axis}[
			xmode=log,
			ymode=log,
			xlabel=\footnotesize $\mathrm{ Latency~(clock~cycles)}$, %
			ylabel=\footnotesize $\mathrm{BER}$,
			xmin = 10,
			xmax=4*10^(3),
			ymax=10^(0),
			ymin=10^(-6),
			mark size=2.5pt,
			legend style={at={(axis cs:4000,1.0)},anchor=north east, draw=none,
            fill=white,
            fill opacity=0.7,}, 
			grid=both,
			minor grid style={gray!25},
			major grid style={gray!25},
			width=\linewidth,
	        height=0.8\linewidth,
			cycle list name=corporate colours markers,
			legend cell align={left},
			line width=0.8pt, %
			]
            
            \addplot+ [mittelgrau, dashed,mark options={solid},mark=+, line width=\lw, mark size= \msize] 
			table[x expr=\thisrowno{0}+0.00 ,y=ber,col sep=comma]{./tikz/results/latency_coded_bpsk/latency_coded_proakis_C_5GLDPC_k66_n132_FANCY_CNN_SPA_10_IDD_10_supplement.txt};
			\label{plot:latency_cnn_bp_big_iters_bp_10_outer_iters_10}
            \addlegendentry{\footnotesize TDD CNN \cite{huang21extrinsic}};
            
            \addplot+ [mittelblau!60!black, dashed,mark options={solid},mark=o, line width=\lw, mark size= \msize] 
			table[x expr=\thisrowno{0}+0.00 ,y=ber,col sep=comma]{./tikz/results/latency_coded_bpsk/latency_gnn_fgnn_iterativ_BPSK_proakisC_it_10_CODED.txt};
			\label{plot:latency_cnn_bp_big_iters_bp_10_outer_iters_10}
            \addlegendentry{\footnotesize TDD FGNN};
            
            \addplot+ [lila, dashed,mark options={solid},mark=star, line width=\lw, mark size= \msize] 
			table[x expr=\thisrowno{0}+0.00 ,y=ber,col sep=comma]{./tikz/results/latency_coded_bpsk/latency_coded_proakis_C_5GLDPC_k66_n132_EP_16_beta_005_llrmax_4_llrout_059_SPA_10_IDD_10_supplement.txt};
			\label{plot:latency_ep_bp_big_iters_bp_10_outer_iters_10}
            \addlegendentry{\footnotesize TDD EP \cite{santos2015block}};
            
            \addplot+ [black, dashed,mark options={solid},mark=x, line width=\lw, mark size= \msize] 
			table[x expr=\thisrowno{0}+0.00 ,y=ber,col sep=comma]{./tikz/results/latency_coded_bpsk/latency_bcjr_bp_10_idd_20.txt};
			\label{plot:latency_bcjr_bp_big_iters_bp_10_outer_iters_20}
            \addlegendentry{\footnotesize TDD BCJR \cite{bahl1974optimal}};
            
            \addplot+ [rot!60!white, dashed,mark options={solid},mark=pentagon, line width=\lw, mark size= \msize] 
			table[x expr=\thisrowno{0}+0.00 ,y=ber,col sep=comma]{./tikz/results/latency_coded_bpsk/latency_coded_proakis_C_5GLDPC_k66_n132_BP_8_SPA_10_IDD_10.txt};
			\label{plot:latency_duidd_3_bp_nn_3_sp_8_snr_15_17_new}
            \addlegendentry{\footnotesize TDD N-SPA \cite{schmid22neuralBP}};

            \addplot+ [mittelblau!60!white, dashed,mark options={solid},mark=triangle, line width=\lw, mark size= \msize] 
			table[x expr=\thisrowno{0}+0.00 ,y=ber,col sep=comma]{./tikz/results/latency_coded_bpsk/latency_gnn_seq.txt};
			\label{plot:latency_gnn_seq}
            \addlegendentry{\footnotesize JDD GNN sequ.};
            
            \addplot+ [mittelblau, dashed,mark options={solid},mark=square, line width=\lw, mark size= \msize] 
			table[x expr=\thisrowno{0}+0.00 ,y=ber,col sep=comma]{./tikz/results/latency_coded_bpsk/latency_gnn_flooding.txt};
			\label{plot:latency_gnn_flooding}
            \addlegendentry{\footnotesize JDD GNN flood};

            \node[] at (axis cs: 22, 7.3e-3) {\footnotesize \textcolor{mittelblau}{($5,[1]$)}};
            \draw[-latex, line width=1pt, color=black] (axis cs: 25, 5.2e-3) --node[below,yshift=0cm, xshift=.1cm, color=black]  {} (axis cs: 50, 1.6e-3);
            
            \node[] at (axis cs: 60, 1e-5) {\footnotesize \textcolor{mittelblau}{($12,[1]$)}};
            \draw[-latex, line width=1pt, color=black] (axis cs: 96, 1e-5) --node[below,yshift=0cm, xshift=.1cm, color=black]  {} (axis cs: 130, 2e-5);

            \node[] at (axis cs: 600, 2e-6) {\footnotesize \textcolor{mittelblau}{($30,[1]$)}};
            \draw[-latex, line width=1pt, color=black] (axis cs: 600, 4e-6) --node[below,yshift=0cm, xshift=.1cm, color=black]  {} (axis cs: 380, 7e-6);

            \node[] at (axis cs: 1600, 2e-4) {\footnotesize \textcolor{black}{($20,[10]$)}};
            \draw[-latex, line width=1pt, color=black] (axis cs: 1600, 1e-4) --node[below,yshift=0cm, xshift=.1cm, color=black]  {} (axis cs: 2700, 9e-6);

            \node[] at (axis cs: 400, 5e-4) {\footnotesize \textcolor{black}{($2,[10]$)}};
            \draw[-latex, line width=1pt, color=black] (axis cs: 400, 3.5e-4) --node[below,yshift=0cm, xshift=.1cm, color=black]  {} (axis cs: 320, 1.4e-4);

		\end{axis}

	\end{tikzpicture}

%% file: tikz/bmi_iterativ_uncoded_bpsk.tikz
\begin{tikzpicture}
		
		\pgfplotsset{compat=1.5}

		\begin{axis}[
			xmode=normal,
			ymode=normal,
			xlabel=\footnotesize $E_\mathrm{b}/N_0~(\mathrm{dB})$, %
			ylabel=\footnotesize $R_\mathrm{TDD}^\mathrm{BMI}$,
			xmin = 3,
			xmax=14,
			ymax=1,
			ymin=0.4,
			mark size=2.5pt,
			legend style={at={(axis cs:6,0.0001)},anchor=south west}, 
			grid=both,
			minor grid style={gray!25},
			major grid style={gray!25},
			width=\linewidth,
	        height=0.77\linewidth,
			cycle list name=corporate colours markers,
			legend cell align={left},
			line width=0.8pt, %
			]

			\addplot+ [black, mark options={solid},mark=x, line width=\lw, mark size= \msize, name path = map] 
			table[x expr=\thisrowno{0}-10*log10(\thisrowno{1}) ,y expr=(\thisrowno{1}),col sep=comma]{./tikz/results/uncoded_rate_bpsk/uncoded_proakis_C_n132_QPSK_BCJR_iter_rate.txt};
			\label{plot:BCJR_bmi_iter}
            
            \plot[name path=f15,thick,opacity=0,samples=100,domain=4:20.0] {5};
            \addplot[opacity=0.3,gray, pattern=north west lines] fill between[of=map and f15];
            \node[align=center] at (axis cs: 5,0.85) {Not\\achievable};

            \addplot+ [mittelblau, mark options={fill=mittelblau, solid},mark=square, line width=\lw, mark size= \msize] 
			table[x expr=\thisrowno{0}-10*log10(\thisrowno{4}) ,y expr=\thisrowno{4},col sep=comma]{./tikz/results/uncoded_bpsk/gnn_FFG_porakis_C_n_132_it_8_bmi_alvarado.txt};
			\label{plot:gnn_ffg_8_bmi_iter}

            \addplot+ [mittelblau!60!white, mark options={ solid},mark=triangle, line width=\lw, mark size= \msize] 
			table[x expr=\thisrowno{0}-10*log10(\thisrowno{4}) ,y expr=\thisrowno{4},col sep=comma]{./tikz/results/uncoded_bpsk/gnn_ungerboek_proakis_C_n132_it_8_alvarado.txt};
			\label{plot:gnn_ung_ffg_8_bmi_iter}

            \addplot+ [mittelblau!60!black, mark options={solid},mark=o, line width=\lw, mark size= \msize] 
			table[x expr=\thisrowno{0}-10*log10(\thisrowno{4}) ,y expr=\thisrowno{4},col sep=comma]{./tikz/results/uncoded_rate_bpsk/gnn_fgnn_iterativ_BPSK_proakisC_it_10_results.txt};
			\label{plot:fgnn_bmi_iter}
            
            \addplot+ [mittelgrau, mark options={solid},mark=+, line width=\lw, mark size= \msize] 
			table[x expr=\thisrowno{0}-10*log10(\thisrowno{4}) ,y expr=\thisrowno{4},col sep=comma]{./tikz/results/uncoded_rate_bpsk/fancy_cnn_results.txt};
			\label{plot:fancy_cnn_bmi_iter}

			\addplot+ [rot!60!white,mark options={solid},mark=pentagon, line width=\lw, mark size= \msize] 
			table[x expr=\thisrowno{0}-10*log10(\thisrowno{1}) ,y expr=(\thisrowno{1}),col sep=comma]{./tikz/results/uncoded_rate_bpsk/uncoded_proakis_C_n132_BPSK_NNBP_8_iter_rate.txt};
			\label{plot:nbp_ffg_8_bmi_iter}

			\addplot+ [rot,mark options={solid},mark=heart, line width=\lw, mark size= \msize]
			table[x expr=\thisrowno{0}-10*log10(\thisrowno{1}) ,y expr=(\thisrowno{1}),col sep=comma]{./tikz/results/uncoded_rate_bpsk/uncoded_proakis_C_n132_BPSK_BP_8_iter_rate.txt};
			\label{plot:bp_ffg_8_bmi_iter}

            \addplot+ [lila!60!white, mark options={solid},mark=Mercedes star, line width=\lw, mark size= \msize]
			table[x expr=\thisrowno{0}-10*log10(\thisrowno{1}) ,y expr=(\thisrowno{1}),col sep=comma]{./tikz/results/uncoded_rate_bpsk/uncoded_proakis_C_n132_BPSK_EP_10_beta002_llrout_05_llrmax_4_iter_rate.txt};
			\label{plot:ep_24_bmi_iter}

            \addplot+ [lila,mark options={solid},mark=star, line width=\lw, mark size= \msize]
			table[x expr=\thisrowno{0}-10*log10(\thisrowno{1}) ,y expr=(\thisrowno{1}),col sep=comma]{./tikz/results/uncoded_rate_bpsk/uncoded_proakis_C_n132_QPSK_LMMSE_iter_rate.txt};
			\label{plot:lmmse_bmi_iter}

        \coordinate (legend) at (axis cs:3,0.15);
		\end{axis}
  \matrix [
draw=none,
fill=white,
fill opacity=0.7,
text opacity=1,
matrix of nodes,
align =left,
column sep = 0,
inner sep= 2,
anchor=south west,
font=\footnotesize,
column 1/.style={anchor=base west},
column 2/.style={anchor=base west},
column 3/.style={anchor=base west},
column 4/.style={anchor=base west},
column 5/.style={anchor=base west},
column 6/.style={anchor=base west},
column 7/.style={anchor=base west},
column 8/.style={anchor=base west},
column 9/.style={anchor=base west},
column 10/.style={anchor=base west},
mark options={solid},
] at (legend) {
	\ref{plot:BCJR_bmi} & BCJR \cite{bahl1974optimal}&  
	\ref{plot:gnn_ffg_8_bmi_iter} & GNN FFG  $N_\mathrm{It}=8$ & 
	\ref{plot:gnn_ung_ffg_8_bmi_iter} & GNN UFG  $N_\mathrm{It}=8$&
	\ref{plot:fgnn_bmi_iter} & FGNN  $N_\mathrm{It}=8$\\
	\ref{plot:fancy_cnn_bmi_iter} & CNN \cite{huang21extrinsic} &  
	\ref{plot:nbp_ffg_8_bmi_iter} & N-SPA FFG \cite{schmid22neuralBP}  $N_\mathrm{It}=8$&
	\ref{plot:bp_ffg_8_bmi_iter} & SPA FFG \cite{colavolpe05bpdetection}  $N_\mathrm{It}=8$&
	\ref{plot:ep_24_bmi_iter} & EP \cite{santos2015block}  $N_\mathrm{It}=16$&
	\ref{plot:lmmse_bmi_iter} & LMMSE \cite{tuchler02mmseapriori} \\
};

	\end{tikzpicture}

%% file: tikz/bmi_iterativ_uncoded_qam16.tikz
\begin{tikzpicture}
		
		\pgfplotsset{compat=1.5}

		\begin{axis}[
			xmode=normal,
			ymode=normal,
			xlabel=\footnotesize $E_\mathrm{b}/N_0~(\mathrm{dB})$, %
			ylabel=\footnotesize $R_\mathrm{TDD}^\mathrm{BMI}$,
			xmin = 5,
			xmax=30,
			ymax=1,
			ymin=0.4,
			mark size=2.5pt,
			legend style={at={(axis cs:6,0.0001)},anchor=south west}, 
			grid=both,
			minor grid style={gray!25},
			major grid style={gray!25},
			width=\linewidth,
	        height=0.77\linewidth,
			cycle list name=corporate colours markers,
			legend cell align={left},
			line width=0.8pt, %
			]

			\addplot+ [black, mark options={solid},mark=x, line width=\lw, mark size= \msize, name path = map] 
			table[x expr=\thisrowno{0}-10*log10(\thisrowno{1}) ,y expr=(\thisrowno{1}),col sep=comma]{./tikz/results/uncoded_QAM16/uncoded_proakis_C_n132_16QAM_BCJR_iter_rate.txt};
			\label{plot:BCJR_qam16_bmi_iter}
            
            \plot[name path=f15,thick,opacity=0,samples=100,domain=4:20.0] {5};
            \addplot[opacity=0.3,gray, pattern=north west lines] fill between[of=map and f15];
            \node[align=center] at (axis cs: 8,0.85) {Not\\achievable};

            \addplot+ [mittelblau, mark options={fill=mittelblau, solid},mark=square, line width=\lw, mark size= \msize] 
			table[x expr=\thisrowno{0}-10*log10(\thisrowno{4}) ,y expr=\thisrowno{4},col sep=comma]{./tikz/results/uncoded_QAM16/gnn_share_edge_ITERATIVE_16QAM_proakisC_it_6_Units_96_F_48_snr_18_it_8_checkpoint_finetuneresults_per_snr.txt};
			\label{plot:gnn_qam16_ffg_8_bmi_iter}

            \addplot+ [mittelblau!60!white, mark options={fill=mittelblau, solid},mark=triangle, line width=\lw, mark size= \msize] 
			table[x expr=\thisrowno{0}-10*log10(\thisrowno{4}) ,y expr=\thisrowno{4},col sep=comma]{./tikz/results/uncoded_QAM16/gnn_ungerboek_results_per_snr.txt};
			\label{plot:gnn_ung_ffg_qam16_bmi_iter}

            \addplot+ [lila!60!white, mark options={solid},mark=Mercedes star, line width=\lw, mark size= \msize]
			table[x expr=\thisrowno{0}-10*log10(\thisrowno{1}) ,y expr=(\thisrowno{1}),col sep=comma]{./tikz/results/uncoded_QAM16/uncoded_n_500_proakis_C_n132_16QAM_EP_24_miv_1e08_maxv_100_llrmax_7_beta_00002_llrout_999_iter_rate.txt};
			\label{plot:ep_16_bmi_iter_qam16}

            \addplot+ [lila,mark options={solid},mark=star, line width=\lw, mark size= \msize]
			table[x expr=\thisrowno{0}-10*log10(\thisrowno{1}) ,y expr=(\thisrowno{1}),col sep=comma]{./tikz/results/uncoded_QAM16/uncoded_proakis_C_n500_16QAM_LMMSE_iter_rate.txt};
			\label{plot:lmmse_qam16_bmi_iter}

            \addplot+ [mittelgrau, mark options={solid},mark=+, line width=\lw, mark size= \msize] 
			table[x expr=\thisrowno{0}-10*log10(\thisrowno{4})  ,y expr=\thisrowno{4},col sep=comma]{./tikz/results/uncoded_QAM16/cnn_results_per_snr.txt};
			\label{plot:cnn_qam16}
			
			\addplot+ [mittelblau!60!black, mark options={solid},mark=o, line width=\lw, mark size= \msize] 
			table[x expr=\thisrowno{0}-10*log10(\thisrowno{4})  ,y expr=\thisrowno{4},col sep=comma]{./tikz/results/uncoded_QAM16/use_5g_0_channel_proakis_c_n_128_Mb_4_const_qam_FGNN_light_Equalizer_shared_ITERresults_per_snr.txt};
			\label{plot:fgnn_qam16}

        \coordinate (legend) at (axis cs:3,0.15);
		\end{axis}
  \matrix [
draw=none,
fill=white,
fill opacity=0.0,
text opacity=0,
matrix of nodes,
align =left,
column sep = 0,
inner sep= 2,
anchor=south west,
font=\footnotesize,
column 1/.style={anchor=base west},
column 2/.style={anchor=base west},
mark options={solid},
] at (legend) {
	\ref{plot:BCJR_bmi} & BCJR \cite{bahl1974optimal}&  
	\ref{plot:gnn_ffg_8_bmi_iter} & GNN FFG  $N_\mathrm{It}=8$ & 
	\ref{plot:gnn_ung_ffg_8_bmi_iter} & GNN UFG  $N_\mathrm{It}=8$&
	\ref{plot:fgnn_bmi_iter} & FGNN  $N_\mathrm{It}=8$\\
	\ref{plot:fancy_cnn_bmi_iter} & CNN \cite{huang21extrinsic} &  
	\ref{plot:nbp_ffg_8_bmi_iter} & NBP FFG \cite{schmid22neuralBP}  $N_\mathrm{It}=8$&
	\ref{plot:bp_ffg_8_bmi_iter} & BP FFG \cite{colavolpe05bpdetection}  $N_\mathrm{It}=8$&
	\ref{plot:ep_24_bmi_iter} & EP \cite{santos2015block}  $N_\mathrm{It}=16$&
	\ref{plot:lmmse_bmi_iter} & LMMSE \cite{tuchler02mmseapriori} \\
};

	\end{tikzpicture}

%% file: tikz/exit_chart.tikz
\begin{tikzpicture}[spy using outlines={rectangle, magnification=4, connect spies}]
		
		\pgfplotsset{compat=1.5}

		\begin{axis}[
			xmode=normal,
			ymode=normal,
			xlabel=\footnotesize $I_\mathrm{A} \text{,} I_\mathrm{E}$, %
			ylabel=\footnotesize $I_\mathrm{E} \text{,}I_\mathrm{A}$,
			xmin = 0,
			xmax=1,
			ymax=1,
			ymin=0.0,
			mark size=2.5pt,
			legend style={at={(axis cs:6,0.0001)},anchor=south west}, 
			grid=both,
			minor grid style={gray!25},
			major grid style={gray!25},
			width=\linewidth,
	        height=0.8\linewidth,
			cycle list name=corporate colours markers,
			legend cell align={left},
			line width=0.8pt, %
			]

            \addplot+ [black, mark options={solid},mark=none, line width=\lw, mark size= \msize, name path=map] 
			table[x=x ,y =y,col sep=comma]{./tikz/results/exit/ldpc_5g_5_snr.txt};
			\label{plot:code}
			
			\addplot+ [mittelblau, dotted, mark options={solid},mark=square, line width=\lw, mark size= \msize, name path=map] 
			table[x=x ,y =y,col sep=comma]{./tikz/results/exit/gnn_proakis_c_16qam_snr_18_rate_4032_4608_idd_4_traj.txt};
			\label{plot:gnn_traj}
			
			\addplot+ [mittelblau, dashed,mark options={solid}, mark=none, line width=\lw, mark size= \msize, name path=map] 
			table[x=x ,y =y,col sep=comma]{./tikz/results/exit/gnn_proakis_c_16qam_snr_18.txt};
			\label{plot:gnn_exit}
			
			\addplot+ [black, mark options={solid},mark=none, line width=\lw, mark size= \msize, name path=map] 
			table[x=x ,y =y,col sep=comma]{./tikz/results/exit/ldpc_5g_8.txt};
			\label{plot:code}
			
			\addplot+ [mittelblau, dotted,mark options={solid},mark=square, line width=\lw, mark size= \msize, name path=map] 
			table[x=x ,y =y,col sep=comma]{./tikz/results/exit/gnn_proakis_c_bpsk_snr_6_rate_8448_25344_idd_10_traj.txt};
			
			\addplot+ [mittelblau, dashed,mark options={solid}, mark=none, line width=\lw, mark size= \msize, name path=map] 
			table[x=x ,y =y,col sep=comma]{./tikz/results/exit/gnn_proakis_c_bpsk_snr_6.txt};
			
			\addplot+ [black, dotted, mark options={solid},mark=o, line width=\lw, mark size= \msize, name path=map] 
			table[x=x ,y =y,col sep=comma]{./tikz/results/exit/bcjr_proakis_c_bpsk_snr_6_n_25344_idd_10_traj.txt};
			\label{plot:trajectory}
			\addplot+ [black,loosely dashed, mark options={solid}, mark=none, line width=\lw, mark size= \msize, name path=map] 
			table[x=x ,y =y,col sep=comma]{./tikz/results/exit/bcjr_proakis_c_bpsk_snr_6_n_25344.txt};
            \label{plot:detector}

            \node[rotate=6] at (axis cs: 0.7, 0.36) {\footnotesize \textcolor{black}{$N=25344$}, $R_\mathrm{C}=\frac{1}{3}$, 5G LDPC};
            
            \node[rotate=3] at (axis cs: 0.7, 0.87) {\footnotesize \textcolor{black}{$N=4608$}, $R_\mathrm{C}=\frac{8}{9}$, 5G LDPC};
            
            \node[rotate=24] at (axis cs: 0.3, 0.52) {  \footnotesize \textcolor{black}{BPSK  at \qty{6}{dB}}};
            
             \node[rotate=0] at (axis cs: 0.32, 0.935) {  \footnotesize \textcolor{black}{16-QAM  at \qty{16}{dB}}};

        \coordinate (legend) at (axis description cs:0.99,0.0);
        \coordinate (spypoint) at (axis cs:9.85,0.9);
        \coordinate (spyviewer) at (axis cs:5,0.8);
		\end{axis}
  \matrix [
draw,
fill=white,
fill opacity=0.7,
text opacity=1,
matrix of nodes,
align =left,
column sep = 0,
inner sep= 2,
anchor=south east,
font=\footnotesize,
column 1/.style={anchor=base west},
column 2/.style={anchor=base west},
column 3/.style={anchor=base west},
column 4/.style={anchor=base west},
mark options={solid},
] at (legend) {
	\ref{plot:detector} & BCJR $T_\mathrm{Det}(I_\mathrm{A},\gamma)\,\quad$& \ref{plot:gnn_exit}& GNN $T_\mathrm{Det}(I_\mathrm{A},\gamma)$ \\
		\ref{plot:trajectory} & BCJR Trajectory& \ref{plot:gnn_traj} & GNN Trajectory\\
};

	\end{tikzpicture}

%% file: tikz/ber_coded_16qam.tikz
\begin{tikzpicture}
		
        \pgfplotsset{compat=1.10}

		\begin{axis}[
			xmode=normal,
			ymode=log,
			xlabel=\footnotesize $E_\mathrm{b}/N_0~(\mathrm{dB})$, %
			ylabel=\footnotesize $\mathrm{BER}$,
			xmin = 15,
			xmax=25,
			ymax=1*10^(-1),
			ymin=10^(-3),
			mark size=2.5pt,
			legend style={at={(axis cs:8,0.1)},anchor=south west}, 
			grid=both,
			minor grid style={gray!25},
			major grid style={gray!25},
			width=\linewidth,
	        height=0.8\linewidth,
			cycle list name=corporate colours markers,
			legend cell align={left},
			line width=0.8pt, %
			xtick distance=1.0
			]

            \addplot+ [black,mark options={solid},mark=x, line width=\lw, mark size= \msize] 
			table[x expr=\thisrowno{0}+0.00 ,y=ber,col sep=comma]{./tikz/results/coded_QAM16/coded_k_4016_n_4608_proakis_C_n132_16QAM_BCJR_spa_10_idd_10.txt};
			\label{plot:coded_proakis_C_BCJR_5GLDPC_k66_n132_SPA_10_IDD_10_16qam}

            \addplot+ [mittelblau,mark options={fill=mittelblau, solid},mark=square, line width=\lw, mark size= \msize] 
			table[x expr=\thisrowno{0}+0.00 ,y=ber,col sep=comma]{./tikz/results/coded_QAM16/gnn_share_edge_ITERATIVE_16QAM_proakisC_it_6_Units_96_F_48_snr_18_it_8_checkpoint_finetuneresults_ldpc_k4032_n4608_bg1_H_clip1_idd10_damped06.txt};
			\label{plot:gnn_16qam_6_idd_10_spa_10flood_best}

            \addplot+ [mittelblau!60!white, mark options={ solid},mark=triangle, line width=\lw, mark size= \msize] 
			table[x expr=\thisrowno{0}+0.00 ,y=ber,col sep=comma]{./tikz/results/coded_QAM16/gnn_ungresults_ldpc_k4032_n4608_bg1_H_clip4_idd10.txt};
			\label{plot:gnn_ung_qam16}

            \addplot+ [mittelgrau, mark options={ solid},mark=+, line width=\lw, mark size= \msize] 
			table[x expr=\thisrowno{0}+0.00 ,y=ber,col sep=comma]{./tikz/results/coded_QAM16/fancy_cnn_ITERATIVE_16QAM_proakisC_snr_20_ldpc_k4032_n4608_bg1_H_clip4_idd10.txt};
			\label{plot:fancy_cnn_qam16_bp_best}

\addplot+ [mittelblau!60!black, mark options={ solid},mark=o, line width=\lw, mark size= \msize] 
			table[x expr=\thisrowno{0}+0.00 ,y=ber,col sep=comma]{./tikz/results/coded_QAM16/use_5g_0_channel_proakis_c_n_128_Mb_4_const_qam_FGNN_light_Equalizer_shared_ITER_results_ldpc_k4032_n4608_bg1_H_idd10_damp_06_clip_4.txt};
			\label{plot:fgnn_qam16_bp_best}

            \addplot+ [lila!60!white,mark options={ solid},mark=Mercedes star, line width=\lw, mark size= \msize] 
			table[x expr=\thisrowno{0}+0.00 ,y=ber,col sep=comma]{./tikz/results/coded_QAM16/coded_k_4016_n_4608_proakis_C_n132_16QAM_EP_24_miv_1e08_maxv_100_llrmax_7_beta_00002_llrout_05_spa_10_idd_10_supplement.txt};
			\label{plot:ber_16qam_ep_24_spa_10_idd_10_best}
            
        \draw[<->, opacity=0.5, line width=1pt, color=black] (axis cs: 18, 3e-3) --node[below,yshift=0cm, xshift=.0cm, color=black, opacity=0.5]  {\qty{6}{dB}} (axis cs: 24, 3e-3);

        \coordinate (legend) at (axis description cs:1.0,1.0);

		\end{axis}

  \matrix [
draw=none,
fill=white,
fill opacity=0.7,
text opacity=1,
matrix of nodes,
align =left,
column sep = -5.,
inner sep= 2,
anchor=north east,
font=\footnotesize,
column 1/.style={anchor=base west},
column 2/.style={anchor=base west},
column 3/.style={anchor=base west},
mark options={solid},
] at (legend) {
	BCJR \cite{bahl1974optimal} & \ref{plot:coded_proakis_C_BCJR_5GLDPC_k66_n132_SPA_10_IDD_10_16qam} ($10,[10]$)\\
    GNN FFG & \ref{plot:gnn_16qam_6_idd_10_spa_10flood_best} ($10,[6,10]$)  \\
    GNN UFG & \ref{plot:gnn_ung_qam16} ($10,[10,10]$)  \\
    CNN \cite{huang21extrinsic} & \ref{plot:fancy_cnn_qam16_bp_best} ($10,[10,10]$) & \\
    FGNN & \ref{plot:fgnn_qam16_bp_best} ($10,[10,10]$)  \\
    EP \cite{santos_turbo_2018} & \ref{plot:ber_16qam_ep_24_spa_10_idd_10_best} ($10,[16,10]$)\\
};

	\end{tikzpicture}

%% file: tables/table_4.tex
\begin{table*}[t]
    \centering
    \caption{Overview of detection schemes and categorization of properties into: good ($\color{apfelgruen}\pmb{+}$), mediocre ($\color{orange}\pmb{\circ}$), and bad ($\color{rot}\pmb{-}$).}
    \begin{tabular}{c|c|c|c|c|c|c}
         Name & Scalability & Latency  & Severe ISI & Type & High rates & Reference\\
         \hline
         SPA-UFG & $\color{apfelgruen}\pmb{+}$& $\color{apfelgruen}\pmb{+}$  &$\color{rot}\pmb{-}$ & APP & $\color{orange}\pmb{\circ}$ & \cite{colavolpe11ungerboeckdetection,schmid22neuralBP}\\
         SPA-FFG & $\color{rot}\pmb{-}$ & $\color{apfelgruen}\pmb{+}$  & $\color{orange}\pmb{\circ}$ & APP & $\color{apfelgruen}\pmb{+}$ & \cite{ffg05covalope,schmid22neuralBP}\\
         BCJR & $\color{rot}\pmb{-}$ & $\color{rot}\pmb{-}$  & $\color{apfelgruen}\pmb{+}$ & APP &  $\color{apfelgruen}\pmb{+}$& \cite{bahl1974optimal,Forney1972MaximumlikelihoodSE}\\
         LMMSE & $\color{apfelgruen}\pmb{+}$ & $\color{apfelgruen}\pmb{+}$  & $\color{rot}\pmb{-}$ & LMMSE & $\color{rot}\pmb{-}$& \cite{proakis2001digital,tuchler02mmseapriori} \\
         EP & $\color{apfelgruen}\pmb{+}$ & $\color{orange}\pmb{\circ}$  & $\color{orange}\pmb{\circ}$ & LMMSE \& APP & $\color{apfelgruen}\pmb{+}$  & \cite{santos_turbo_2018,santos2015block} \\
         CNN & $\color{orange}\pmb{\circ}$ & $\color{apfelgruen}\pmb{+}$ & $\color{apfelgruen}\pmb{+}$ & Neural &  $\color{apfelgruen}\pmb{+}$  & \cite{xu18nnjed,huang21extrinsic} \\
         GNN & $\color{orange}\pmb{\circ}$ & $\color{apfelgruen}\pmb{+}$  & $\color{apfelgruen}\pmb{+}$ & Neural & $\color{apfelgruen}\pmb{+}$   & proposed \\
    \end{tabular}
    \label{tab:my_label_new}
\end{table*}

%% file: tables/table_5.tex
\begin{table}[h]
  \caption{\neu{Rounded number of operations (multiplications, additions, activations, $\max^\star$) for different modulation orders $M$ and channel memories $L$ for $N_\mathrm{It}=8$ iterations for \ac{EP}, \ac{SPA} and the \ac{GNN}.}}
  \label{tab:numop}
  \centering
  \begin{tabular}{l l l l l l}
    Algorithm & \multicolumn{2}{c}{QPSK ($M=4$)} & \multicolumn{2}{c}{$16$-QAM ($M=16$)}& \# weights\\
    \hline
    & $L=4$& $L=6$& $L=4$& $L=6$&\\
    \hline
    BCJR & $2\cdot10^3$ & $7\cdot10^4$ & $7\cdot10^7$ & $2\cdot10^{10}$ & - \\
    \neur{EP} & \neur{$4\cdot10^4$ }& \neur{$1\cdot10^5$ }& \neur{$4\cdot10^4$ }& \neur{$1\cdot10^5$ }& - \\
    SPA-FFG & $5\cdot10^4$ & $1\cdot10^5$ & $5\cdot10^7$ & $2\cdot10^{10}$ & - \\
    SPA-UFG & $1\cdot10^3$ & $2\cdot10^3$ & $2\cdot10^4$ & $4\cdot10^4$ & - \\
    GNN & $6\cdot10^5$ & $8\cdot10^5$ & $6\cdot10^5$ & $8\cdot10^5$  & $2\cdot10^4$ \\
    CNN & $5\cdot10^5$ & $5\cdot10^5$ & $5\cdot10^5$ & $5\cdot10^5$  & $5\cdot10^5$  \\
  \end{tabular}
\end{table}

%% file: training/training.tex
\begin{tikzpicture}
		
		\pgfplotsset{compat=1.5}

		\begin{axis}[
			xmode=normal,
			ymode=log,
			each nth point=1,
			xlabel=\footnotesize Epoch, %
			ylabel=\footnotesize $\mathcal{L}_\mathrm{BCE}$,
			xmin = 0,
			xmax=80000,
			ymax=1,
			ymin=0.01,
			mark size=2.5pt,
			legend style={at={(axis cs:80000,1)},anchor=north east}, 
			grid=both,
			minor grid style={gray!25},
			major grid style={gray!25},
			width=0.999\linewidth,
	        height=0.88\linewidth,
			cycle list name=corporate colours markers,
			legend cell align={left},
			line width=0.8pt, %
			]

			\addplot+ [dashed,forget plot,mittelblau, mark options={solid},mark=x, line width=\lw, mark=none] 
			table[x expr=\thisrowno{0} ,y expr=(\thisrowno{1}),col sep=comma]{./training/training_JDD_normal_5.txt};

			\addplot+ [dashed,forget plot, mittelblau,mark options={solid},mark=x, line width=\lw, mark=none] 
			table[x expr=\thisrowno{0} ,y expr=(\thisrowno{1}),col sep=comma]{./training/training_JDD_normal_1_5.txt};

			\addplot+ [dashed, mittelblau,mark options={solid},mark=x, line width=\lw, mark=none] 
			table[x expr=\thisrowno{0} ,y expr=(\thisrowno{1}),col sep=comma]{./training/training_JDD_normal_0_5.txt};
			\addlegendentry{\footnotesize Standard (Tab.~\ref{tab:table_example})};
			
			\addplot+ [ mark options={solid},mark=x, line width=\lw, mark=none] 
			table[x expr=\thisrowno{0} ,y expr=(\thisrowno{1}),col sep=comma]{./training/training_JDD_more_units_5.txt};
			\addlegendentry{\footnotesize  \# hidden units $128$};
			
			\addplot+ [ mark options={solid},mark=x, line width=\lw, mark=none] 
			table[x expr=\thisrowno{0} ,y expr=(\thisrowno{1}),col sep=comma]{./training/training_JDD_more_layers_5.txt};
			\addlegendentry{\footnotesize \# hidden layers $3$};
			
			\addplot+ [ mark options={solid},mark=x, line width=\lw, mark=none] 
			table[x expr=\thisrowno{0} ,y expr=(\thisrowno{1}),col sep=comma]{./training/training_JDD_lower_lr_5.txt};
			\addlegendentry{\footnotesize Learning rate $10^{-5}$};
			
			\addplot+ [ mark options={solid},mark=x, line width=\lw, mark=none] 
			table[x expr=\thisrowno{0} ,y expr=(\thisrowno{1}),col sep=comma]{./training/training_JDD_lower_bs_5.txt};
			\addlegendentry{\footnotesize Batch size $64$};
			
			\addplot+ [ mark options={solid},mark=x, line width=\lw, mark=none] 
			table[x expr=\thisrowno{0} ,y expr=(\thisrowno{1}),col sep=comma]{./training/training_JDD_less_units_5.txt};
			\addlegendentry{\footnotesize\# hidden units $32$};
			
			\addplot+ [ mark options={solid},mark=x, line width=\lw, mark=none] 
			table[x expr=\thisrowno{0} ,y expr=(\thisrowno{1}),col sep=comma]{./training/training_JDD_less_layers_5.txt};
			\addlegendentry{\footnotesize \# hidden layers $1$};
			
			\addplot+ [ mark options={solid},mark=x, line width=\lw, mark=none] 
			table[x expr=\thisrowno{0} ,y expr=(\thisrowno{1}),col sep=comma]{./training/training_JDD_higher_lr_5.txt};
			\addlegendentry{\footnotesize Learning rate $10^{-3}$};
			
			\addplot+ [ mark options={solid},mark=x, line width=\lw, mark=none] 
			table[x expr=\thisrowno{0} ,y expr=(\thisrowno{1}),col sep=comma]{./training/training_JDD_higher_bs_5.txt};
			\addlegendentry{\footnotesize Batch size $512$};
			
            \addplot+ [black,dotted, mark options={solid},mark=x, line width=\lw, mark=none] 
			table[x expr=\thisrowno{0} ,y expr=(\thisrowno{1}),col sep=comma]{./training/training_use_5g_1_channel_proakis_c_n_132_Mb_2_const_qam_GNN_attention_head_4_working_50.txt};
			\addlegendentry{\footnotesize \neur{ Graph Attention}};

		\end{axis}

	\end{tikzpicture}

%% file: main.bbl
\begin{thebibliography}{10}
\providecommand{\url}[1]{#1}
\csname url@samestyle\endcsname
\providecommand{\newblock}{\relax}
\providecommand{\bibinfo}[2]{#2}
\providecommand{\BIBentrySTDinterwordspacing}{\spaceskip=0pt\relax}
\providecommand{\BIBentryALTinterwordstretchfactor}{4}
\providecommand{\BIBentryALTinterwordspacing}{\spaceskip=\fontdimen2\font plus
\BIBentryALTinterwordstretchfactor\fontdimen3\font minus
  \fontdimen4\font\relax}
\providecommand{\BIBforeignlanguage}[2]{{%
\expandafter\ifx\csname l@#1\endcsname\relax
\typeout{** WARNING: IEEEtran.bst: No hyphenation pattern has been}%
\typeout{** loaded for the language `#1'. Using the pattern for}%
\typeout{** the default language instead.}%
\else
\language=\csname l@#1\endcsname
\fi
#2}}
\providecommand{\BIBdecl}{\relax}
\BIBdecl

\bibitem{jddisit}
J.~Clausius, M.~Geiselhart, D.~Tandler, and S.~ten Brink, ``{Graph Neural
  Network-Based Joint Equalization and Decoding},'' in \emph{2024 IEEE
  International Symposium on Information Theory (ISIT)}, 2024, pp. 1203--1208.

\bibitem{proakis2001digital}
J.~Proakis, \emph{Digital Communications}.\hskip 1em plus 0.5em minus
  0.4em\relax McGraw-Hill, 2001.

\bibitem{You2020}
{Xiaohu You et al.}, ``Towards {6G} wireless communication networks: vision,
  enabling technologies, and new paradigm shifts,'' \emph{Science China
  Information Sciences}, vol.~64, Nov. 2020.

\bibitem{gibson89mlpequalizers}
G.~Gibson, S.~Siu, and C.~Cowen, ``Multilayer perceptron structures applied to
  adaptive equalisers for data communications,'' in \emph{International
  Conference on Acoustics, Speech, and Signal Processing,}, 1989, pp.
  1183--1186 vol.2.

\bibitem{schlezinger2023model}
N.~Shlezinger, J.~Whang, Y.~C. Eldar, and A.~G. Dimakis, ``{Model-based Deep
  Learning},'' \emph{Proceedings of the IEEE}, 2023.

\bibitem{huang21extrinsic}
X.~Huang, J.~Cho, K.~Hashemizadeh, and R.-R. Chen, ``{Extrinsic Neural Network
  Equalizer for Channels with High Inter-Symbol-Interference},'' in \emph{ICC
  2021 - IEEE International Conference on Communications}, 2021, pp. 1--6.

\bibitem{xu18nnjed}
W.~Xu, Z.~Zhong, Y.~Be'ery, X.~You, and C.~Zhang, ``Joint neural network
  equalizer and decoder,'' in \emph{2018 15th International Symposium on
  Wireless Communication Systems (ISWCS)}, 2018, pp. 1--5.

\bibitem{kechriotis94rnn}
G.~Kechriotis, E.~Zervas, and E.~Manolakos, ``Using recurrent neural networks
  for adaptive communication channel equalization,'' \emph{IEEE Transactions on
  Neural Networks}, vol.~5, no.~2, pp. 267--278, 1994.

\bibitem{farsad2018neural}
N.~Farsad and A.~Goldsmith, ``{Neural Network Detection of Data Sequences in
  Communication Systems},'' \emph{IEEE Trans. on Signal Process.}, vol.~66,
  no.~21, pp. 5663--5678, 2018.

\bibitem{plabst24sicrnn}
D.~Plabst, T.~Prinz, F.~Diedolo, T.~Wiegart, G.~Böcherer, N.~Hanik, and
  G.~Kramer, ``Neural network-based successive interference cancellation for
  non-linear bandlimited channels,'' \emph{IEEE Transactions on
  Communications}, pp. 1--1, 2024.

\bibitem{bahl1974optimal}
L.~{Bahl}, J.~{Cocke}, F.~{Jelinek}, and J.~{Raviv}, ``Optimal decoding of
  linear codes for minimizing symbol error rate (corresp.),'' \emph{IEEE
  Transactions on Information Theory}, vol.~20, no.~2, pp. 284--287, March
  1974.

\bibitem{shlezinger2020bcjrnet}
N.~Shlezinger, N.~Farsad, Y.~C. Eldar, and A.~J. Goldsmith, ``Data-driven
  factor graphs for deep symbol detection,'' in \emph{2020 IEEE International
  Symposium on Information Theory (ISIT)}.\hskip 1em plus 0.5em minus
  0.4em\relax IEEE, 2020, pp. 2682--2687.

\bibitem{forney72observationmodel}
G.~Forney, ``{Lower Bounds on Error Probability in the Presence of Large
  Intersymbol Interference},'' \emph{IEEE Transactions on Communications},
  vol.~20, no.~1, pp. 76--77, 1972.

\bibitem{ungerboeck74observationmodel}
G.~Ungerboeck, ``{Adaptive Maximum-Likelihood Receiver for Carrier-Modulated
  Data-Transmission Systems},'' \emph{IEEE Transactions on Communications},
  vol.~22, no.~5, pp. 624--636, 1974.

\bibitem{kschischang01factorgraph}
F.~Kschischang, B.~Frey, and H.-A. Loeliger, ``Factor graphs and the
  sum-product algorithm,'' \emph{IEEE Transactions on Information Theory},
  vol.~47, no.~2, pp. 498--519, 2001.

\bibitem{colavolpe05bpdetection}
G.~Colavolpe and G.~Germi, ``{On the application of factor graphs and the
  sum-product algorithm to ISI channels},'' \emph{IEEE Transactions on
  Communications}, vol.~53, no.~5, pp. 818--825, 2005.

\bibitem{colavolpe11ungerboeckdetection}
G.~Colavolpe, D.~Fertonani, and A.~Piemontese, ``{SISO Detection Over Linear
  Channels With Linear Complexity in the Number of Interferers},'' \emph{IEEE
  Journal of Selected Topics in Signal Processing}, vol.~5, no.~8, pp.
  1475--1485, 2011.

\bibitem{liu21nnfactornode}
B.~Liu, S.~Li, Y.~Xie, and J.~Yuan, ``{A Novel Sum-Product Detection Algorithm
  for Faster-Than-Nyquist Signaling: A Deep Learning Approach},'' \emph{IEEE
  Transactions on Communications}, vol.~69, no.~9, pp. 5975--5987, 2021.

\bibitem{schmid22neuralBP}
L.~Schmid and L.~Schmalen, ``{Low-Complexity Near-Optimum Symbol Detection
  Based on Neural Enhancement of Factor Graphs},'' \emph{IEEE Transactions on
  Communications}, vol.~70, no.~11, pp. 7562--7575, 2022.

\bibitem{nachmani2016learning}
E.~Nachmani, Y.~Be'ery, and D.~Burshtein, ``Learning to decode linear codes
  using deep learning,'' in \emph{Allerton Conf.}\hskip 1em plus 0.5em minus
  0.4em\relax IEEE, 2016, pp. 341--346.

\bibitem{gnn}
F.~Scarselli, M.~Gori, A.~C. Tsoi, M.~Hagenbuchner, and G.~Monfardini, ``The
  graph neural network model,'' \emph{IEEE Transactions on Neural Networks},
  vol.~20, no.~1, pp. 61--80, 2009.

\bibitem{zhang2023factor}
Z.~Zhang, M.~H. Dupty, F.~Wu, J.~Q. Shi, and W.~S. Lee, ``Factor graph neural
  networks,'' \emph{Journal of Machine Learning Research}, vol.~24, no. 181,
  pp. 1--54, 2023.

\bibitem{satorras2021neural}
V.~G. Satorras and M.~Welling, ``{Neural Enhanced Belief Propagation on Factor
  Graphs},'' in \emph{International Conference on Artificial Intelligence and
  Statistics}.\hskip 1em plus 0.5em minus 0.4em\relax PMLR, 2021, pp. 685--693.

\bibitem{cammerer2022gnn}
S.~Cammerer, J.~Hoydis, F.~A. Aoudia, and A.~Keller, ``{Graph Neural Networks
  for Channel Decoding},'' in \emph{IEEE Global Communications Conference
  (GLOBECOM) Workshop}, 2022.

\bibitem{Scotti2020GraphNN}
A.~Scotti, N.~N. Moghadam, D.~Liu, K.~Gafvert, and J.~Huang, ``{Graph Neural
  Networks for Massive MIMO Detection},'' \emph{ArXiv}, vol. abs/2007.05703,
  2020.

\bibitem{Cammerer2023ANR}
S.~Cammerer, F.~A. Aoudia, J.~Hoydis, A.~Oeldemann, A.~Roessler, T.~Mayer, and
  A.~Keller, ``{A Neural Receiver for 5G NR Multi-User MIMO},'' \emph{2023 IEEE
  Globecom Workshops (GC Wkshps)}, pp. 329--334, 2023.

\bibitem{Kosasih22GnnMIMO}
A.~Kosasih, V.~Onasis, V.~Miloslavskaya, W.~Hardjawana, V.~Andrean, and
  B.~Vucetic, ``{Graph Neural Network Aided MU-MIMO Detectors},'' \emph{IEEE
  Journal on Selected Areas in Communications}, vol.~40, no.~9, pp. 2540--2555,
  2022.

\bibitem{Douillard95turbo}
C.~Douillard, M.~Jézéquel, C.~Berrou, D.~Electronique, A.~Picart, P.~Didier,
  and A.~Glavieux, ``Iterative correction of intersymbol interference:
  Turbo-equalization,'' \emph{European Transactions on Telecommunications},
  vol.~6, no.~5, pp. 507--511, 1995.

\bibitem{ye17NNjed}
H.~Ye and G.~Y. Li, ``{Initial Results on Deep Learning for Joint Channel
  Equalization and Decoding},'' in \emph{2017 IEEE 86th Vehicular Technology
  Conference (VTC-Fall)}, 2017, pp. 1--5.

\bibitem{tsai20JEDbcjrNet}
W.-C. Tsai, C.-F. Teng, H.-M. Ou, and A.-Y.~A. Wu, ``{Neural Network-Aided BCJR
  Algorithm for Joint Symbol Detection and Channel Decoding},'' in \emph{2020
  IEEE Workshop on Signal Processing Systems (SiPS)}, 2020, pp. 1--6.

\bibitem{henkel19jointequalLDPCdecoding}
W.~Henkel, N.~S. Islam, and M.~A. Leghari, ``{Joint Equalization and LDPC
  Decoding},'' in \emph{2019 11th International Congress on Ultra Modern
  Telecommunications and Control Systems and Workshops (ICUMT)}, 2019, pp.
  1--5.

\bibitem{Robertson95suboptimalMAP}
P.~Robertson, E.~Villebrun, and P.~Hoeher, ``A comparison of optimal and
  sub-optimal {MAP} decoding algorithms operating in the log domain,'' in
  \emph{Proceedings IEEE International Conference on Communications ICC '95},
  vol.~2, 1995, pp. 1009--1013 vol.2.

\bibitem{brink01exit}
S.~ten Brink, ``Convergence behavior of iteratively decoded parallel
  concatenated codes,'' \emph{IEEE Transactions on Communications}, vol.~49,
  no.~10, pp. 1727--1737, 2001.

\bibitem{bran05conv}
F.~Brannstrom, L.~Rasmussen, and A.~Grant, ``Convergence analysis and optimal
  scheduling for multiple concatenated codes,'' \emph{IEEE Transactions on
  Information Theory}, vol.~51, no.~9, pp. 3354--3364, 2005.

\bibitem{hagenauer2002turbo}
J.~Hagenauer, ``The turbo principle in mobile communications,'' \emph{Proc.
  ISITA, XI'AN, Peoples Republic of China, Oct. 2002}, 2002.

\bibitem{shannon48theory}
C.~E. Shannon, ``A mathematical theory of communication,'' \emph{The Bell
  System Technical Journal}, vol.~27, no.~3, pp. 379--423, 1948.

\bibitem{ashikhmin04area}
A.~Ashikhmin, G.~Kramer, and S.~ten Brink, ``Extrinsic information transfer
  functions: model and erasure channel properties,'' \emph{IEEE Transactions on
  Information Theory}, vol.~50, no.~11, pp. 2657--2673, 2004.

\bibitem{bocherer2017achievable}
G.~B{\"o}cherer, ``Achievable rates for probabilistic shaping,'' \emph{arXiv
  preprint arXiv:1707.01134}, 2017.

\bibitem{cammerer2019tcom}
S.~Cammerer, F.~Ait~Aoudia, S.~D{\"o}rner, M.~Stark, J.~Hoydis, and S.~ten
  Brink, ``{Trainable Communication Systems: Concepts and Prototype},''
  \emph{IEEE Transactions on Communications}, vol.~68, no.~9, pp. 5489--5503,
  2020.

\bibitem{Glorot2010UnderstandingTD}
X.~Glorot and Y.~Bengio, ``Understanding the difficulty of training deep
  feedforward neural networks,'' in \emph{International Conference on
  Artificial Intelligence and Statistics}, 2010.

\bibitem{veličković2018graphattentionnetworks}
\BIBentryALTinterwordspacing
P.~Veličković, G.~Cucurull, A.~Casanova, A.~Romero, P.~Liò, and Y.~Bengio,
  ``Graph attention networks,'' 2018. [Online]. Available:
  \url{https://arxiv.org/abs/1710.10903}
\BIBentrySTDinterwordspacing

\bibitem{falconer_adaptive_1973}
D.~D. Falconer and F.~R. Magee, ``Adaptive channel memory truncation for
  maximum likelihood sequence estimation,'' \emph{The Bell System Technical
  Journal}, vol.~52, no.~9, pp. 1541--1562, 1973.

\bibitem{rusek12optimalshortening}
F.~Rusek and A.~Prlja, ``{Optimal Channel Shortening for MIMO and ISI
  Channels},'' \emph{IEEE Transactions on Wireless Communications}, vol.~11,
  no.~2, pp. 810--818, 2012.

\bibitem{ffg05covalope}
G.~Colavolpe and G.~Germi, ``On the application of factor graphs and the
  sum-product algorithm to isi channels,'' \emph{IEEE Transactions on
  Communications}, vol.~53, no.~5, pp. 818--825, 2005.

\bibitem{santos2015block}
I.~Santos, J.~J. Murillo-Fuentes, and P.~M. Olmos, ``{Block expectation
  propagation equalization for ISI channels},'' in \emph{2015 23rd European
  Signal Processing Conference (EUSIPCO)}.\hskip 1em plus 0.5em minus
  0.4em\relax IEEE, 2015, pp. 379--383.

\bibitem{lian19bploss}
M.~Lian, F.~Carpi, C.~Häger, and H.~D. Pfister, ``{Learned Belief-Propagation
  Decoding with Simple Scaling and SNR Adaptation},'' in \emph{2019 IEEE
  International Symposium on Information Theory (ISIT)}, 2019, pp. 161--165.

\bibitem{kingma2014adam}
D.~P. Kingma and J.~Ba, ``{Adam: A Method for Stochastic Optimization},''
  \emph{arXiv preprint arXiv:1412.6980}, 2014.

\bibitem{roumy99proakiscloss}
A.~Roumy, I.~Fijalkow, and D.~Pirez, ``Joint equalization and decoding: why
  choose the iterative solution?'' in \emph{Gateway to 21st Century
  Communications Village. VTC 1999-Fall. IEEE VTS 50th Vehicular Technology
  Conference (Cat. No.99CH36324)}, vol.~5, 1999, pp. 2989--2993 vol.5.

\bibitem{szczecinski2015bit}
L.~Szczecinski and A.~Alvarado, \emph{Bit-interleaved coded modulation:
  fundamentals, analysis and design}.\hskip 1em plus 0.5em minus 0.4em\relax
  John Wiley \& Sons, 2015.

\bibitem{Forney1972MaximumlikelihoodSE}
J.~G. Forney, ``Maximum-likelihood sequence estimation of digital sequences in
  the presence of intersymbol interference,'' \emph{IEEE Trans. Inf. Theory},
  vol.~18, pp. 363--378, 1972.

\bibitem{explainableGNN}
H.~Yuan, H.~Yu, S.~Gui, and S.~Ji, ``{Explainability in Graph Neural Networks:
  A Taxonomic Survey},'' \emph{IEEE Transactions on Pattern Analysis and
  Machine Intelligence}, vol.~45, no.~5, pp. 5782--5799, 2023.

\bibitem{schmid2023local}
L.~Schmid, J.~Brenk, and L.~Schmalen, ``Local message passing on frustrated
  systems,'' in \emph{Uncertainty in Artificial Intelligence}.\hskip 1em plus
  0.5em minus 0.4em\relax PMLR, 2023, pp. 1837--1846.

\bibitem{schlezinger20viterbinet}
N.~Shlezinger, N.~Farsad, Y.~C. Eldar, and A.~J. Goldsmith, ``{ViterbiNet: A
  Deep Learning Based Viterbi Algorithm for Symbol Detection},'' \emph{IEEE
  Transactions on Wireless Communications}, vol.~19, no.~5, pp. 3319--3331,
  2020.

\bibitem{sionna}
J.~Hoydis, S.~Cammerer, F.~{Ait Aoudia}, A.~Vem, N.~Binder, G.~Marcus, and
  A.~Keller, ``{Sionna: An Open-Source Library for Next-Generation Physical
  Layer Research},'' \emph{arXiv preprint}, Mar. 2022.

\bibitem{bellman61curseofdimensionality}
R.~Bellman, \emph{{Adaptive Control Processes: A Guided Tour}}.\hskip 1em plus
  0.5em minus 0.4em\relax Princeton University Press, 1961.

\bibitem{donoho00blessingofdimensionality}
D.~Donoho, ``{High-Dimensional Data Analysis: The Curses and Blessings of
  Dimensionality},'' \emph{AMS Math Challenges Lecture}, pp. 1--32, 01 2000.

\bibitem{bicm98caire}
G.~Caire, G.~Taricco, and E.~Biglieri, ``Bit-interleaved coded modulation,''
  \emph{IEEE Transactions on Information Theory}, vol.~44, no.~3, pp. 927--946,
  1998.

\bibitem{douillard95turboequalization}
C.~Douillard, M.~Jézéquel, C.~Berrou, D.~Electronique, A.~Picart, P.~Didier,
  and A.~Glavieux, ``{Iterative correction of intersymbol interference:
  Turbo-equalization},'' \emph{European Transactions on Telecommunications},
  vol.~6, no.~5, pp. 507--511, 1995.

\bibitem{wiesmayr2022duidd}
R.~Wiesmayr, C.~Dick, J.~Hoydis, and C.~Studer, ``{DUIDD: Deep-Unfolded
  Interleaved Detection and Decoding for MIMO Wireless Systems},'' in
  \emph{2022 56th Asilomar Conference on Signals, Systems, and
  Computers}.\hskip 1em plus 0.5em minus 0.4em\relax IEEE, 2022, pp. 181--188.

\bibitem{pretrain20koike}
T.~Koike-Akino, Y.~Wang, D.~S. Millar, K.~Kojima, and K.~Parsons, ``{Neural
  Turbo Equalization: Deep Learning for Fiber-Optic Nonlinearity
  Compensation},'' \emph{Journal of Lightwave Technology}, 2020.

\bibitem{clausius21serialAE}
J.~Clausius, S.~D{\"o}rner, S.~Cammerer, and S.~ten Brink, ``{Serial vs.
  Parallel Turbo-Autoencoders and Accelerated Training for Learned Channel
  Codes},'' in \emph{2021 11th International Symposium on Topics in Coding
  (ISTC)}, 2021.

\bibitem{tuchler02mmseapriori}
M.~Tuchler, A.~Singer, and R.~Koetter, ``Minimum mean squared error
  equalization using a priori information,'' \emph{IEEE Transactions on Signal
  Processing}, vol.~50, no.~3, pp. 673--683, 2002.

\bibitem{santos_turbo_2018}
I.~Santos, J.~J. Murillo-Fuentes, E.~Arias-de Reyna, and P.~M. Olmos, ``{Turbo
  EP-Based Equalization: A Filter-Type Implementation},'' \emph{IEEE
  Transactions on Communications}, vol.~66, no.~9, pp. 4259--4270, 2018.

\bibitem{martinsson_fast_2005}
P.~G. Martinsson, V.~Rokhlin, and M.~Tygert, ``A fast algorithm for the
  inversion of general {Toeplitz} matrices,'' \emph{Computers \& Mathematics
  with Applications}, vol.~50, no.~5, pp. 741--752, Sept. 2005.

\bibitem{turboep17santos}
I.~Santos, J.~J. Murillo-Fuentes, R.~Boloix-Tortosa, E.~Arias-de Reyna, and
  P.~M. Olmos, ``{Expectation Propagation as Turbo Equalizer in ISI
  Channels},'' \emph{IEEE Transactions on Communications}, vol.~65, no.~1, pp.
  360--370, 2017.

\bibitem{doerner2022jointautoencoder}
S.~Dörner, J.~Clausius, S.~Cammerer, and S.~ten Brink, ``{Learning Joint
  Detection, Equalization and Decoding for Short-Packet Communications},''
  \emph{IEEE Transactions on Communications}, vol.~71, no.~2, pp. 837--850,
  2023.

\bibitem{han23SNNjed}
C.~Han, H.~Zhao, Z.~Chen, and F.~Wang, ``{Sparse Neural Network for Detection
  and Decoding of Non-Binary Polar-Coded SCMA},'' \emph{IEEE Transactions on
  Wireless Communications}, vol.~22, no.~7, pp. 4475--4488, 2023.

\bibitem{wiesmayr2024design}
R.~Wiesmayr, S.~Cammerer, F.~A. Aoudia, J.~Hoydis, J.~Zakrzewski, and
  A.~Keller, ``{Design of a Standard-Compliant Real-Time Neural Receiver for 5G
  NR},'' \emph{arXiv preprint arXiv:2409.02912}, 2024.

\bibitem{ney23cnn}
J.~Ney, C.~F{\"u}llner, V.~Lauinger, L.~Schmalen, S.~Randel, and N.~Wehn,
  ``{From Algorithm to Implementation: Enabling High-Throughput CNN-Based
  Equalization on FPGA for Optical Communications},'' in \emph{Embedded
  Computer Systems: Architectures, Modeling, and Simulation}, C.~Silvano,
  C.~Pilato, and M.~Reichenbach, Eds.\hskip 1em plus 0.5em minus 0.4em\relax
  Cham: Springer Nature Switzerland, 2023, pp. 158--173.

\bibitem{freire23fpgacnn}
P.~J. Freire~et al., ``Implementing neural network-based equalizers in a
  coherent optical transmission system using field-programmable gate arrays,''
  \emph{Journal of Lightwave Technology}, vol.~41, no.~12, pp. 3797--3815,
  2023.

\end{thebibliography}
